\documentclass[12pt]{iopart}

\usepackage{iopams} 
\usepackage{epsfig}
\usepackage{graphicx}
\usepackage{makeidx}
\usepackage{comment}
\usepackage{cite}
\usepackage{hyperref}

\makeindex

\begin{document}

\title[STM imaging of vortex core and lattice]{Imaging superconducting vortex core and lattice with the scanning tunneling microscope}

\author{H. Suderow$^{1,2}$, I. Guillam\'on$^{1,2,3}$,  J.G. Rodrigo$^{1,2}$, S. Vieira$^{1,2}$}

\address{$^1$ Laboratorio de Bajas Temperaturas, Departamento de F\'isica de la Materia Condensada, Instituto de Ciencia de Materiales Nicol\'as Cabrera and Condensed Matter Physics Center (IFIMAC), Universidad Aut\'onoma de Madrid, E-28049 Madrid, Spain}

\address{$^2$ Unidad Asociada de Bajas Temperaturas y Altos Campos Magn\'eticos, UAM, CSIC, Cantoblanco E-28049 Madrid, Spain}

\address{$^3$ H.H. Wills Physics Laboratory, University of Bristol, Tyndall Avenue, Bristol, BS8 1TL, UK}

\ead{hermann.suderow@uam.es}
\begin{abstract}
The observation of vortices in superconductors was a major breakthrough in developing the conceptual background for superconducting applications. Each vortex carries a flux quantum, and the magnetic field radially decreases from the center. Techniques used to make magnetic field maps, such as magnetic decoration, give vortex lattice images in a variety of systems. However, strong type II superconductors allow penetration of the magnetic field over large distances, of order of the magnetic penetration depth $\lambda$. Superconductivity survives up to magnetic fields where, for imaging purposes, there is no magnetic contrast at all. Static and dynamic properties of vortices are largely unknown at such high magnetic fields. Reciprocal space studies using neutron scattering have been employed to obtain insight into the collective behavior. But the microscopic details of vortex arrangements and their motion remain difficult to obtain. Direct real space visualization can be made using scanning tunneling microscopy and spectroscopy (STM/S). Instead of using magnetic contrast, the electronic density of states describes spatial variations of the quasiparticle and pair wavefunction properties. These are of order of the superconducting coherence length $\xi$, which is much smaller than $\lambda$. In principle, individual vortices can be imaged using STM up to the upper critical field where vortex cores, of size $\xi$, overlap. In this review, we describe recent advances in vortex imaging made with scanning tunneling microscopy and spectroscopy. We introduce the technique and discuss vortex images which reveal the influence of the Fermi surface distribution of the superconducting gap on the internal structure of vortices, the collective behavior of the lattice in different materials and conditions, and the observation of vortex lattice melting. We consider challenging lines of work, which include imaging vortices in nanostructures, multiband and heavy fermion superconductors, single layers and van-der-Waals crystals, studying current driven dynamics and the liquid vortex phases.
\end{abstract}

%Uncomment for PACS numbers title message
% Keywords required only for MST, PB, PMB, PM, JOA, JOB? 
%\vspace{2pc}
%\noindent{\it Keywords}: Article preparation, IOP journals
% Uncomment for Submitted to journal title message
%\submitto{\JPA}
% Comment out if separate title page not required

\submitto{\SUST}
\maketitle
%\printindex

\section{Introduction}\index{1. Introduction}

Microscopy opens the world of real space imaging of small sized structures and behaviors. Real space images are sought because it is felt that they improve our understanding, and convey a sense of beauty which is often more difficult to obtain otherwise. They often provide new means of manipulation and device fabrication. The invention of the Scanning Tunneling Microscope (STM) in 1981 by Binnig and Rohrer in IBM Z\"urich is a milestone of microscopy\cite{RohrerNobel}. STM is based on two simple ideas: that electronic vacuum tunneling current depends exponentially on the distance between tip and sample, and that the position of a tip on top of a flat surface can be controlled with subatomic resolution using piezoelectrics\cite{Binnig,BinnigAFM}. When operated at cryogenic temperatures, thermal motion of atoms is reduced or blocked, resolution in energy is improved, and sampled electronic systems show interesting coherent quantum condensate or insulating behaviors. To use STM at low temperatures, it is important that piezoelectrics retain a sizeable motion when cooling. This was not so obvious during the first days of STM \cite{VieiraPiezo}. Now it is clear that, although piezos reduce their range of motion at low temperatures, they are still useful. Past decades have seen an increasing number of STM experiments at low temperatures\cite{Pan00,Song10,Fischer07,Hoffman11}. The field of superconductivity is particularly interesting for STM, because the superconducting gap provides a clear-cut property which can be observed and traced as a function of the position. Of course, the pioneering STM groups realized this point and made first spectroscopic measurements in conventional low T$_c$ and in cuprate high T$_c$ superconductors rather early\cite{deLozanne85,Kirtley87,Kirk88,Vieira88}. However, it was not until the observation of the vortex lattice with beautiful detail in Bell Labs in the early 90’s  that the possibilities of this technique for superconductivity became clear\cite{Hess89,Hess90,Hess91,Hess94}.

Vortices in superconductors were proposed far earlier, when A.A. Abrikosov found that Ginzburg-Landau equations for superconductors under a magnetic field slightly above the lower critical field have periodic solutions if the Ginzburg-Landau parameter exceeds a certain value\cite{A57}. Another prestigious soviet physicist (V.V. Schmidt) writes about this discovery: \textit{Here we come across a rare case in which the development of an extensive branch of the physics of superconductors was initiated by a single theoretical development}\cite{Schmidt}. Abrikosov succesfully compared in his seminal work his predictions to a previous experiment, made far earlier by Shubnikov in 1936\cite{Shubnikow36,Shubnikov37}. The experiment of Shubnikov is the actual breaking ground for applications, because it shows that type II superconductors can carry a supercurrent at high magnetic fields. This lifted the early deception of Kamerlingh Onnes, who pioneeringly suggested high fields as the main application of superconductors\cite{Onnes13}. He could not realize his dream because he was studying most simple elemental Pb and other type I materials, which lose superconductivity at very low magnetic fields\cite{KOChicago,vanDelft10}.

\begin{figure}[ht]
\includegraphics[width=0.9\columnwidth,keepaspectratio, clip]{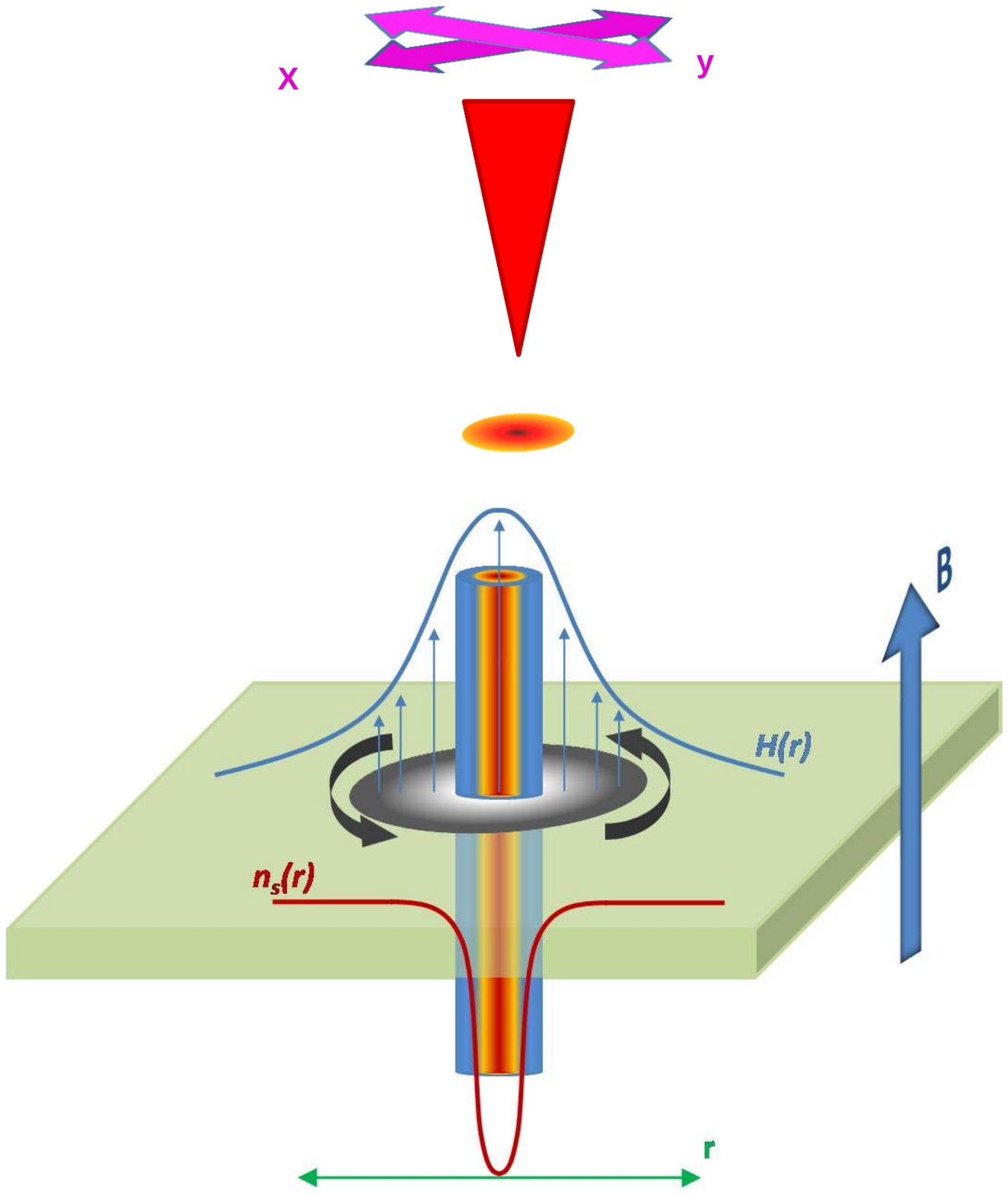}
\vskip -0cm \caption{A vortex consists of a core, where the pair wavefunction (or the superfluid density $n_s$) drops to zero at a scale of the coherence length, which is much smaller than the magnetic penetration depth. Thus, the magnetic field profile extends far above the vortex core. In many strongly type II superconductors, the scale of the magnetic field variation is several orders of magnitude larger than the actual core. Most experiments image vortices using the magnetic field profile. When increasing the magnetic field in type II superconductors, the spatial variation of the magnetic field is smaller than the spatial variation of the superconducting density of states in the core. STM (schematically shown in the figure by the red tip and the x-y scanning mechanism) probes the vortex core by measuring the spatial variation of the electronic density of states. Thus, it is the only real space superconducting vortex imaging technique which can be used at all magnetic fields and temperatures.} \label{Fig1}
\end{figure}

Type II superconductivity survives high magnetic fields by allowing the field to enter the superconductor in the form of vortices, each one carrying a single flux quantum. The vortices are composed of circular supercurrents around a core (Fig.\ 1). The field of type II superconductivity has grown, since the work of A.A. Abrikosov, into one of the most dynamic scientific areas, with technological applications\cite{BookVictor}. The term “vortex matter” was coined and developed during past decades into a thriving area involving fundamentals as well as applications\cite{Brandt95,Blatter94}. The quest to obtain dissipationless current flow is leading to improved metallurgy of superconducting materials\cite{Kunzler61,DewHughes01,Larbalestier01,Llordes12}. In superconducting cables, vortices are firmly pinned through nanosize inclusions of other materials, or by the resulting changes induced by diverse, and often intricate, thermal and mechanical treatment procedures. On the other hand, fluxonics emerges as a new kind of flux based “electronics“, where the control over flux motion is obtained by nanostructuring superconductors\cite{BookVictor,Anders10}. Controlled vortex motion can be used to fabricate pumps, diodes or lenses of quantized flux. Vortex ratchet effects are, for example, being studied to manipulate vortices\cite{Villegas03,Silhanek10}. Nanostructuring allows applications in photonics \cite{Dienst13,Gabay13,Ricci07} and the coexistence with ferromagnetic order unveils vortices in spin-triplet systems, with parallels to helium 3 \cite{Flouquet02,Bulaevski85,Buzdin05,Aoki2001}.

\begin{figure}[ht]
\includegraphics[width=0.9\columnwidth,keepaspectratio, clip]{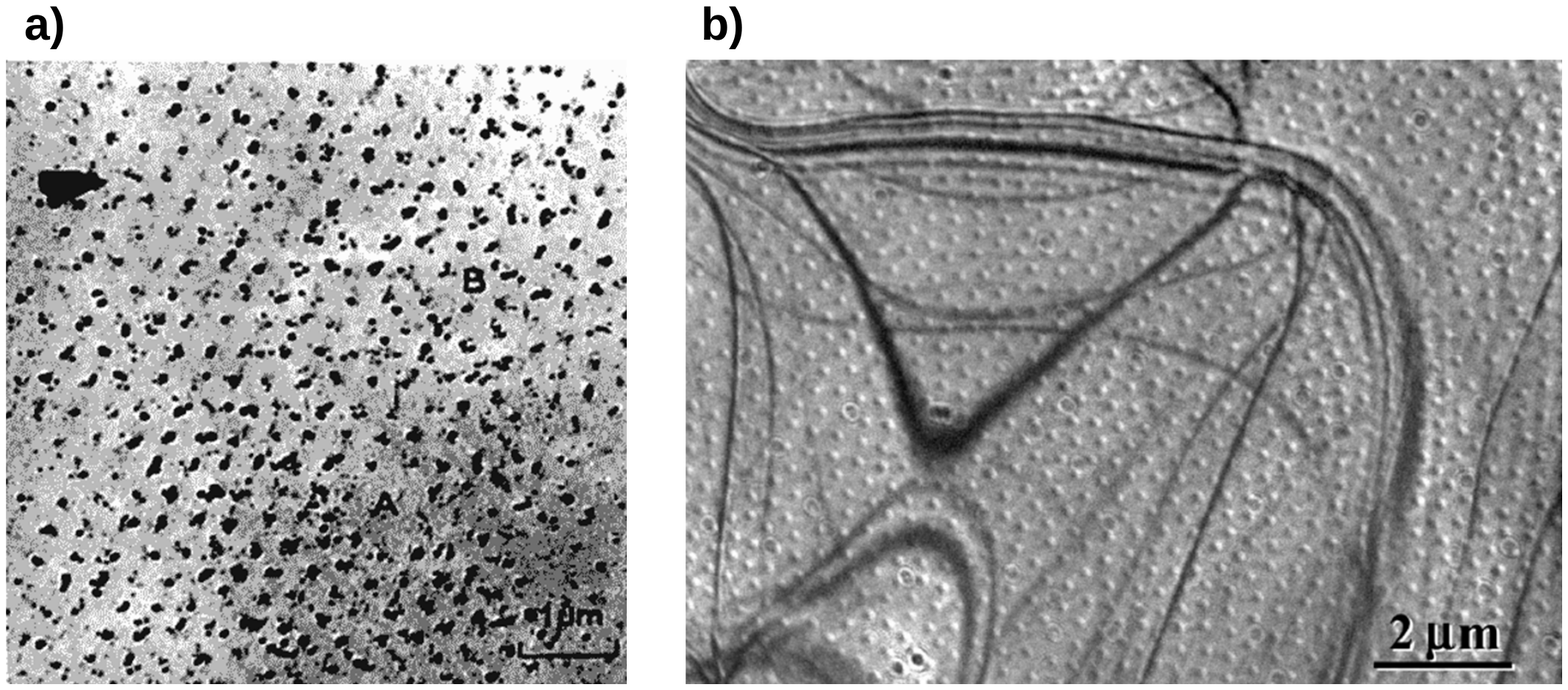}
\vskip -2cm \caption{In a) we show one of the first vortex lattice images taken by magnetic decoration in Pb with 4\% In at 1.1 K (from \protect\cite{Essman67,Trauble68}). A and B are, respectively, a hole and a dislocation. In b) we show Lorentz microscopy image of vortices in Nb at 10 mT and 4.2 K (reprinted with permission of Macmillan Publishers\cite{Harada92}), see also \protect\cite{Harada92,Tonomura99,video}.} \label{Fig2}
\end{figure}

The observation of isolated vortices was pioneered by Essman and Tr\"auble in 1967\cite{Essman67,Trauble68}. They used decoration of surfaces of different superconductors with small magnetic particles at low temperatures and under an applied magnetic field (Fig.\ 2a). In their first work, they observe "elastic" and "plastic" distortions of the vortex lattice in images made at 500 G. Because they work at low magnetic fields, they observe vortex free areas together with distorted lattices. They find dislocations, stacking faults, defects and holes in the vortex lattice, and point out that it interacts strongly with the crystalline defects of the sample. Vortex lattices are a property of the type II superconductor, and as such they appear in all kinds of samples, from single crystals without defects to amorphous materials. The interaction of the vortex lattice with the crystalline lattice, defects, dopants or sample geometry modifies vortex positions and creates pinning. Understanding pinning is fundamental to create functional devices with superconductors.

Decoration experiments have given insight into the behavior of the vortex lattice at low magnetic fields\cite{Fasano08}. This includes flux arrangements, their order and associated transitions, as well as interaction with crystal defects in cuprate superconductors\cite{Fasano08}. The technique is also used to image the dynamic behavior of the lattice under current flow. Complex phases such as the smectic moving phase and re-entrance of hexagonal patterns at high currents have been observed \cite{Pardo98,Marchevsky97,Marchevsky99}. These experiments provide the local magnetic field averaged during the deposition of the magnetic particles.

Time resolved low field imaging of vortices is made using Lorentz-microscopy\cite{Harada96,Tonomura99,Buzdin09} (Fig.\ 2b). The vivid observation of the flux-line lattice dynamics and its interaction with nanostructures have provided unique insight into vortex motion. For instance, the flow along preferential channels, determined by the stiffness of the lattice and the pinning centers, has been directly observed \cite{Harada96}.

\begin{figure}[ht]
\includegraphics[width=0.9\columnwidth,keepaspectratio, clip]{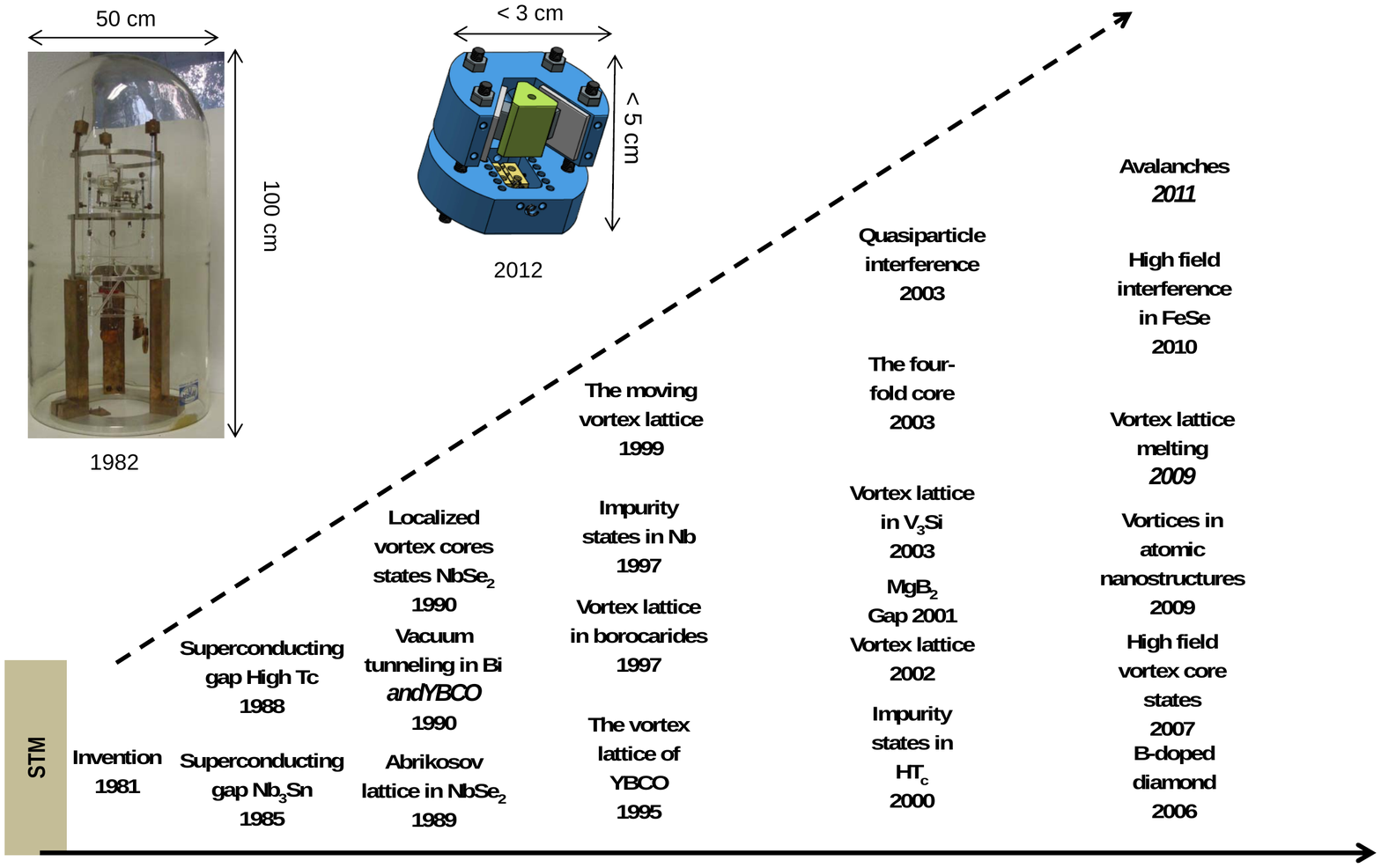}
\vskip -1cm \caption{Since the invention of the STM in 1981 (in the upper left photograph we show one of the first models built), the imaging possibilities have been continuously improved. The second half of the first decade of the century has witnessed the wide application of STM to many subjects. The increased data acquisition and treatment capability, and the improvements in the technique have brought about a change, from the application to a few specific problems, to extended use. Nowadays, smaller microscopes (upper design) are less sensitive to external influences. Cryogenic STM has been applied in several subjects, such as quantum Hall structures or heavy fermions\protect\cite{Song10}. The figure shows a few achievements in superconductivity obtained with a STM. It lists work made in Refs.\protect\cite{Binnig,Vieira88,deLozanne85,Kirtley87,Hess89,Hess90,Kirk88,Troyanovski99,Yazdani97,deWilde97,Maggio95,Hoffman02,Sakata00,Sosolik03,Rubio01,Eskildsen02,Pan00,Guillamon11b,Hanaguri10,Guillamon09Nat,Day09,Guillamon08c,Kaneko12,Sacepe06,Zhang10,Nishio08,Cren09,Cren11,Ning09,Ning10,Tominaga12,Tominaga13} providing results relevant to vortex matter. Further work is described in the text.} \label{Fig3}
\end{figure}

Another interesting time resolved technique is magneto-optics. It can be used to understand field and current distributions over large sized areas, still viewing single vortices. Flux entry has been studied with great detail, from isolated vortices to the whole specimen\cite{Jooss02,Koblischka95}. The interaction between magnetic domains and the vortex lattice has been studied by moving domain walls on magnetic thin film deposited on top of the superconductor \cite{Altshuler04}.

Further low field microscopies such as Scanning SQUID microscopy \cite{Tsuei00,Kirtley10,Finkler12}, Scanning Hall probe microscopy\cite{Bending99} and Magnetic Force Microscopy \cite{Auslaender09} are used to study geometries, dynamics and interactions of vortex matter in different systems. For instance, in multiband superconductors and in nanostructures\cite{Moshchalkov09,Gutierrez12,Banerjee04}. Of particular interest are recent advances in size reduction of SQUIDs. The spatial resolution has been considerably increased allowing vortex observation close to the Tesla range, much higher than any other real space magnetically sensitive vortex imaging technique \cite{Zeldov13}.

STM is, in principle, the only technique able to observe individual vortices at all magnetic fields (Fig.\ 3). The application of STM to superconductivity has been reviewed in several excellent papers\cite{Fischer07,Hoffman11}, focusing on the cuprates and the pnictides respectively. The aim of the present review is to discuss vortex observation with STM, in particular results found in a few single crystaline systems and in thin films. We focus here on materials where vortex observation is particularly enlightening regarding the features of vortex cores and the collective behavior of the whole lattice. 

\section{Theory behind imaging}\index{2. Theory behind imaging}

\subsection{Tunneling.}\index{2. Theory behind imaging!2.1. Tunneling}

The tunneling experiment between two metallic atomically sharp tips is rather complicated to understand in detail, because it requires solving quantum transport phenomena at atomic level. To discuss this problem, let us start by discussing the electrical conduction through a wire of a simple metal with a constriction (Fig.\ 4a). If the constriction is smaller than the electron mean free path, the conduction through such a system is ballistic, and follows Sharvin's conductance formula\cite{RevAgrait,Deutscher05}:

\begin{equation}
G=\frac{2e^2}{h}\left(\frac{k_F a}{2}\right)^2
\end{equation}

with $G$ the conductance, $e$ electron charge, $h$ Planck's constant, $k_F$ Fermi wavevector and $a$ the radius of the contact. When the wire is pulled, the contact becomes smaller and smaller, until it reaches atomic size (Fig.\ 4b). Then, quantum effects dominate conduction. The conduction can be described using Landauer's formula and the concept of conduction channels, each one with a given transparency $T_i$. The amount of channels available in a single atom point contact and their transparency can be related to the electronic structure of the atom or molecule making up the contact\cite{Setal97,Setal98,Suderow00c,Rodrigo04b}. $T_i$ is of order one for metallic elements, such as gold, lead or aluminum\cite{CML98,BG02}. The conductance is written as 

\begin{equation}
G=\frac{2e^2}{h}\sum\limits_i{T_i}
\end{equation}

Conductance vs. distance curves give well-defined steps at $\frac{2e^2}{h}$ in case of a monovalent simple metal as Au (Fig.\ 5), and structured features which can be related to the atomic orbitals (Fig.\ 4b) in case of more complex metals as Al, Pb or Nb\cite{RevAgrait}.

When the wire is broken into two pieces, the conduction process is no longer determined by the single contacting atom or molecule, but by the outermost coupling between electronic wavefunctions of the two parts (Fig.\ 4c). The effect on each other is nearly perturbative, resulting in an exponential dependence of the current with respect to the distance between both atoms at the apex\cite{Bardeen61,Chen}. The transparency of the junction is thus small, and it seems reasonable to assume that only one channel contributes to the conduction process. Bardeen's transfer Hamiltonian formalism can be used to calculate the resulting conductance, which depends on the way the outermost atom's wavefunctions couple to each other\cite{Bardeen61}.

An STM tip can be viewed as one part of such a sharp broken wire which is moved on top of the flat surface of a sample (Fig.\ 4d). The junction is then made between two different materials, and the conduction properties are determined by the interaction of the last few atoms of the tip with the surface.
 
\begin{figure}[ht]
\includegraphics[width=1.1\columnwidth,keepaspectratio, clip]{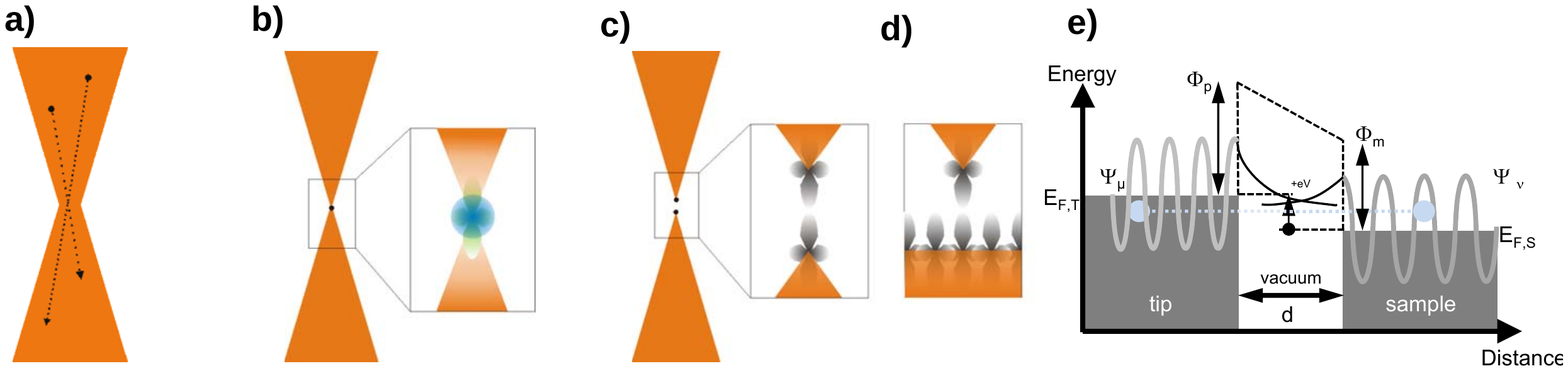}
\vskip -2.5cm \caption{In a) we schematically represent ballistic transport through a junction of size smaller than the bulk electron mean free path. In b), we show schematically a single atom junction. Transport is quantum-mechanical and dominated by the shape of the atomic orbitals of the contacting atom. When both ends are separated (c), transport is from the perturbative coupling between wavefunctions of outermost atoms of both ends. In a STM, the latter is realized (d) by moving an atom on the apex of a tip over the surface. Positional changes in the transport can be understood by tip-sample wavefunction coupling, as shown in e.} \label{Fig4}
\end{figure}

The measured tunneling current depends thus on the way outermost tip atom couples to sample's wavefunctions at each position (Fig.\ 4d). Using Bardeen's transfer Hamiltonian formalism, Tersoff and Hamann found, assuming s-wavefunctions, that the current is proportional to the sample local density of states at the position of the tip\cite{Tersoff83,Tersoff85}, provided the bias voltage is very close to zero. For non-zero bias voltages, one finds

\begin{equation}
I(V) =\frac{G_N}{e} \int^{eV}_{0} dE [f(E-eV)-f(E)] N_T(E-eV) N_S(E) T(E,eV)
\end{equation}

with $f(E)$ the Fermi function, $N_T$ and $N_S$, respectively, the tip and sample electronic densities of states and $T$ the tunneling matrix element. $N_T(E)=constant$ in many surfaces of simple metals close to the Fermi level. Thus, for measurements made at low bias voltages, $N_T(E)$ can be taken out of the integral. It has been shown that localized states formed at the junction can be neglected\cite{Tekman88}, as well as the energy dependence of $T$. Then, the conductance is proportional to the sample's density of states\cite{Sacks91,Yang93}, convoluted by the derivative of the Fermi function :

\begin{equation}
G(V) \propto \int^{eV}_{0} dE \frac{df}{dV} N_S(E)
\end{equation}

Until now, except in a few cases which we detail below, nearly all authors working on vortex observation in superconductors with STM base the interpretation of their data on this model. The model holds often when stuyding low energy phenomena. However, it gives a simplified account of the tunneling experiment. We believe that future vortex observation will  take advantage of the full information that one can obtain using tunneling spectroscopy. Therefore, we give an extensive account of other tunneling spectroscopy methods and discuss briefly the first experiments where these have been used.

Sometimes, broken wires of Au are used to make normal tips. The mechanical annealing method described in \cite{Rodrigo04} leads to sharp Au pyramids which are useful for scanning topography and spectroscopy. There are of course many other methods to obtain suitable STM tips. Popular tips are of W or Pt-Ir, which are relatively hard materials that can be shaped using field emission. These methods are well described in literature\cite{Fischer07,Song10}.
 
\begin{figure}[ht]
\includegraphics[width=0.9\columnwidth,keepaspectratio, clip]{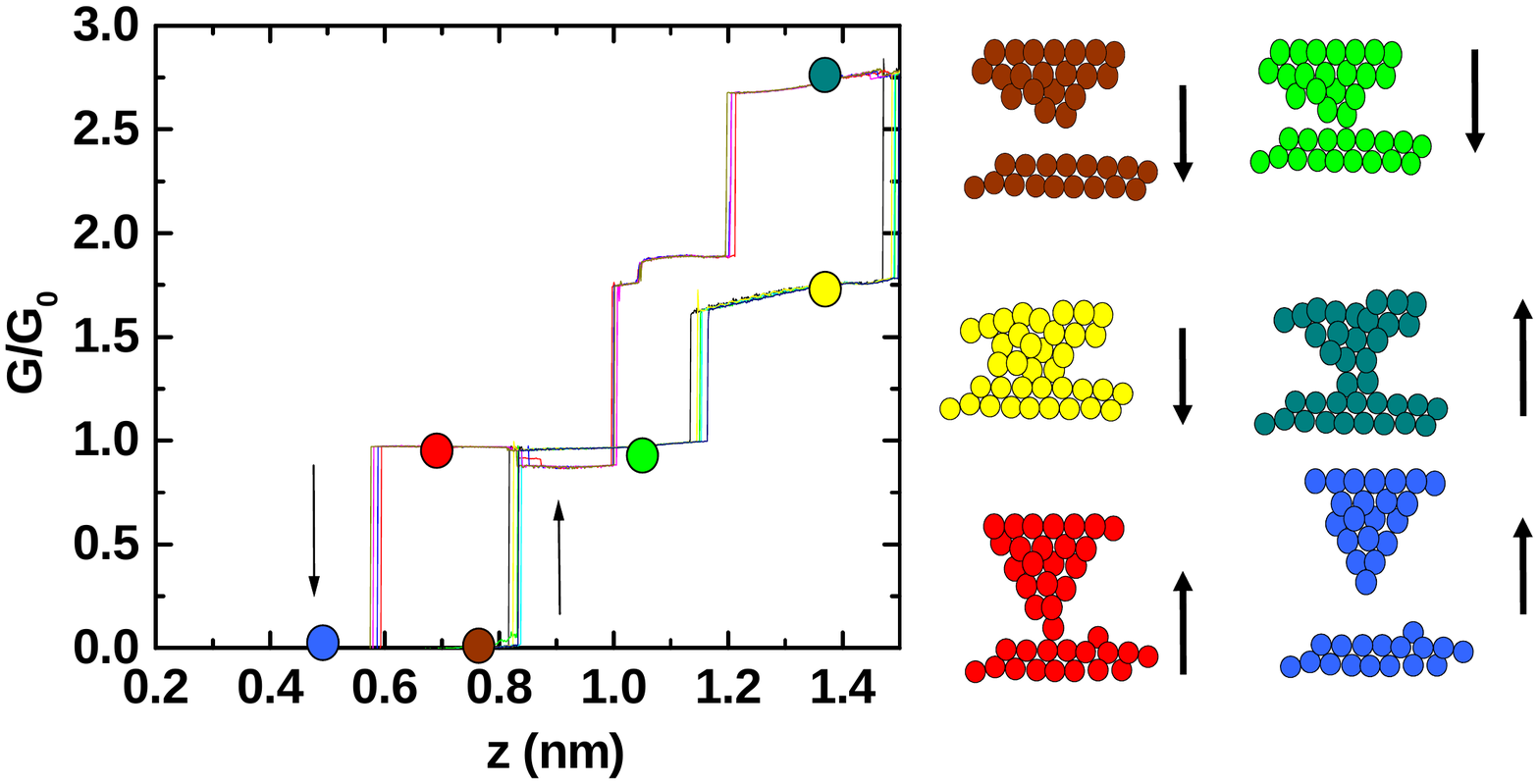}
\vskip -2cm \caption{Conductance ($G$ in units of $G_0=\frac{2e^2}{h}$) as a function of the tip's displacement when entering and leaving the few atom point contact regime in Au. Possible atomic arrangements are shown in schematically at the right of the figure. Contacts larger than the single atom case give Sharvin's resistance, see Ref.\protect\cite{Rodrigo04b}, and are needed to fabricate long and sharp tips for tunneling.} \label{Fig5}
\end{figure}

\subsection{Superconducting tips.}\index{2. Theory behind imaging!2.2. Superconducting tips}

Superconducting tips can be obtained by breaking wires of Al or Pb  and mechanically annealing them as shown in Ref.\cite{Rodrigo04}. Other methods use evaporation of Pb on W or Pt-Ir tips or a small superconducting single crystal attached to the tip apex\cite{Stip1,Stip2,Noat10,Bergeal08,Ji08,Franke12}.

When the tip is superconducting with a critical temperature T$_c$ of the same order as sample's T$_c$, the energy dependence of its density of states $N_T(E)$ cannot be neglected, and one has to use Equ.3. $G(V)$ is not proportional to $N_S(E)$, and the Fermi function does not enter as a smearing parameter. If $N_T(E-eV)$ is just the Bardeen Cooper Schrieffer (BCS\cite{BCS57}) density of states with a gap of the same order but smaller than the sample's gap $\Delta_T<\Delta_S$, then the current will be zero for $eV<\Delta_1=|\Delta_T-\Delta_S|$, except for thermally excited quasiparticles. Exactly at $eV=\Delta_1$, and at a finite temperature, the current peaks due to the quasiparticle peak of the tip's density of states. For $eV>\Delta_1$, the current decreases, and then it increases abruptly at $|eV|>\Delta_2=\Delta_T+\Delta_S$. The size of the corresponding jump in current decreases when increasing temperature, but the conductance remains practically divergent at $\Delta_2/e$. The in-gap conductance shows accordingly a temperature induced exponential increase for $eV<\Delta_1$, a peak-dip structure, with negative conductance for $eV\approx\Delta_1$ and a sharp, roughly temperature independent, peak at $eV=\Delta_2$. These features are reviewed in several books and papers\cite{Wolf,T96,Rodrigo04b,Rodrigo03,Rodrigo04,Guillamon07}.

The effect on the tunneling current of the quasiparticle peak of the superconducting tip is close to the effect obtained by substituting $N_T(E-eV)$ with a $\delta$ function located at the gap edge of the tip. With a $\delta$ function like tip density of states, the current  traces the density of states of the sample. Temperature induced smearing is absent, as long as the $\delta$ function divergence remains\cite{Crespo12}. A superconducting tip is similar to such a tip, in that the quasiparticle peak at the gap edge is usually high and narrow (the width depends on quasiparticle lifetime and gap anisotropy\cite{Dynes78}). Of course, the resulting current is also affected by the quasiparticle continuum above the gap edge, and one obtains the bias voltage dependence discussed in Ref.\cite{Crespo12}.

The increase in resolution in the spectroscopy has been used to obtain results obtain with superconducting tips in Refs.\cite{Crespo09,Rodrigo04PhysC,Ji08,Shimizu10,Franke12,Noat10,Petal98,Xu03,Stip1,Stip2,Stip3,Stip4,Stip5,Rodrigo03,Rodrigo04,Martinez03,Kohen06,Maldonado12URu2Si2,Saha10}. For example, in Ref.\cite{Crespo09}, the reduction of temperature induced smearing was relevant to determine the temperature dependence of the sample's density of states in ErRh$_4$B$_4$. In 2H-NbSe$_2$, it was used to determine the peculiar temperature dependence of the multigap structure. Smallest gaps are the most affected by temperature and they close well below $T_c$\cite{Rodrigo04PhysC}. This result has explained the apparent discrepancy between angular resolved photoemission (ARPES) and London penetration measurements about the presence or not of a superconducting gap in the small 3D pocket derived form Se bands. Whereas ARPES does not observe gap opening in this band at temperatures very close to $T_c$\cite{Yokoya01}, London penetration depth measurements at much lower temperatures \cite{Fletcher07} did show evidence for the presence of a sizeable gap opening in the Se band. Furthermore, the spatial variation of the quasiparticle peaks obtained using superconducting tips give a length scale, termed Doppler shift length scale, related to the screening currents around vortex cores, and thus more to $\lambda$ than to $\xi$ \cite{Kohen06}. The increased energy resolution is also useful when it is difficult to achieve very low temperatures\cite{Noat10}. In single magnetic adatoms and deposited molecules, superconducting tips have helped to determine local spectroscopic features\cite{Ji08,Franke12}. 

The magnetic field dependence of the superconducting tip has been explored in Refs.\cite{Rodrigo04b,Misko01}. Most of the tips used are type I superconductors, such as Pb or Al. Generally, it is found that the pyramidal structure of the tip gives an upper critical field exceeding by several times the bulk critical field of the tip's material\cite{Metal95}. If the pyramid is long enough, superconductivity disappears in the bulk and remains at the tip's apex\cite{Petal98,Misko01,Suderow02,Guillamon08}.
 
Below the superconducting gap the current is not exactly zero, even at zero temperature (Fig.\ 6). Multiple Andreev reflection leads to a finite conductivity. Exactly at zero bias, the Josephson current appears. Both transport phenomena require multiple passage of the tunnel barrier, and are thus further exponentially reduced with respect to the single particle tunneling. Nevertheless, these effects have been observed by approaching the tip to the sample and reducing the tunneling barrier to the minimum possible prior to enter into contact\cite{Stip2,Stip3,Stip4,Kohen06,Guillamon08,Crespo12}. Both effects have been used to image the vortex lattice, in particular in NbSe$_2$ (see also section 4.1.2).

Andreev reflection spectroscopy has been made at a single location between tips and samples of superconducting materials. For instance, multiple Andreev reflections between electrons in $\pi$ and $\sigma$ bands of MgB$_2$ were observed in Ref.\cite{Martinez03}. Andreev current is also observed below the gap in a superconducting tip on top of a normal region of the sample. Andreev spectroscopy imaging has been made in NbSe$_2$ (see below Fig.15) and is a new imaging method which is been developed. It can give information about local Andreev conduction behavior, and in particular lead in future to spin sensitive imaging, as the Andreev process with a conventional s-wave superconductor at the tip requires spin flip \cite{Crespo12}.

\begin{figure}[ht]
\includegraphics[width=0.9\columnwidth,keepaspectratio, clip]{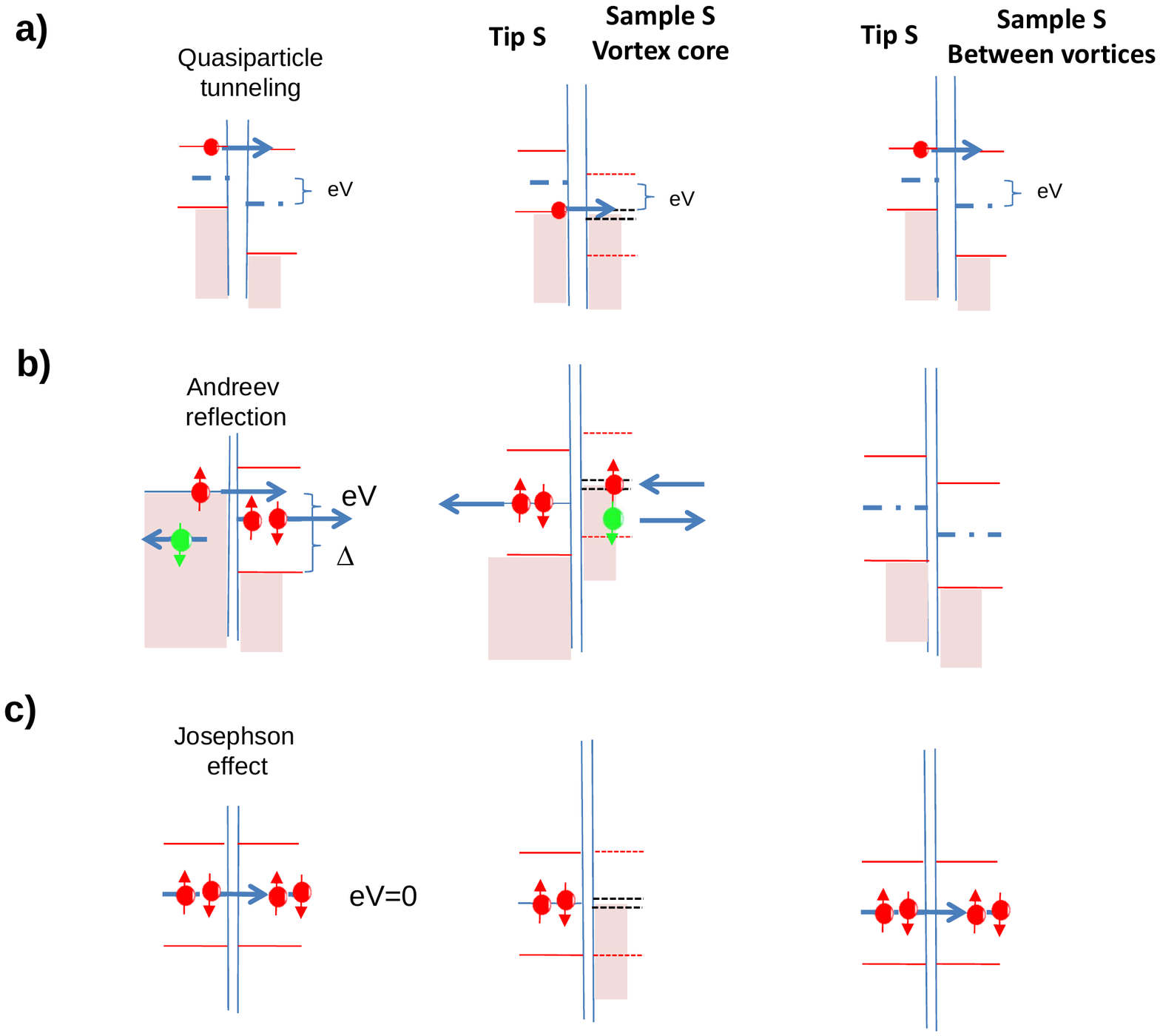}
\vskip -0cm \caption{Cartoon figure of possible tunneling phenomena as observed with a superconducting tip on a superconducting sample with vortices at zero temperature. a) represents possible quasiparticle tunneling process (left panel). Middle panel represents the situation when the superconducting tip is on top of a vortex core. Occupied states below the gap can tunnel into the sample's core states (blue arrow). Right panel shows the situation between vortices. Only tunneling for voltages larger than the sum of both gaps is possible. In b) we show the case of Andreev reflection (left panel). Within the core, states at the core levels enter the superconductor below the gap through Andreev reflection (blue arrows, red circles represent filled and green empty states). In between vortex cores, in the supeconductor (right panel), no Andreev reflection is possible. Multiple Andreev S-S reflections are of course allowed, but they imply higher order tunneling processes which give a smaller contribution to the current. Finally, Josephson tunneling (c) is only possible outside the core (right panel).} \label{Fig6}
\end{figure}

In Ref.\cite{Stip2}, the Josephson effect at a single location, and its dependence as a function of the tip-sample distance, was studied in detail. It was shown that the Josephson effect is considerably reduced due to fluctuations with respect to the values in junctions of larger size\cite{Goffman00,Proslier06}. Similarly reduced Josephson current is generally present in later STM work using superconducting tips. Vortex lattice observations by mapping Josephson signal was made in V$_3$Si using a tip of MgB$_2$\cite{Bergeal08}, and as a function of the position under magnetic fields in a thin film of MgB$_2$ Ref.\cite{Proslier06}. It was shown that, in V$_3$Si, the length scales involved in vortex lattice observation from Josephson effect and quasiparticle conductance are very similar. Experiments in NbSe$_2$, described below, point out that in NbSe$_2$ there may be some relevant differences.

All this work is building up a new spatially dependent spectroscopic mode, Scanning Josephson Spectroscopy\cite{Rodrigo04,Proslier06}, which can be used to map Cooper pair density, instead of the quasiparticle density of states. Its further development and application to nanostructures and unconventional superconductors, such as heavy fermions and non-centrosymmetric systems, is to be followed and can provide insight which seems impossible to obtain otherwise. For example, the Josephson coupling between an s-wave tip and a p-wave or d-wave superconductor is strongly position dependent\cite{Strand10}, and is one of the few direct probes for the occurence of such complex superconducting wavefunctions.

\subsection{Multiple tunneling.}\index{2. Theory behind imaging!2.3. Multiple tunneling}

When the sample holds resonant excitations close to the surface, such as adsorbed Kondo ions, single particle tunneling does no longer lead to a correct description of the experiment, because the tunneling probability into states with different kinds of electronic wavefunctions is of the same order\cite{Ternes09,Ernst11}. Losely speaking, tunneling into a magnetic ion on a conducting lattice can occur either into the resonant magnetic states, or into the delocalized electron background. Tunneling electrons into the two different channels interfere, producing a Fano lineshape which is close to the energy level of the resonant states. If this level is close to the Fermi level, it produces asymmetry in the tunneling conductance and the apperance of peaks or dips at energies related to the resonant state's position and width. The preferential tunneling channel and the nature of the interference determine the form of the Fano resonance\cite{Ternes09}. A tip coupling preferentially to de-localized states gives a flat conductance with a dip at the resonant level due to destructive interference. A tip coupling to resonant levels gives a sharp peak at the resonant level. The work of \cite{Bork11} shows a nice example of the control over the resonant levels and their coupling to the itinerant electronic properties by changing the tip-sample distance.

To our knowledge, vortex observations taking into account multiple tunneling have not been made until now. Interesting candidates are vortices in systems showing Coulomb blockade; proximity induced structures with organic compounds having low energy levels \cite{Cottet12}; or heavy fermions where the Kondo energy is of order of the superconducting critical temperature\cite{FlouquetRoad}. 

\subsection{Asymmetric tips.}\index{2. Theory behind imaging!2.4. Asymmetric tips}

Metals show at the surface electronic wave patterns\cite{Fischer07,Hoffman11,Crommie93,Burgi02,Simon11} due to interference effects associated with scattering by defects or steps at the surface. The properties of the associated charge modulations at the surface can be obtained by Fourier transform of the real-space images. Wavevectors are generally incommensurate with the lattice and dispersive. Their changes as a function of the bias voltage allow tracing the electron dispersion relation\cite{Crommie93,Burgi02,Simon11}. 

Often, a small atomic size effect due to tunneling matrix elements is also found\cite{Fischer07,Hoffman11}. Superconductivity is expected to be homogeneous below the coherence length. But atomic size changes of superconducting local density of states can appear if the sample shows anisotropic gap structure. In Ref.\cite{Guillamon08PRB}, an atomic size modulation in the density of states was found in 2H-NbSe$_2$. Small changes of the form of the tunneling density of states are detected, depending on the atomic location of the tip. It was shown that anisotropic and multiband superconducting properties together with an anisotropic tip-sample coupling can explain the atomic size modulation of the superconducting density of states. This can be used to study the gap anisotropy of the sample. The anisotropic coupling was modelled by assuming an ellipsoidal tip apex (Fig.\ 7). Depending on the relative orientation between tip's wavefunctions and sample's gap anisotropy, a modulated superconducting gap appears when describing a circle around a given position. This leads to the observed atomic size modulation. Authors of Ref.\cite{daSilva13} have revised the effect of an anisotropic tip in pnictide and heavy fermion superconductors. They have shown that the conductance maps can vary depending on the anisotropy of the interaction between tip and substrate.

As we discuss below, the atomic dependence of vortex core tunneling density of states features depends on the asymmetry of the tip \cite{Guillamon08PRB} and carries information about band dependent electronic vortex core properties.

\begin{figure}[ht]
\includegraphics[width=0.9\columnwidth,keepaspectratio, clip]{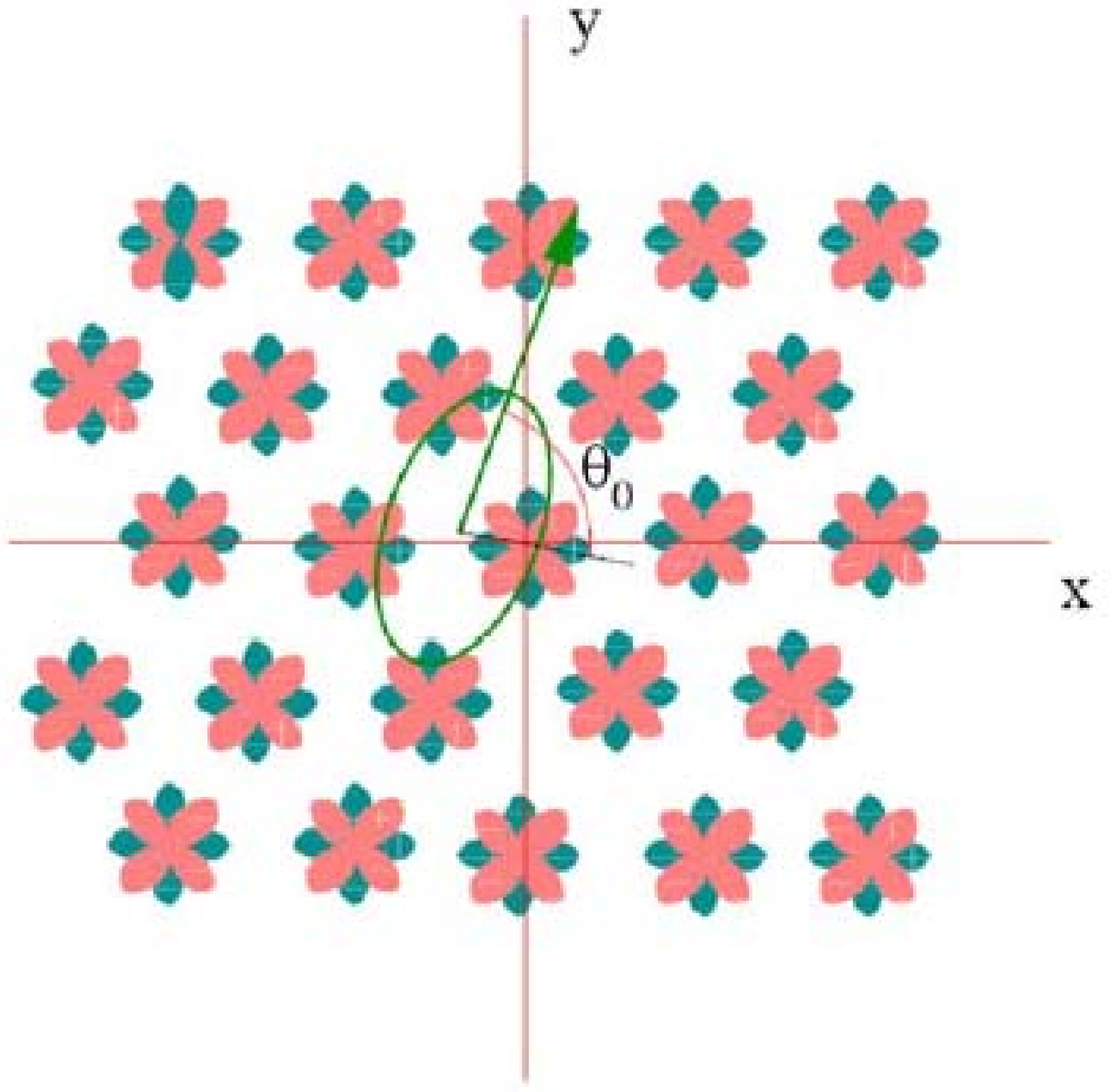}
\vskip -0cm \caption{Schematics of the model used to obtain an atomically modulated superconducting local density of states. The ellipse shows the anisotropic of the tip-sample interaction, and Nb orbitals ($d_{x^2-y^2}$ and $d_{xy}$ are shown as green and red structures). Adapted from Ref.\protect\cite{Guillamon08PRB}. Copyright (2008) by The American Physical Society.} \label{Fig7}
\end{figure}

\section{Experimental}\index{3. Experimental}

\subsection{Design and construction.}\index{3. Experimental!3.1. Design and construction}

A. Tonomura, who developed transmission electron microscopes into impressive probes of fundamental physics, stated \textit{you have to develop new equipment when you attack a new problem}\cite{Harada12}. Efforts in scanning probe microscopes are widespread across many groups, often specifically adapting designs for the particular problem to be addressed. For vortex imaging at high magnetic fields, instrumental work has focused on achieving the lowest temperatures, being able to measure the conductance with precision and over large areas, and controlling the size and orientation of the magnetic field. The present low temperature limitation of scanning probe microscopy has not fully reached the actual lowest temperature where superconductors are generally studied using macroscopic techniques (a few tens of mK\cite{Suderow98,Seyfarth06}), athough it has come close to it. Several STM designs for millikelvin temperatures have been reported\cite{Song10,Assig13,Singh13,Zhang11,Suderow11,Moussy01,Pan99,Marz10,HanaguriWeb}. Dilution refrigeration is unique in providing a cold point stable in time that can cool down large devices to temperatures which are one or two orders of magnitude below 1 K. The environment of dilution refrigeration is, however, rather inadequate for STM, due to the vibrations usually associated to its operation\cite{Kirichek02,Gorla04,Hudson96,Assig12}. Some improvements in the cryogenics have been proposed, such as eliminating the 1 K pot bubbling \cite{Pan99,Libioulle03}, by substituting it with a capillary or by using a large 1 K pot which can be closed using a needle valve. But mechanical vibrations remain unavoidable in operating a STM in a dilution refrigerator, and they lead to changes in the position of the probe with respect to the sample.

\begin{figure}[ht]
\includegraphics[width=0.9\columnwidth,keepaspectratio, clip]{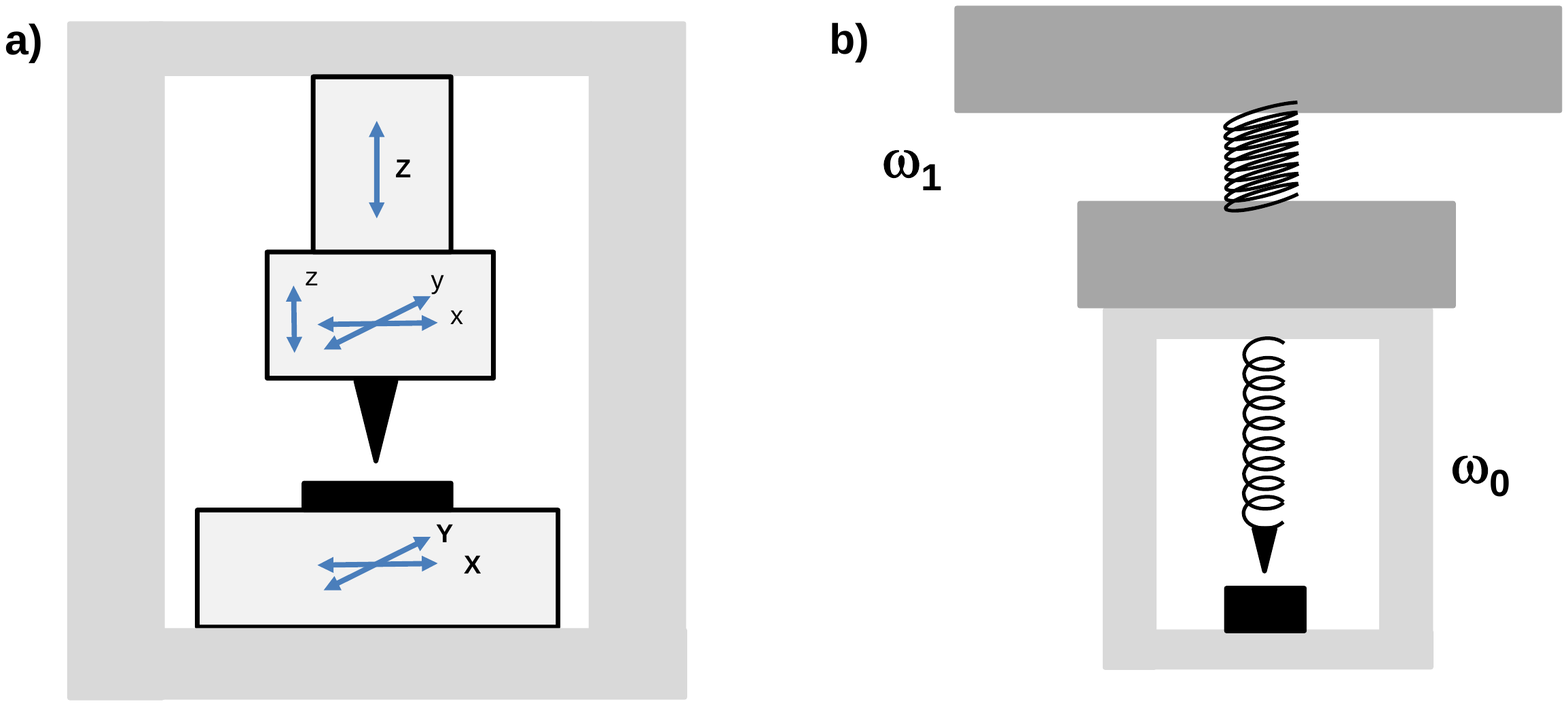}
\vskip -3cm \caption{Schematic arrangement of a STM. It consists (a) of a coarse positioning system, firmly fixed through a frame to a scanning system. Ideally, the whole set-up should have a high resonance frequency (b, $\omega_0$) and be coupled to the outside using a low resonance frequency system (b, $\omega_1$). In this way, outside vibrations minimally influence the tip-sample system.} \label{Fig8}
\end{figure}

A STM requires a piezotube\cite{Chen93,Pohl} and a coarse motion system allowing to change the scanning window\cite{SuderowPrague}. Both positioning systems are firmly fixed to the same frame (Fig.\ 8). Noise induced by mechanical vibrations should be of the order of or less than noise levels in the measurement of the current. Typical current noise levels correspond to height changes of the order of the pm. Vibrations should be of the order or below this value\cite{Libioulle03,Smit07}. The resonant frequency of the STM head should be as large as possible, and the one of the supporting assembly as low as possible. Then, low frequency vibrations, unavoidable in practice, move the tip-sample system in phase, but their relative position remains stable. To reach high resonance frequencies, the frame of the STM head can be built of Ti or Al alloys, or in Macor, with large stiffness and small weight (see table I).

\begin{table}[ht]
     \begin{tabular}{ | p{4cm}| p{2.5cm} | p{2.5cm} | p{2.5cm} |}
     \hline
     \textbf{Material} & \textbf{Density} (kg/m$^3$) & \textbf{Young Modulus} (GPa) & \textbf{Ratio $\times$ 100}\\ \hline
     Ti Grade 5 (Ti6Al4V) & 4420 & 110  & 2.4 \\ \hline
     Al 7075 & 2700 & 70 & 2.5\\
     \hline
     Macor & 2520 & 66 & 2.6 \\ \hline
     Shapal & 2900 & 190 & 6.6 \\ \hline
     Sapphire ($\alpha$-Al$_2$O$_3$) & 3980 & 340 & 8.5 \\
     \hline
     WC & 15500 & 550 & 3.5\\
     \hline 
\end{tabular}
\caption{Materials often used in STM construction. Ideally, low weight and high strength lead to a high resonance frequency. Data are from different sources. The first two materials are good metals, which are easily machinable. Rest are ceramics. Macor bears problems with de-gasing, which are solved in Shapal. But both are bad thermal conductors at low temperatures. This is solved either by using the wires to thermalize, or using sapphire, which is a good thermal conductor. Sapphire is difficult to work with, although it has highest Young modulus vs density ratio.} 
\end{table}

\begin{figure}[ht]
\includegraphics[width=0.9\columnwidth,keepaspectratio, clip]{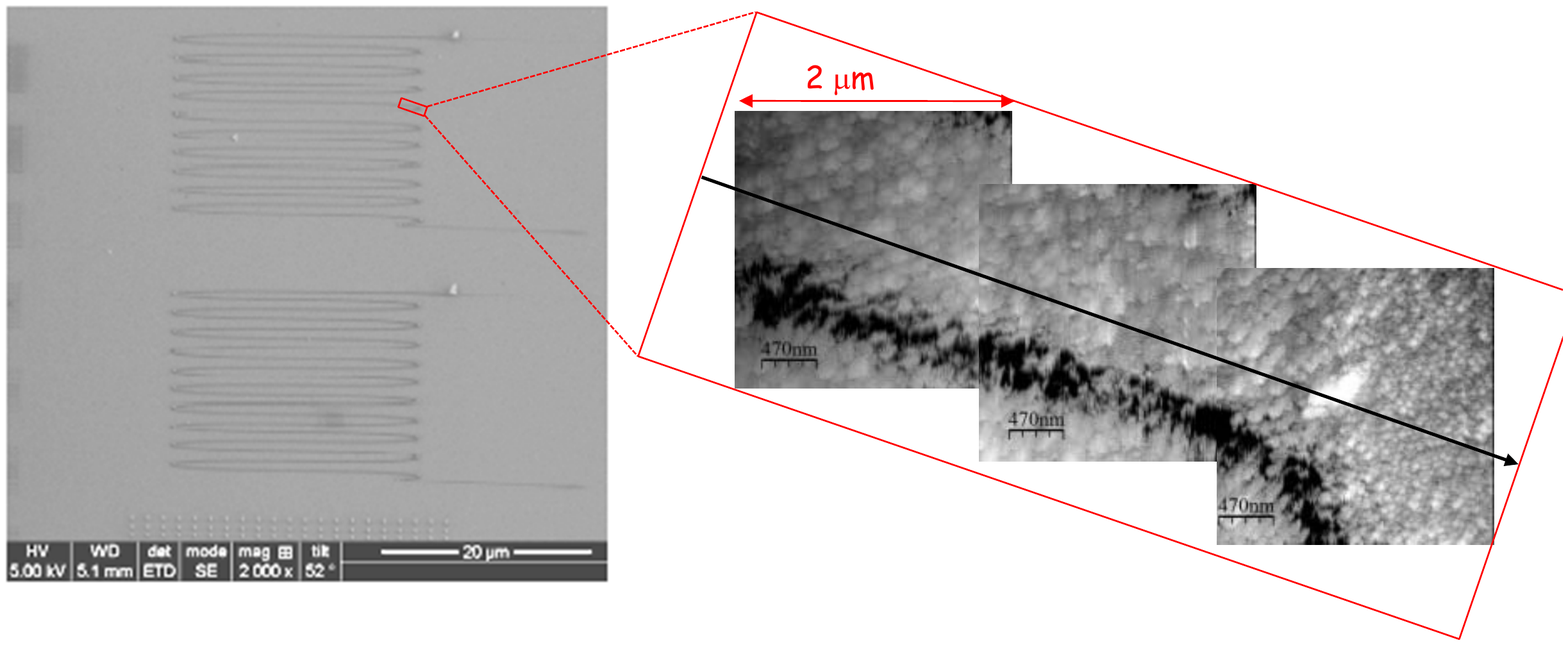}
\vskip -0cm \caption{Scanning electron micrograph of a path milled using focused ion beam on an Au sample (left panel). The path is found, at 100 mK, using STM, and imaged in the right panel. Scanning window is of nearly $2\mu$m. The subsequent images are taken after moving the sample holder in-situ. Adapted from Ref.\protect\cite{Suderow11}.} \label{Fig9}
\end{figure}

The energy resolution in the tunneling spectroscopy is limited by the voltage resolution of the bias voltage source. At 100 mK, the thermal energy limits the resolution to 8.7$\mu$eV. But this is rather difficult to achieve for the bias voltage source, considering the requirement of maintaining whole bandwidth, from DC up to tens or hundreds of kHz. Reports in literature range from 15$\mu$eV\cite{Suderow11}, to down to 2$\mu$eV \cite{LeSueur06}, or 11$\mu$eV\cite{Assig13}. It is not easy, however, to determine accurately the actual resolution in energy, and thus the lowest temperature of the spectroscopic measurements. The best method needs a clean junction between tip and sample of the same superconducting material, ideally a most simple s-wave superconductor, and temperatures far below its critical temperature. This gives the highest quasiparticle peak height and narrowest peak width which can be obtained using superconducting tunneling\cite{Wolf}. When coming to ultimate resolution, gap anisotropies or lifetime broadening effects due to phonons produce additional smearing\cite{Rodrigo03,Rodrigo04,Rodrigo04b}. Thus, the obtained resolution is always a conservative estimate. Al-Al tunneling in Ref.\cite{Rodrigo04} gives, for example, the 15$\mu$eV mentioned above. Combinations of normal tip and superconducting sample or viceversa, are not adequate, because it is difficult to be sure that the normal electrode is free from small pieces of the superconducting counterelectrode. This can artificially increase the resolution and lead to optimistic measurements. It is thus difficult to make a quantitative analysis of the obtained resolution. However, by now several experiments in widely different systems give sharp features in the tunneling conductance which show that an energy resolution in the spectroscopy close to ten $\mu$eV has been reached.

Sample and tip preparation methods are described in several articles\cite{Song10,Assig13}. Using a connection to a UHV preparation chamber one can study systems which can only be grown and handled in UHV. An increasing number of groups is working on UHV systems to address interesting vortex properties of systems which can only be grown in UHV conditions, such as few atom thick islands of Pb\cite{Nishio08,Cren09,Ning10,Tominaga12,Tominaga13}. In many other materials, among them most superconducting compounds, one can devise methods to obtain fresh surfaces in-situ. A cold enviroment below 1 K gives best possible vacuum conditions, because the vapour pressure of all chemically active elements and molecules is zero for practical purposes (the one of helium 4 and helium 3 remains of course finite but drops very fast). One option to prepare sample's surface is a stick catching the sample holder out of the STM and moving it, from outside the vacuum enclosure into a place (also cold or in UHV) where the sample can be broken or cleaved\cite{Song10,Assig13,Singh13,Zhang11,Pan99}. Such a system requires large samples. The holder has to be moved large distances, and it is difficult to make sure that it will be inserted in the STM at exactly the same position. Usually, samples have to be larger than a hundred micron or so.

Small samples are better broken or cleaved using a system which moves only a few mm. This allows precise positioning of the tip over the sample. The movable sample holder described in Ref.\cite{Suderow11} is one option. It consists of a sliding sample holder. A pulling rope moves the holder and breaks the crystal in-situ, exposing a fresh surface. The sample holder is then released to its original position below the tip \cite{Suderow11,Crespo06a,Guillamon10}. Samples of several tens of $\mu$m have been measured using this technique. Moreover, the same system is used to change the scanning window. The precision is in the $\mu$m range, and can be used to locate nanostructures on the sample by imaging at different positions(Fig.\ 9).

\subsection{Basic vortex imaging techniques.}\index{3. Experimental!3.1. Vortex imaging techniques}

The intervortex distance in the triangular Abrikosov lattice is given by

\begin{equation}
d=(0.75)^{1/4}\left(\frac{\Phi_0}{B}\right)^{1/2}
\end{equation}

where $\Phi_0$ is the flux quantum and $B$ the magnetic field (in Tesla). As a rule of thumb, vortices are separated by 50 nm at 1 T. Generally, experiments have been made with the aim either to study vortex cores or to identify vortex positions as a function of the magnetic field or temperature. Corresponding technical requirements are different.

To study vortex cores, highest energy and spatial resolution is needed. Measurements need to be made at lowest temperatures, aquiring the full I-V curve with atomic or even sub-atomic resolution. For example, if a single vortex is measured, in a square of size $50 nm \times 50 nm$, a mesh of $256 \times 256$ points gives a point each 0.2 nm. To obtain at the same time atomic resolution, the mesh needs to be doubled at least, which implies two times more points and a corresponding increase in the time needed to acquire the image.

To obtain the position of a large amount of vortices at many magnetic fields and temperatures, it is useful to find the bias voltage range where the conductance measurement in and out the vortex core gives maximum contrast. Early work carried out measurements of the quasiparticle peak height by tracing the conductance at the gap edge as a function of the position\cite{deWilde97} in the nickel borocarbides. Also, the work of Ref.\cite{Troyanovski99,Troyanovski02} has used the STM in topography mode at the bias voltage where maximum contrast is obtained in the current, which is close to the quasiparticle peak. As mentioned by these authors, when increasing the magnetic field, the contrast is reduced and images are washed out. A compromise must be found between imaging speed and resolution\cite{Troyanovski99,Uchiyama10}, depending on the experiment to be made. Taking data at a single position as a function of time may be useful to study some dynamic behavior\cite{Kohen05}. For imaging, highest resolution is obtained by taking the whole I-V curve, which gives good images up to T$_c$\cite{Guillamon09Nat}. Highest speed is obtained by imaging at fixed bias voltage\cite{Troyanovski99,Uchiyama10}.

\section{Imaging vortex matter}\index{4. Imaging vortex matter}

\subsection{The vortex core.}\index{4. Imaging vortex matter!4.1. The vortex core}

\subsubsection{Introduction.} Tracing the spatial dependence of the tunneling conductance unveils local electronic properties of vortices. The vortex is in itself a rather complex object, where the electronic properties of the superconductor are dominated by phase locked circular Cooper pair currents, which vanish at the center in favor of localized states. The latter are created through multiple Andreev reflection(Fig.\ 10). The spatial distribution of these localized states and of the Cooper pair currents depends on the material under study, i.e. on Fermi surface gap anisotropy and symmetry of the order parameter. The \textit{vortex core} is thus an interesting laboratory on its own, and it can only be imaged in real space using STM.

\begin{figure}[ht]
\includegraphics[width=1.2\columnwidth,keepaspectratio, clip]{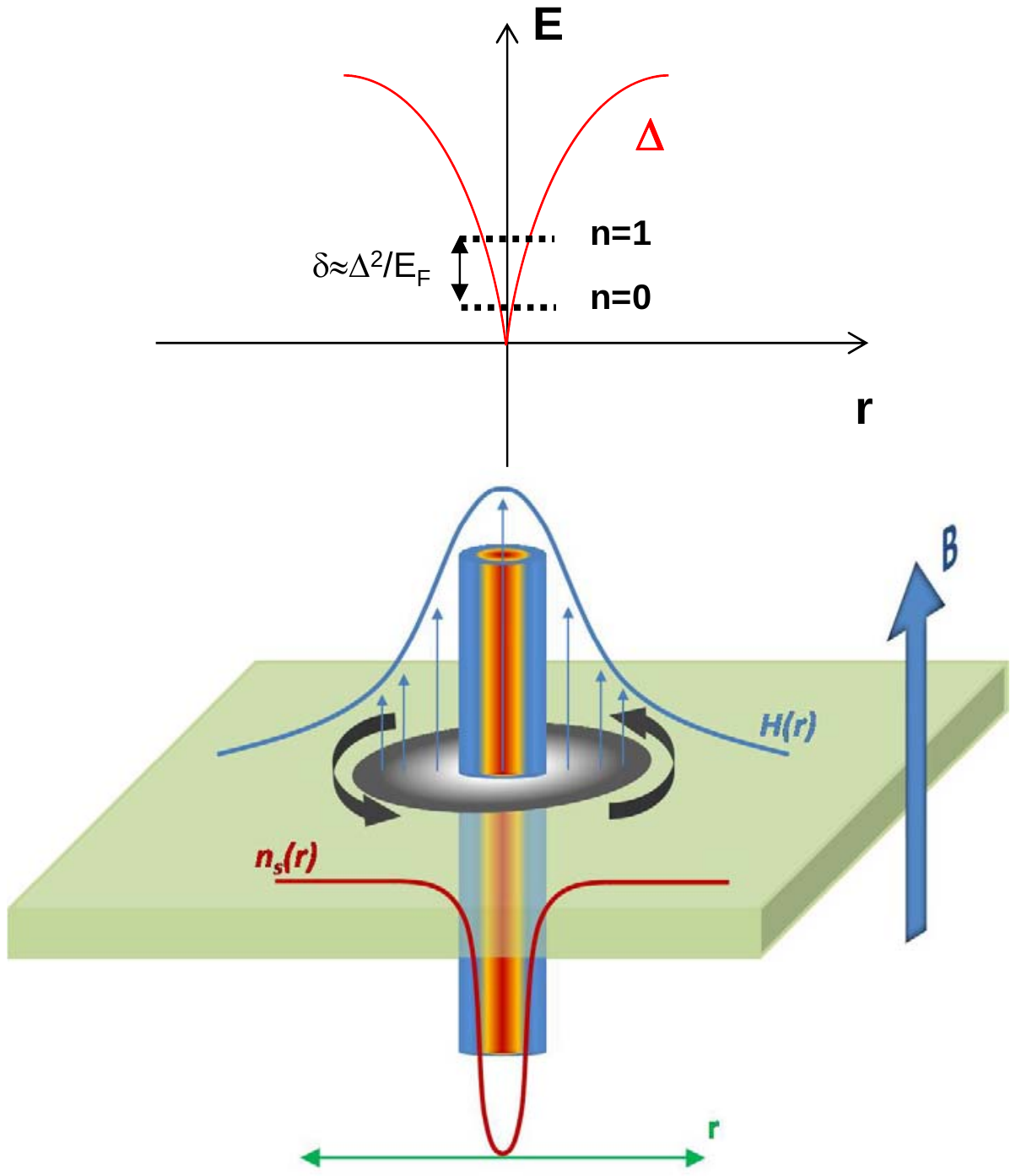}
\vskip -0cm \caption{Vortex is shown together with a scheme of the vanishing superconducting gap at the vortex core and the establishment of Andreev bound states inside. Level spacing is of order $\delta\approx \frac{\Delta^2}{E_F}$.} \label{Fig10}
\end{figure}

Caroli, deGennes and Matricon showed theoretically that the vortex line in a superconductor leads to the creation of quantized bound states\cite{Caroli64}. They realized that the spatially varying pair potential inside the core should lead to the formation of quasiparticles with mixed electron-hole character, and be thus Andreev reflected when leaving the core. This requires in principle clean superconductivity, with a mean free path larger than the superconducting coherence length $\xi$. Constructive interference between phases of multiply reflected quasiparticles gives bound states. In the 90's Hess et al. discovered, by making STM measurements of the vortex lattice in 2H-NbSe$_2$ at very low temperatures, a peak in the density of states at the vortex center\cite{Hess89,Hess90,Hess91}. They showed how the peak splits when leaving the core, pointing out a non-trivial quantization phenomenon, which was interpreted as the experimental observation of the Caroli, deGennes and Matricon bound states\cite{Gygi91,Hayashi98,Melnikov01}. Since then, these bound states have been observed in different compounds.

There is one relevant difference between bound states in a quantum well and Andreev levels in a superconducting vortex. Whereas in a quantum well, the states are truly confined in the well, in a superconducting vortex, the pair wavefunction vanishes only at a single point. Excitations are not "normal", but are quasiparticles created on top of a spatially varying finite superconducting background through Andreev exchange. Quantized levels arise as a consequence more of the phase winding than as the formation of a true well for quasiparticles\cite{Virtanen99,Berthod05,Berthod13}. Macroscopic measurements, in particular thermal conductivity, show some features of this relationship\cite{Seyfarth06,Suderow98,Izawa01}.

\subsubsection{2H-NbSe$_2$.} The core has been investigated in greatest detail in the transition metal dichalchogenide 2H-NbSe$_2$. Authors have used normal tips, superconducting tips exploring enhanced resolution, Andreev reflection and Josephson effect, and asymmetric tips.

\paragraph{Normal tips.}

Hess et al. measured the full bias voltage dependence of the tunneling conductance, and found vortex core images which change as a function of the bias voltage\cite{Hess89,Hess90,Hess91}.

\begin{figure}[ht]
\includegraphics[width=0.9\columnwidth,keepaspectratio, clip]{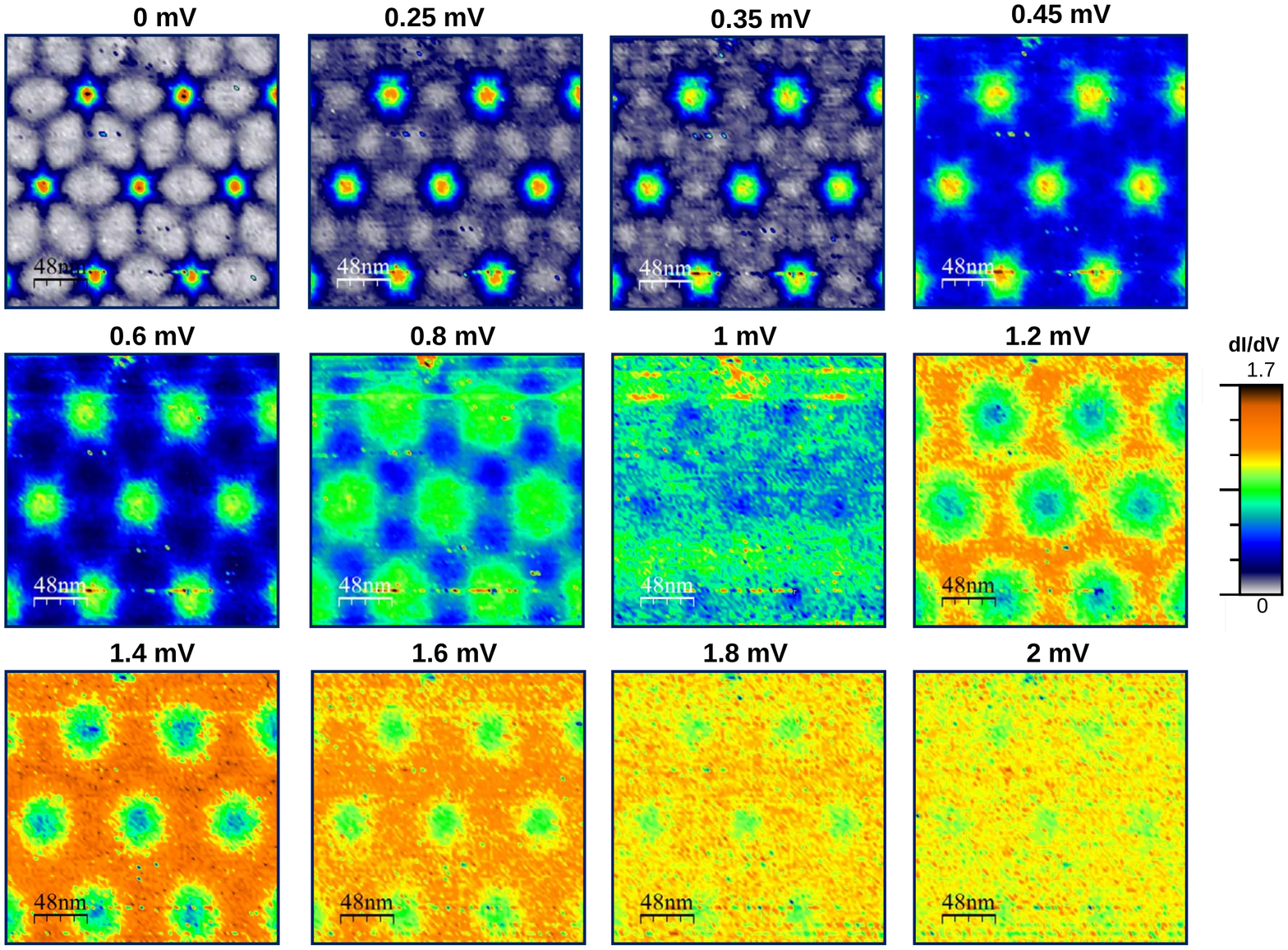}
\vskip -0cm \caption{Vortex core structure in 2H-NbSe$_2$ as a function of the bias voltage. Data are from Ref.\protect\cite{GuillamonPhD}. A normal tip of Au is used, and the applied magnetic field is of 0.15 T.} \label{Fig11}
\end{figure}

\begin{figure}[ht]
\includegraphics[width=0.9\columnwidth,keepaspectratio, clip]{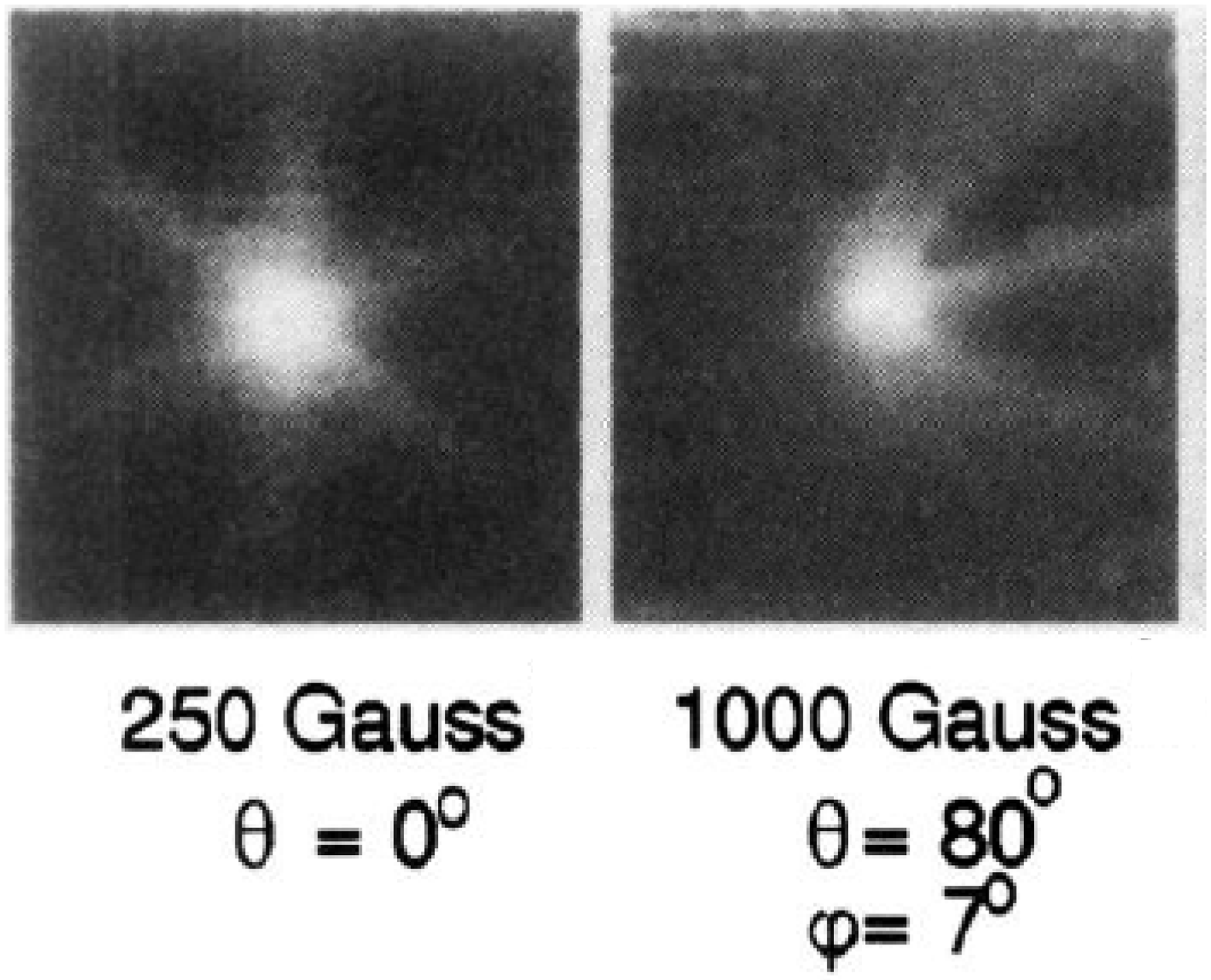}
\vskip -0cm \caption{Vortex cores in a magnetic field perpendicular to the surface (left panel) and at an angle to the surface (right panel). Note that the core shape significantly distorts, transforming the star into a comet like structure. Size of the images are of 150 nm. Data are taken using a normal tip of Au. Further data at other angles and bias voltages are found in Refs.\protect\cite{Hess92,Hess94}. Copyright (1992) by The American Physical Society.} \label{Fig12}
\end{figure}

At zero bias, a peak is observed in the tunneling conductance. The peak splits towards higher bias voltage when leaving the vortex core. The spatial dependence of this splitting is strongly asymmetric. When tracing maps of the zero bias conductance, a six fold star shape is observed (Fig.\ 11). The star shape is oriented at 30$^\circ$ with respect to the hexagonal vortex lattice, and is thus not a simple consequence of greater vortex core overlap along the high symmetry directions of the hexagonal vortex lattice. Interestingly, at the center between three vortices, the density of states at the Fermi level increases. This occurs at the point where the rays of the star join, which is at 30$^\circ$ to the nearest neighbor direction of the hexagonal lattice. It shows how vortex core states can interact with each other and hybridize\cite{Melnikov09}. The rays of the star double when increasing the bias voltage and then the star turns at 0.5 mV into a star whose rays are oriented along the high symmetry directions of the hexagonal vortex lattice. The complex vortex structure was numerically modelled by assuming a six-fold anisotropic superconducting gap in Ref.\cite{Hayashi98}. The conductance maps of 2H-NbSe$_2$ are still, more than 20 years after its discovery, unique in the richness of the observed features.

Hess et al. also examined the form of the vortex core when turning the magnetic field out of the plane\cite{Hess92,Hess94}. They showed the formation of a distorted hexagonal lattice in agreement with the theory of anisotropic superconductors in presence of a tilted field\cite{Campbell88}. They observed several remarkable features, such as a transition in the lattice symmetry when inclining the magnetic field for some azimuthal angles. The lattice is always oriented in such a way as to have one of its principal axis at 30$^\circ$ to the surface, which implies that the hexagonal vortex lattice is oriented such that one of the faces of the hexagon stays parallel to the surface. Other features remain however unexplained. For instance, the appearance of stripes when the magnetic field is close to being parallel to the surface. Also, the form of vortices crossing the sample surface, which are rather elongated\cite{Buzdin09,Samokhvalov10}. The core bound state features were observed to change from a star shape into a comet like form (Fig.\ 12), for reasons which are yet unknown.

\begin{figure}[ht]
\includegraphics[width=0.9\columnwidth,keepaspectratio, clip]{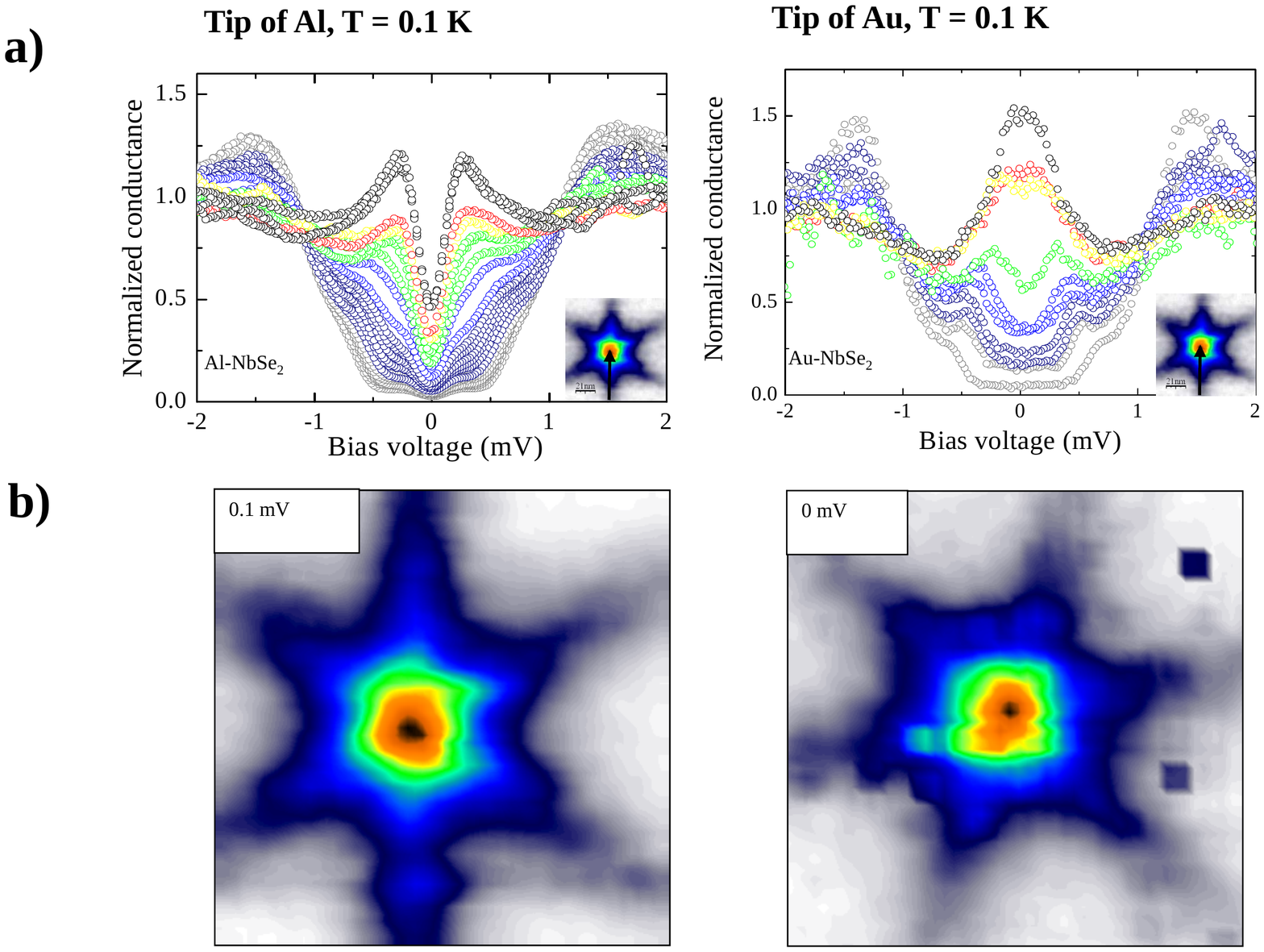}
\vskip -0cm \caption{A vortex core, as observed with a tip of Al (left panels) and a tip of Au (right panels) at 0.03 T. Size of the image is of 70 nm. In a) we show tunneling conductance curves as a function of the position when entering the core. In b) we show an image of the core at 0.1 mV (left) and 0mV (right). The gap of the tip displaces features of the vortex core to higher energies. In this experiment, the tunneling conductance is of about 1 $\mu S$, which is too low to observe further tunneling features, such as Andreev reflection or Josephson effect. Adapted from Ref.\protect\cite{Guillamon07}} \label{Fig13}
\end{figure}

Lateral imaging of the surface of 2H-NbSe$_2$ was made in Refs.\cite{Fridman11,Fridman13}, with a STM mounted in such a way that the magnetic field lies parallel to the surface. No sign of Andreev bound states was reported. Authors observe stripe patterns due to spatial modulations of the Doppler shift induced by surface screening currents.

\begin{figure}[ht]
\includegraphics[width=0.9\columnwidth,keepaspectratio, clip]{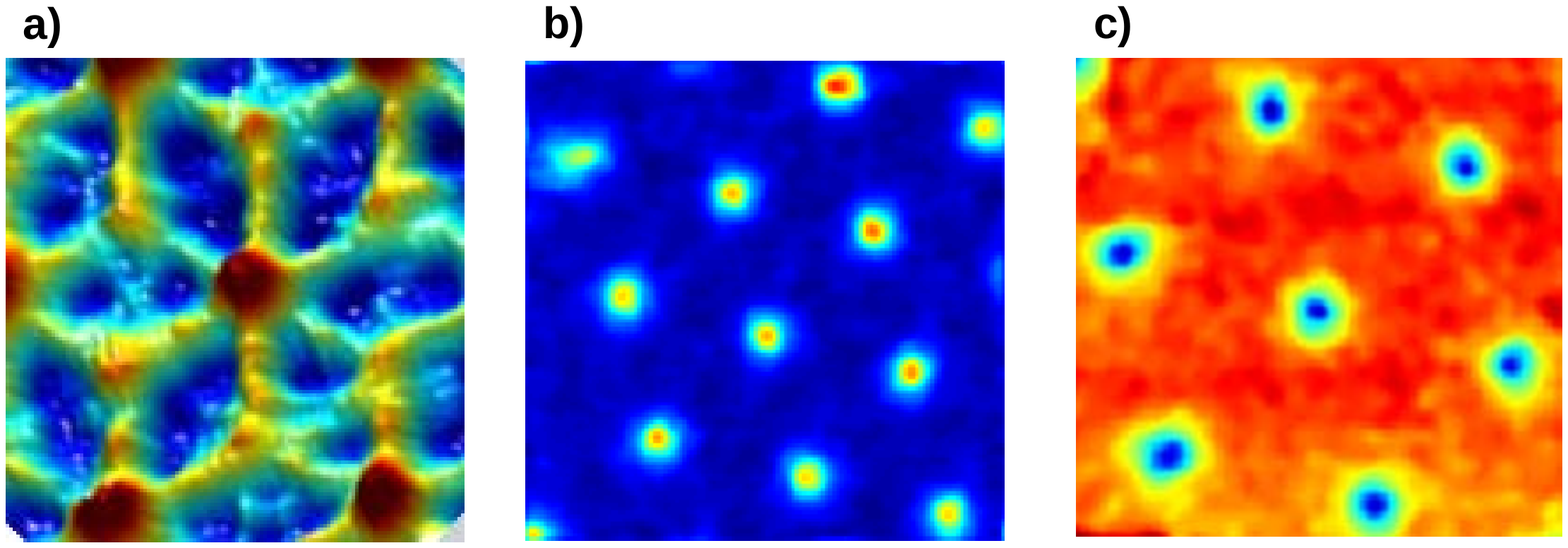}
\vskip -0cm \caption{Vortices as viewed with tips of Pb at 0.1 T (a, size of 300 nm), 0.15 T (b, size of 390 nm) and 0.1 T (c, size of 390 nm). This gives an experiment showing the tunneling processes discussed in Fig.6. Mapping the quasiparticle density of states (a), mapping Andreev reflection (b) and Josephson effect (c). Details about color code and size of the effects are found in Refs.\protect\cite{CrespoPhD,Crespo12}} \label{Fig14}
\end{figure}

\begin{figure}[ht]
\includegraphics[width=1.1\columnwidth,keepaspectratio, clip]{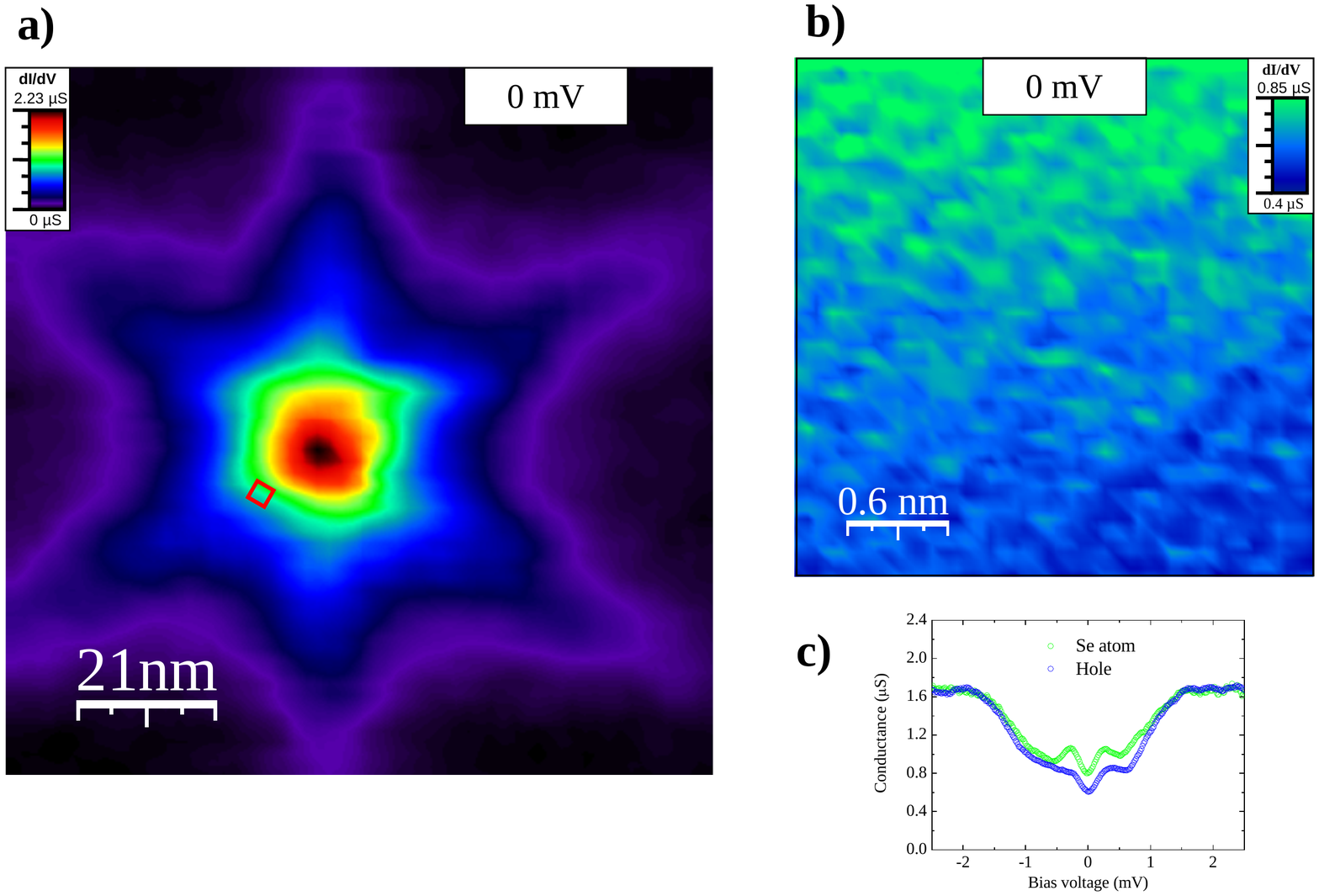}
\vskip -0cm \caption{Atomic size modulations of the vortex core, imaged with a normal tip of Au at a magnetic field of 0.03 T. The vortex core star shape is shown in a). The red square gives roughly the position where the image in b) was taken. Note the difference in scale. An atomic size modulation of the density of states, as shown in representative curves in c) is observed. Adapted from Ref.\protect\cite{Guillamon08PRB}. Copyright (2008) by The American Physical Society.} \label{Fig15}
\end{figure}

The effects of impurities was studied in Ref.\cite{Renner91,Miranovic04} and it was found that addition of impurities into 2H-NbSe$_2$ leads to a decrease of the peak at the center of the vortex cores.

\paragraph{Superconducting tips.}

Several vortex lattice observations as a function of the magnetic field and temperature have been made in 2H-NbSe$_2$ using superconducting tips. Authors of Ref.\ \cite{Guillamon08} use a tip of Al (with $T_c=1.1 K$ as compared to $T_c=7.1 K$ of 2H-NbSe$_2$). The density of states of the Al tip was characterized on a normal sample. At the center of the vortex core, instead of the zero bias peak due to the lowest level Andreev bound state, a dip is measured (Fig.\ 13). The tunneling conductance is the convolution between the vortex core zero bias peak and the superconducting density of states of the Al tip. The density of states of the Al tip at the center of the vortex is slightly modified, with respect to the density of states outside the vortex, because of the magnetic field dependence around the vortex. By de-convoluting the known dependence of the density of states of 2H-NbSe$_2$, one can obtain a map of the Al tip's density of states as a function of the position. This gives the magnetic field profile of the vortex. Measurements are compatible with expectations from the penetration depth value in 2H-NbSe$_2$, and provide spatial resolution which is promising when compared to magnetic imaging probes\cite{Bending99,Finkler12}.

In Ref.\cite{Crespo08,CrespoPhDV,Crespo12}, a tip of Pb, and in Ref.\cite{Kohen06} a tip of Nb is used. The superconducting gap of both elements is of same order than the superconducting gap of the sample. Superconductivity in the Pb tip remains at magnetic fields well above the critical field of Pb, as shown in Refs.\cite{Petal98,Suderow02,Rodrigo04,Rodrigo04b}. Thus, the positional variation of the tip's features can be independent of the magnetic field variations created on the surface of the sample by the vortex lattice. Observations of vortices from tunneling conductance maps made at voltages between the gap of Pb $\Delta_{Pb}$ and $\Delta_2=\Delta_{Pb}+\Delta_{NbSe_2}$ give nicely the same localized state features observed with normal tips (Fig.\ 14a). The energy resolution is somewhat improved due to the sharp quasiparticle edge of Pb.

In Ref.\cite{Kohen06}, it was shown that a detailed analysis of the tunneling conductance curves close to the quasiparticle peaks can be used to obtain, at the same time, length scale associated to screening ($\lambda$), which Doppler shifts the density of states, and to the vortex core ($\xi$), which governs the spatial dependence of the quasiparticle anomalies.

When reducing the tip-sample distance, Andreev reflection between the superconducting tip and the vortex core is observed through conductance maps made at bias voltages below $\Delta_1=|\Delta_T-\Delta_S|$ (Fig.\ 14b)\cite{Crespo12}. Furthermore, when measuring at exactly zero bias, a Josephson effect between tip and sample is observed outside the vortex cores (Fig.\ 15c).

In the case of Josephson tunneling, the size of the Josephson effect depends on the Cooper pair density of the sample, although other factors, such as energy resolution and Josephson dynamics of small junctions play a relevant role\cite{Stip2,Bergeal08,Rodrigo04}. The complex spatial dependence of the vortex core localized state features described above is not observed, at least with the same clarity, in the spatial dependence of the Cooper pair density(Fig.\ 15c). Thus, there is a difference between the spatial dependence of the star shaped vortex core Andreev bound states as observed in quasiparticle tunneling and the vortex as observed with techniques tracing Cooper pair density or supercurrent distribution. Future studies will probably resolve this difference, and provide new information about the core properties of NbSe$_2$.

\paragraph{Asymmetric tips.}

Authors of Ref.\cite{Guillamon08PRB} made atomic size conductance maps of the vortex core, showing that an atomic size modulation of the superconducting density of states appears within the vortex core close to the center (Fig.\ 15). The resulting structure does not change its symmetry when varying the bias voltage, but the size of the modulations is maximal on the peaks in the density of states due to the Andreev bound states. The vortex core also holds detailed information about the Fermi surface features of the superconducting gap. In particular, the formation of Andreev states inside the core is Fermi-surface dependent. It is shown that Andreev bound states due to smaller sized superconducting gaps appear with more intensity in the tunneling conductance when approaching the vortex core.

\subsubsection{2H-NbS$_2$.} Authors of Ref.\cite{Guillamon08c} showed that vortex core states can be also observed in the compound 2H-NbS$_2$. This material is similar to 2H-NbSe$_2$. It has a T$_c=5.7 K$ (as compared to 7.1 K in 2H-NbSe$_2$) and has no charge density wave. A localized state is observed at the center of the vortex core in 2H-NbS$_2$ as a strong zero bias peak (Fig.\ 16). The zero bias peak splits into two peaks when leaving the center. Farther away from the center, the conductance varies smoothly as a function of the bias voltage and shows superconducting gap features. However, the spatial anisotropy of the vortex core observed in 2H-NbSe$_2$ is lost. Furthermore, the zero field tunneling conductance shows that 2H-NbS$_2$ is a two-band superconductor. Subsequent macroscopic thermal and magnetic penetration depth measurements are in agreement with this point\cite{Kacmarcik10,Kacmarcik10b,Diener11}.
The vortex core observations in 2H-NbS$_2$ show that the star shape and spatial anisotropy observed in the vortex core of 2H-NbSe$_2$ is due to the presence of a charge density wave, which shapes the vortex core through the introduction of a strong in-plane gap anisotropy. Thus, the charge density wave induces the in-plane gap anisotropy found in the material 2H-NbSe$_2$.

\begin{figure}[ht]
\includegraphics[width=0.9\columnwidth,keepaspectratio, clip]{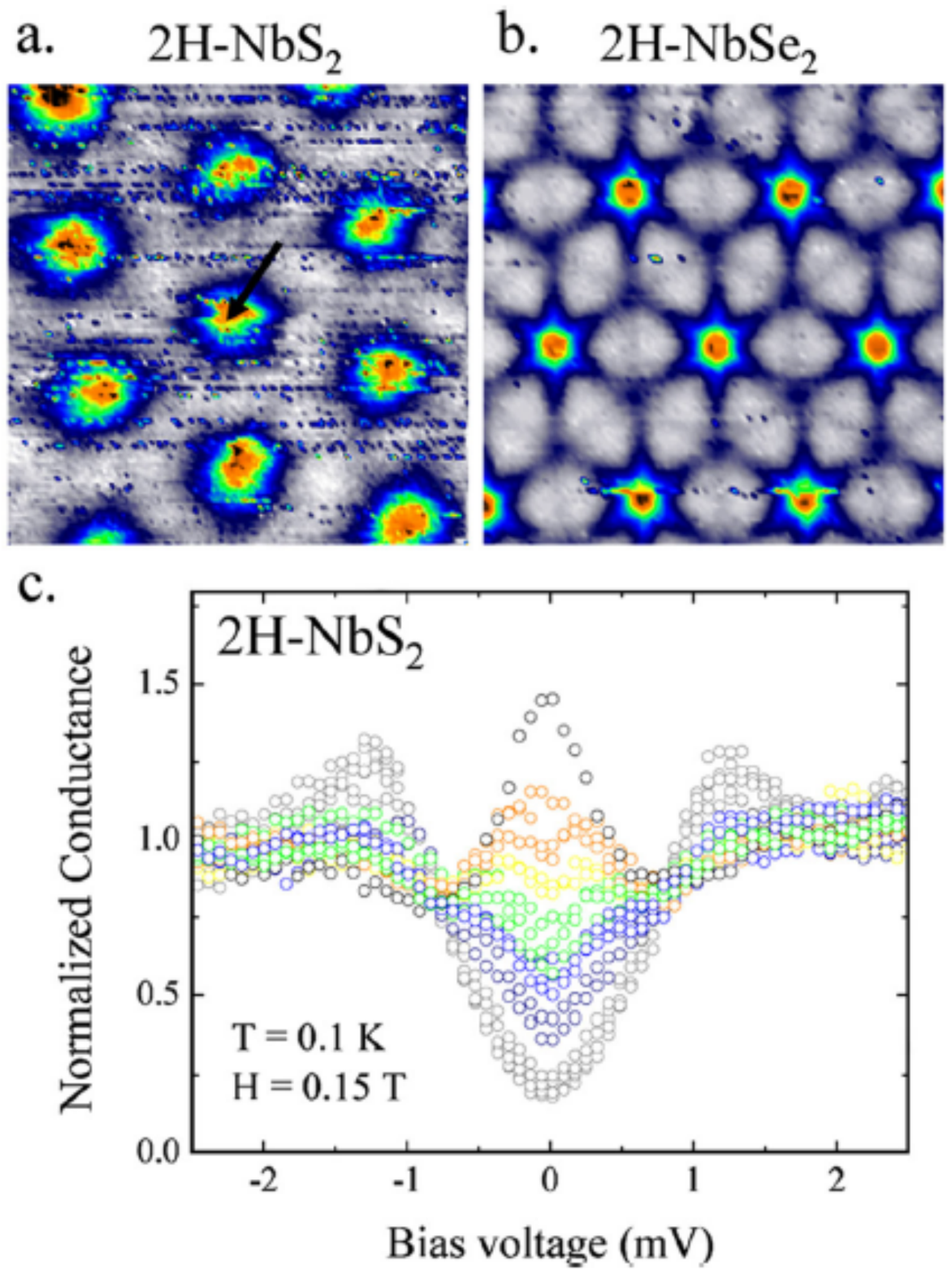}
\vskip -0cm \caption{Vortices in 2H-NbS$_2$ observed through STM at zero bias made with a normal tip of Au, compared to results in 2H-NbSe$_2$ (upper panels, a and b, magnetic field of 0.15 T; size of the image is of 360 nm). Note the presence of a localized state (c), but the absence of the star shape core as in 2H-NbSe$_2$ (b). Adapted from Ref.\protect\cite{Guillamon08c}. Copyright (2008) by The American Physical Society.} \label{Fig16}
\end{figure}

\begin{figure}[ht]
\includegraphics[width=0.9\columnwidth,keepaspectratio, clip]{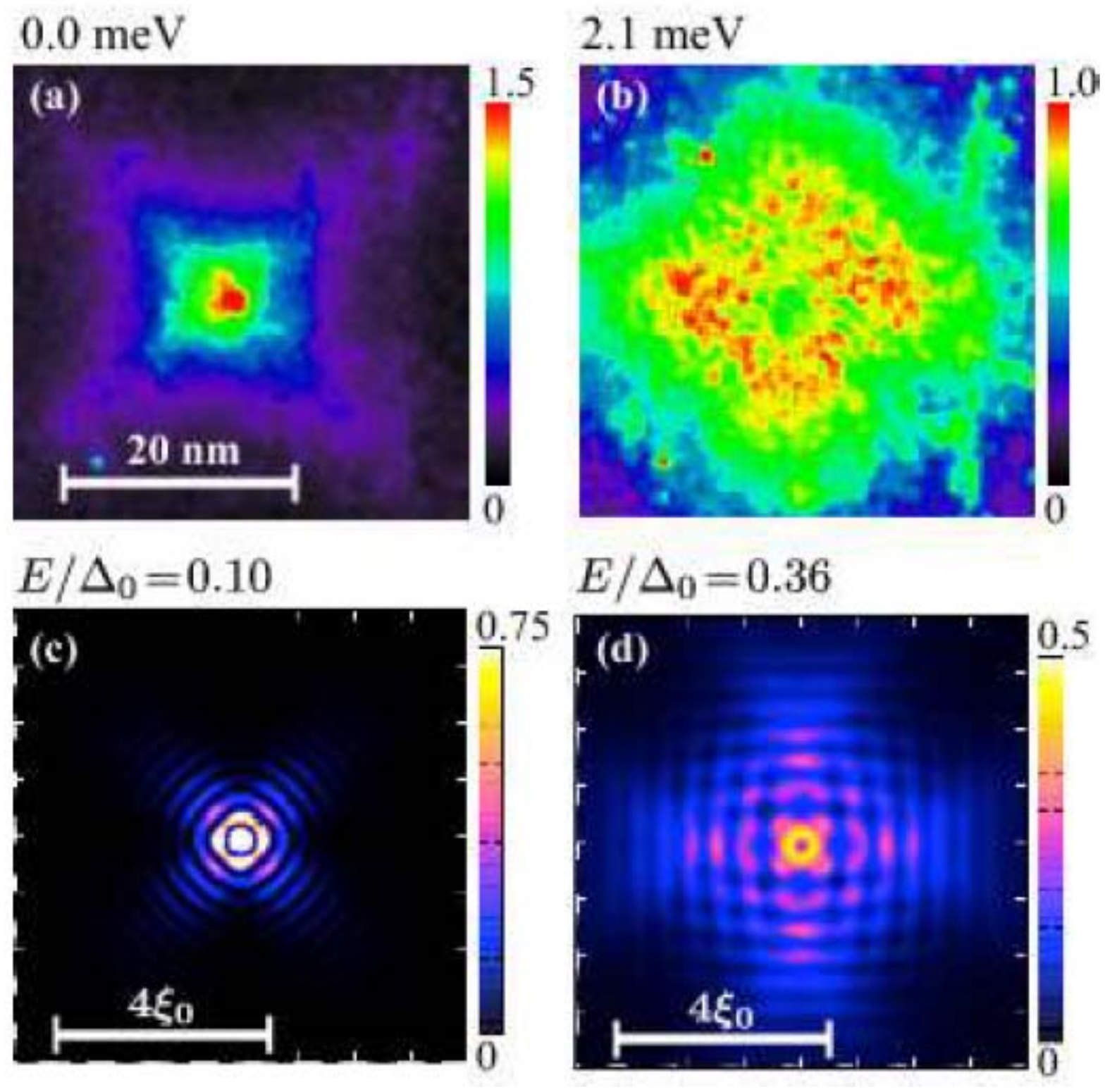}
\vskip -0cm \caption{Vortex cores in YNi$_2$B$_2$C at two different bias voltages, taken with a normal tip. Top panels (a,b) show experimental conductance maps taken at 180 mK and 0.3T. Lower panels (d, c) show corresponding maps of the calculated density of states. The four-fold star shape is oriented along one of the high symmetry directions of the square atomic surface lattice. YNi$_2$B$_2$C is tetragonal and was, in this experiment, cleaved in such a way as to expose the surface parallel to the basal plane. Magnetic field is along the c-axis. Adapted from Ref.\protect\cite{Kaneko12}} \label{Fig17}
\end{figure}

\subsubsection{YNi$_2$B$_2$C.} YNi$_2$B$_2$C is a crystalline system with a relatively high T$_c$ (15.5 K\cite{Canfield98}) which superconducts via an anisotropic phonon mechanism\cite{Martinez03b}. This produces shaped tunneling conductance curves, with an in-gap shoulder similar to the one observed in 2H-NbSe$_2$. The symmetry of the vortex lattice is by contrast four-fold due to non-local effects and the four-fold symmetry of the Femi surface\cite{Kogan96,Kogan97}. In plane atomic lattice was observed in Ref.\cite{Nishimori04}, where they also nicely show four-fold star shaped vortices in zero bias tunneling conductance maps. Further experiments with greater spatial resolution \cite{Kaneko12} unveiled in more detail the fourfold structure (Fig.\ 17). The tunneling conductance curves are rather anisotropic, and the vortex core shape is modified when measuring at higher bias voltages. The four-fold star opens into a rough square and turns in tunneling conductance maps made above zero bias. The tunneling conductance is rather asymmetric as a function of the sign of the bias voltage. Such asymmetries are further pronounced in pnictide and cuprates, as discussed below. Authors of Ref.\cite{Kaneko12} also discuss the influence of the anisotropic gap structure of YNi$_2$B$_2$C, with possible nodes\cite{Maki02,Izawa01} giving a rather continuum vortex core spectrum superimposed to discrete core Andreev bound states from regions on the Fermi surface with a large gap amplitude.

\subsubsection{Boron doped diamond and thin films.} Authors of Ref.\cite{Sacepe06} have studied single crystalline boron doped diamond. They find a fully disordered vortex lattice, and at the vortex cores a series of peaks located at bias voltages different from zero bias, with a strong asymmetry, both in the sign of the bias voltage and the location with respect to the vortex core. The presence of these in-gap states and local dispersion seems to be a feature of disordered systems. It bears some ressemblance with the findings of the cuprates and related intrinsically inhomogeneous materials, where gapped vortex cores are also found\cite{Fischer07}. 

\subsubsection{Cuprates.} Vortex core states in the cuprates were reviewed in detail in Ref.\cite{Fischer07}. No single peak is observed at the Fermi level\cite{Maggio95,Hoffman02a,Matsuba07}, although such a peak clearly appears around non-magnetic impurities and has been associated to the creation of Andreev bound states from scattering in a d-wave order parameter\cite{Balatsky06}. Interesting recent measurements of Ref.\cite{Yoshizawa13} observe striped vortex core states which are largely influenced by patches of different short range order. These results show the interplay between the core-states and the underlying electronic inhomogeneity which appears to dominate the physics of cuprates.

\begin{figure}[ht]
\includegraphics[width=0.9\columnwidth,keepaspectratio, clip]{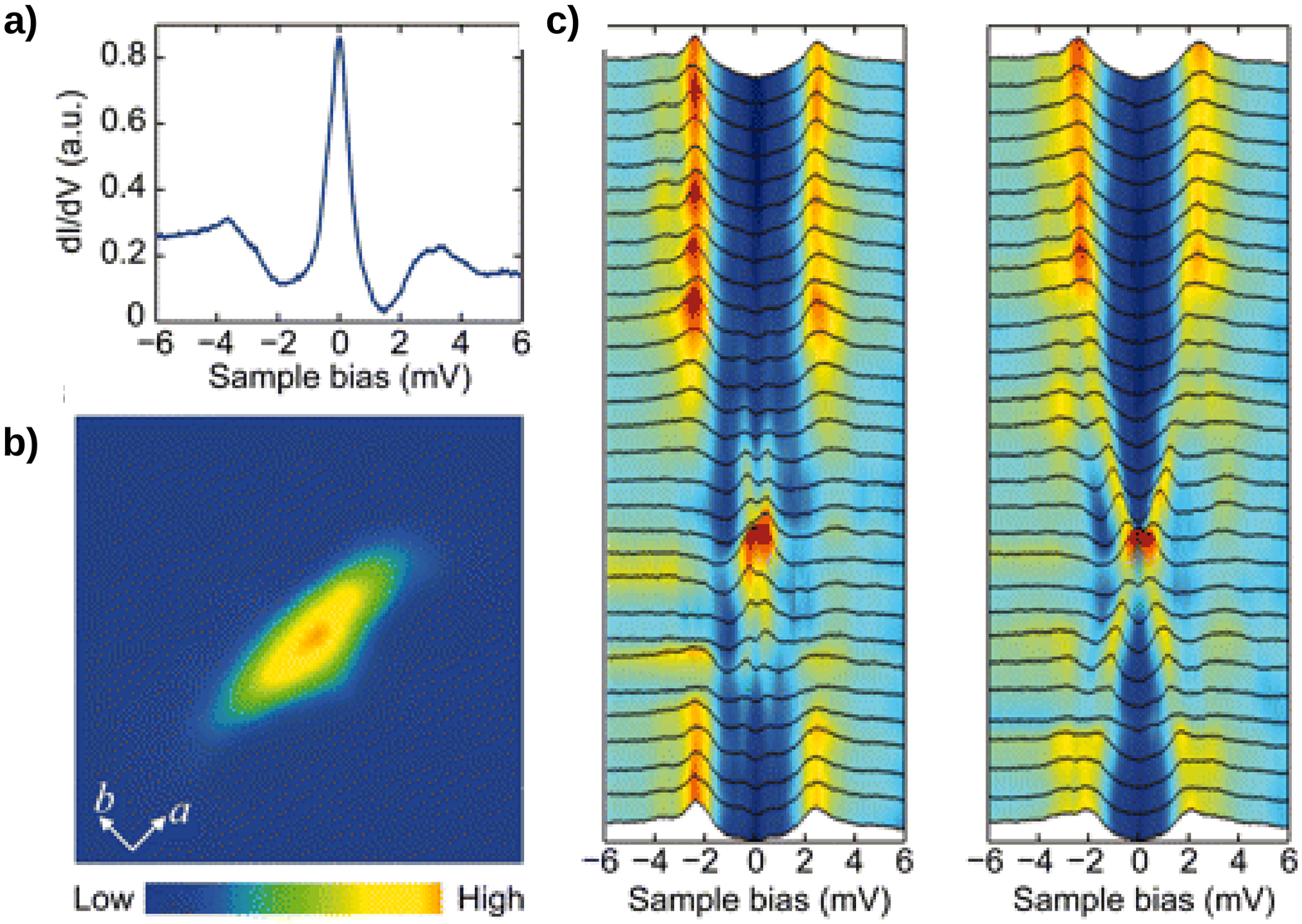}
\vskip -0cm \caption{Vortex core in FeSe. The localized state is observed at the center (a) and has a marked two-fold in-plane anistropy (b, image taken at zero bias using a normal tip, size of 40 nm, T=0.4 K and H=1 T). Tunneling conductance curves in c) are taken along the two main directions each 2 nm. Note the difference in the bias voltage scale, which is here significantly larger than in 2H-NbSe$_2$. Note also that the core is aligned with the crystal axis. Adapted from Ref.\protect\cite{Song11}} \label{Fig18}
\end{figure}

\subsubsection{Pnictides and related materials.} Results in these materials have been reviewed in Ref.\cite{Hoffman11,Hirschfeld11}. In-situ grown FeSe films were investigated in Ref.\cite{Song11}. Authors find a pronounced zero bias peak at the center of the vortex and a uniaxially elongated vortex core (Fig.\ 18). The peak splits into two when leaving the core. The orientation of the cores are found to vary close to twin boundaries. The form of the core agrees with present models for two-fold anisotropic gap structure in this compound. In other pnictide superconductors, such as BaFe$_{1.8}$Co$_{0.2}$As$_2$, cores are featureless possibly due to the reduced mean free path\cite{Yin09}. Further vortex bound states have been observed in Ref.\cite{Shan11} in the compound Ba$_{0.6}$K$_{0.4}$Fe$_2$As$_2$. The peak in the tunneling conductance is not located exactly at zero bias, and there is a strong bias voltage dependence, also with a bias voltage asymmetric peak splitting when leaving the center of the core. The form of the Andreev peak appears to be somewhat linked to the surface topography.

Further work in LiFeAs shows a four-fold vortex core in the clean limit\cite{Hanaguri12}. The four-fold star shape opens and turns in conductance maps made out of the bias voltage. The lowest Andreev level is observed. It is proposed that the shift with respect to zero voltage appears because the temperature is sufficiently low to reach the quantum limit. The lowest energy Andreev resonance is located at $\Delta^2/2E_F$ and thus appears at $V\approx \Delta^2/2eE_F$ in the tunneling conductance, provided that the temperature is far below $\Delta^2/E_F$. This condition needs large gap values and can realized in these systems.

\subsubsection{Heavy fermions.} Recently, vortices were observed in the heavy fermion superconductor CeCoIn$_5$\cite{Zhou13,Allan13}. The core shows bound states located at zero bias, which have a in-plane four-fold anisotropy and an asymmetric spatial dependence\cite{Zhou13}. The vortex lattice is also four-fold, possibly due to the Fermi surface anisotropy. The bias voltage dependence is further asymmetric, and its relationship with the gap structure is yet to be determined (Fig.\ 19). The four-fold core symmetry holds strong relation to the anisotropic gap structures found in macroscopic measurements\cite{Izawa01b}.

\begin{figure}[ht]
\includegraphics[width=0.9\columnwidth,keepaspectratio, clip]{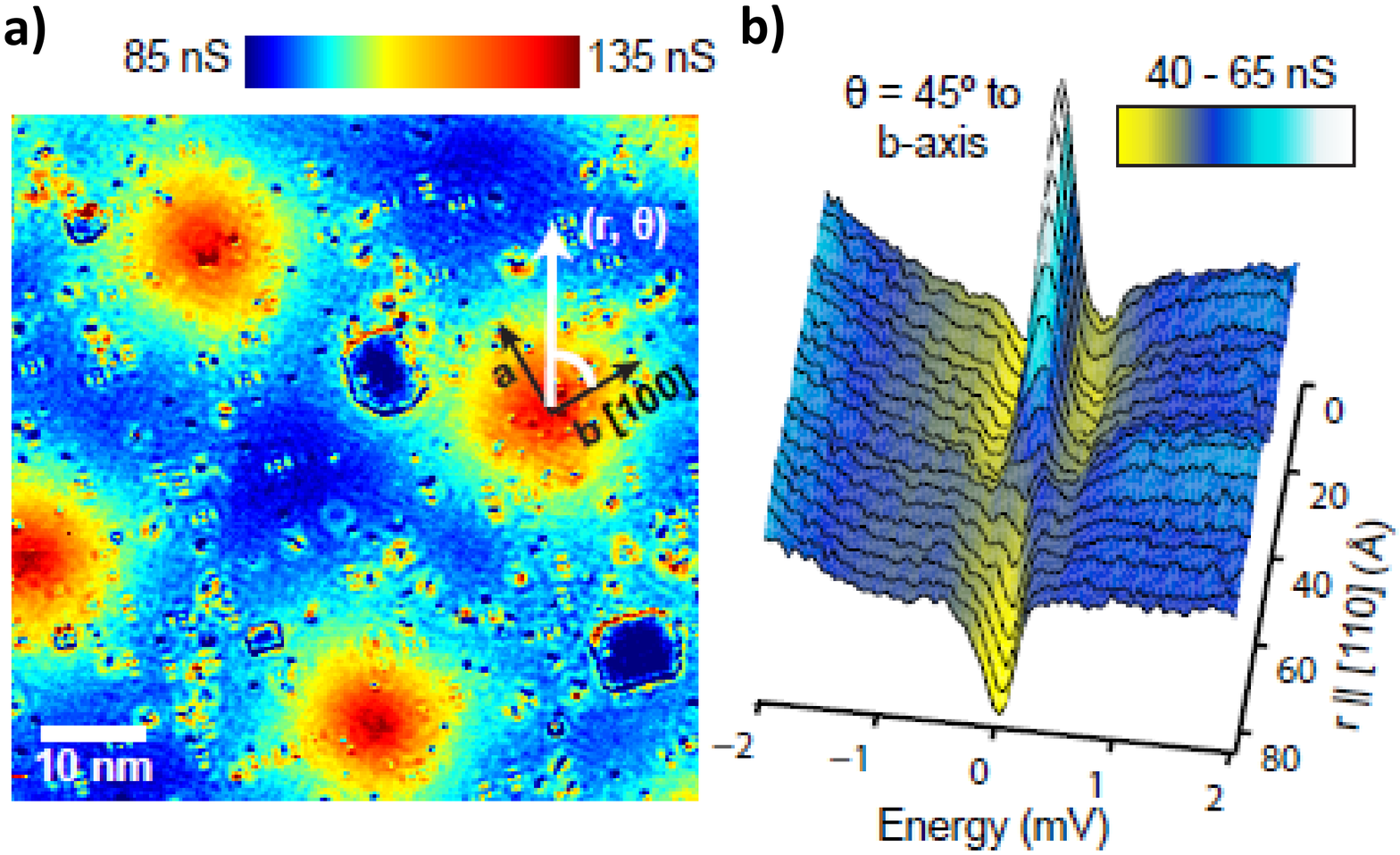}
\vskip -0cm \caption{Vortex core in CeCoIn$_5$. The core and the vortex lattice have, at this magnetic field, a marked four fold anisotropy (a), which has been associated to the unconventional superconducting properties of this material (T=245 mK and H=1.5 T). The anisotropy of the core gives states extended along the crystal axis, as shown in a. Data taken with a normal tip. From Ref.\protect\cite{Zhou13}} \label{Fig19}
\end{figure}

\subsubsection{Summary.} The anisotropy of the vortex core is largely determined by the gap symmetry and its anisotropy over the Fermi surface. Its relation to the crystal structure can be non trivial. In 2H-NbSe$_2$, the vortex lattice is oriented with the hexagonal crystal lattice, but the rays of the star shaped core (at zero bias) are misoriented by 30$^\circ$ to the crystal lattice. In YNi$_2$B$_2$C, in CeCoIn$_5$ and in some iron-based superconductors, the shape of the vortex core follows the crystalline axis, except in FeSe where the core is elongated and two-fold\cite{Nishimori04,Zhou13,Allan13,Song11,Shan11,Kaneko12}.

In 2H-NbSe$_2$, the origin of the star shape is related to the CDW opening, whose in-plane spatial anisotropy is not yet fully clear. Authors of refs.\cite{Yokoya01,Kiss07} point out that the maximum of the superconducting gap occurs at reciprocal space points joined by CDW wavevectors, coinciding with the star shape 30$^\circ$ misorientation with respect to the crystal lattice. But in Ref.\cite{Borisenko08}, they obtain the opposite result and find that the CDW gap becomes zero at those directions where the superconducting gap is maximal. Taking the full gap-anisotropy, with a minimum superconducting gap misoriented 30$^\circ$ to the crystal lattice, the authors of Refs.\cite{Hayashi96,Ichioka97,Hayashi97,Hayashi98} reproduce the zero bias star shape as well as its bias voltage dependence. This gap distribution is opposite to what is shown in ARPES measurements of refs.\cite{Yokoya01,Kiss07} which find at these directions the maximum value for the superconducting gap. Thus, the connection between Fermi surface features and vortex core is not fully determined in 2H-NbSe$_2$.

On the other hand, cores in YNi$_2$B$_2$C and in CeCoIn$_5$ show the underlying gap anisotropy. In YNi$_2$B$_2$C because of the anisotropic electron-phonon coupling and in CeCoIn$_5$ possibly because of reduced symmetry superconductivity\cite{Martinez03,Nishimori04,Zhou13,Allan13}.

Finally, let us note that many other materials show essentially a flat or featureless tunneling conductance inside the vortex core (see e.g.\cite{Sosolik03,Petrovic09,Guillamon08b}). It is unclear if this is just related to a short mean free path or if other materials related aspects, such as surface corrugation or lattice symmetry have some influence.

\subsection{The collective behavior of vortices: materials and conditions}\index{4. Imaging vortex matter!4.2. Collective behavior}

\begin{figure}[ht]
\includegraphics[width=0.9\columnwidth,keepaspectratio, clip]{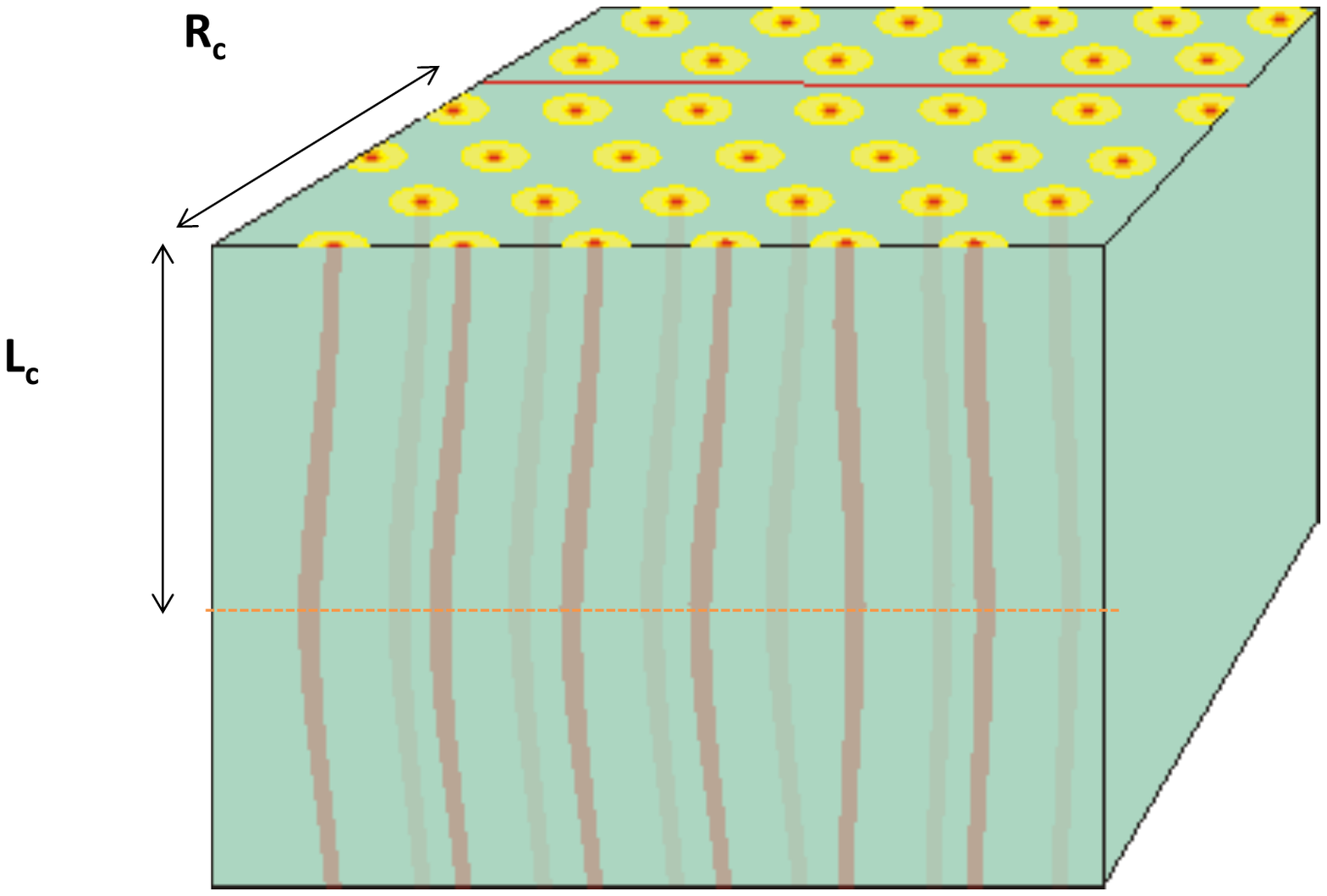}
\vskip -0cm \caption{Vortices arrange themselves in many cases in a hexagonal lattice. Their position is determined by local pinning arrangements, due to the interaction between the vortex and the underlying pinning potential. The concept of vortex bundle is useful to understand the collective behavior of the lattice. Within a vortex bundle, of size L$_c$ and R$_c$, vortices are ordered and behave coherently. Defects in the vortex lattice or vortex bending define the size of the vortex bundle. If the sample is thinner than L$_c$, which is of order of the magnetic penetration depth, the vortex lattice behaves as a 2D solid of vortex disks. As the magnetic penetration depth of type II superconductors exceeds several hundred nanometers in many cases, the thickness required to obtain a 2D lattice is relatively large, sometimes up to a hundred nanometer.} \label{Fig20}
\end{figure}

\subsubsection{Introduction.} Usually, flux enters the sample from the edge in the form of vortices, which are driven towards the interior by the Lorentz force due to the Meissner shielding currents. As proposed by Bean\cite{Bean64}, motion toward the sample center is hindered by the pinning centers, which tend to pin any vortex that passes by. A nonequilibrium state, the critical state, is created where the vortex density is largest in the regions where magnetic flux enters the sample. The metastable landscape of vortices can be altered in several ways, as, e.g., increasing temperature or the applied magnetic field, or by an external current circulation. Then, as stated by Anderson, flux leaks through the material to return to the critical state\cite{Anderson62,Kim62,Anderson64}. The concept of bundles of flux lines was introduced. Bundles, aided by the Lorentz forces, move over a potential barrier landscape formed by the different kinds of pinning acting in the material. Anderson associates to these bundles a typical size of the order of the magnetic penetration length $\lambda$. Vortices inside of a bundle are assumed to be bound together by the interaction of their fields and wave functions. Longitudinal size of the bundle is $L_c$ and its lateral size $R_c$ (Fig.\ 20)\cite{Blatter94,Brandt93,LO68}. Both are of order of $\lambda$, unless pinning introduces a smaller scale. Along the last decades, mainly following the discovery of superconductors with higher critical temperatures, much experimental and theoretical work has focused on the understanding of the critical state, and the mechanisms behind the organization of vortices in pinning landscapes\cite{Blatter94,Brandt93}. The magnetic vortex system of superconductors is highly tunable and allows to investigate issues concerning different regimes of flow in a wide range of flux densities and pinning landscapes. The real space observation of superconductors using STM on the scale of single vortices has opened a field which transcends in many aspects the physics of superconductors. The playground covers from fundamental issues such as dynamical phase transition or critical states\cite{Guillamon11b}, and applications, such as the improvement of the pinning action by intelligent design of materials and pinning centers distribution\cite{Cordoba13}. Recently, a new and promising tool has been developed which combines the passage of an applied current trough the sample and the STM capabilities\cite{Maldonado11,Maldonado13}.

One of the most important difficulties for this kind of studies is that the surface of the materials has to be clean and very flat in a wide area. Topographic and spectroscopic images have to be obtained in short times, in order to get relevant information about time dependent modifications of the vortex lattices. To avoid these problems is one the main reasons for which the layered superconductor 2H-NbSe$_2$ has attracted the attention of many researchers for some time. It can be exfoliated easily and it presents large atomically flat areas, which show, in addition, high stability even in ambient conditions.

Recently flat amorphous W-based thin films have been fabricated by focused ion beam deposition. These films show high quality surfaces in which the vortex lattice is nicely observed and studied. Amorphous superconducting thin films are of interest because of the absence of crystalline symmetry and associated defects. This significantly simplifies the available vortex pinning mechanisms, which are then largely dominated by variations in the film thickness.

\subsubsection{2H-NbSe$_2$.} Thermal effects and dynamics of the vortex lattice have been studied in depth in the compound 2H-NbSe$_2$. In Refs.\cite{Troyanovski99,Troyanovski02}, authors use a smartly designed STM set-up working on a liquid helium transport container with a small superconducting coil. They measure thus at 4.2 K and magnetic fields below a Tesla. In Ref.\cite{Troyanovski99}, authors generate a magnetic field of 1 T and then decrease the magnetic field to some hundreds mT. After waiting for some minutes, authors take images at a fast rate (6.5 s per image, Fig.\ 21). Above 0.6 T, vortex motion is observed in both pristine and in heavy ion irradiated 2H-NbSe$_2$ single crystalline samples (Fig.\ 21). In the irradiated sample, the disordered flowing lattice consists of pinned vortices which wiggle isotropically around the center of their pinning potential well and moving vortices which move either in one direction or are pushed by neighbors. The ordered vortex lattice in the pristine crystal is observed to move along one of the principal directions of the lattice in bundles formed by several tens of vortices. Velocity of vortex motion is around $1 nm/s$ and corresponding electric field $0.6 nV$ per meter. From this, authors estimate an energy barrier for vortex creep of $U=77 K$. The corresponding current density is of $2 \times 10^6 A/m^2$. Interestingly, authors observe that motion shows traces of the periodic lattice potential through a velocity which is modulated by the intervortex distance $a_0$. Thus, there is a periodic washboard lattice potential, fixed by the surrounding pinning landscape, which influences motion of bundles.

In Ref.\cite{Lee11,Yoon12}, authors observe vortex motion from the decay in the magnetic field of the superconducting coil. They also work at 4.2 K and at fields around a Tesla. The vortices are followed over a week, and their speed is extremely small, of the order of 1 pm  per second, which is three orders of magnitude below the one found in Refs.\cite{Troyanovski99,Troyanovski02}. The vortex lattice acts as a sensitive probe of current leaks in the switch of the superconducting coil. Vortex motion was observed to be coherent and also non-uninform.

Similar techniques are used to trace vortex behavior close to the critical temperature\cite{Troyanovski02}. Authors are able to follow the vortex lattice even when the STM signal is small. Close to T$_c$, their sample enters the peak effect regime, where the critical current strongly increases\cite{Blatter94}. They observe that the lateral correlation length $R_c$ drops down to the intervortex distance, but they find no evidence for the vortex liquid. Instead, they observe trapped vortices which behave individually and not as large bundles, pointing out a change in the pinning properties that can be related to the peak effect.

\begin{figure}[ht]
\includegraphics[width=0.9\columnwidth,keepaspectratio, clip]{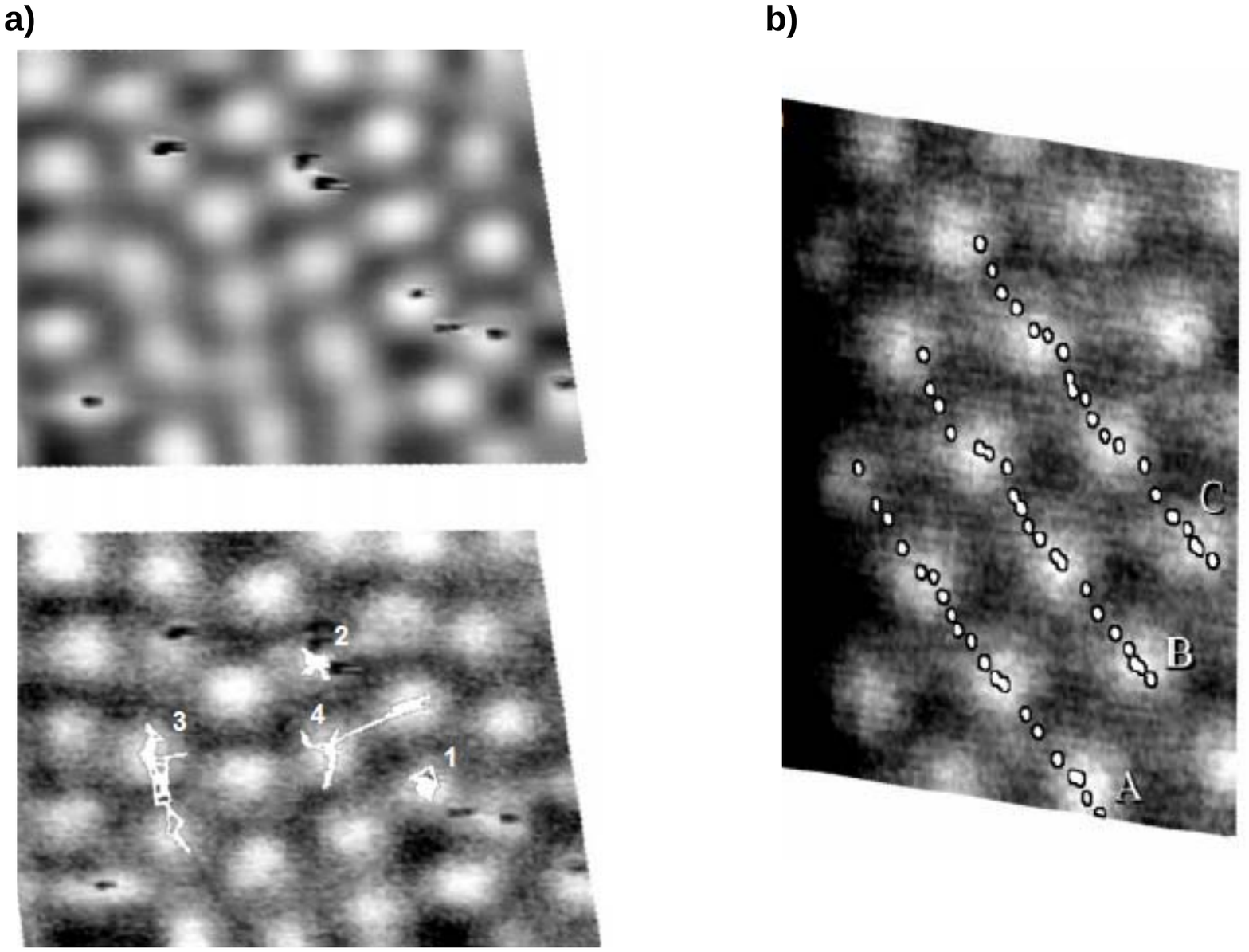}
\vskip -0cm \caption{In a) we show a STM image (420$\times$280 nm) at 0.6 T of 2H-NbSe$_2$ with columnar defects taken with a normal tip. Top panel is the average over 128 images, and bottom panel is a single image, with the trajectories of four typical vortices. In b) we show the trajectories of a few vortices in a pristine 2H-NbSe$_2$ sample. Images from \protect\cite{Troyanovski99}.} \label{Fig21}
\end{figure}

\begin{figure}[ht]
\includegraphics[width=1\columnwidth,keepaspectratio, clip]{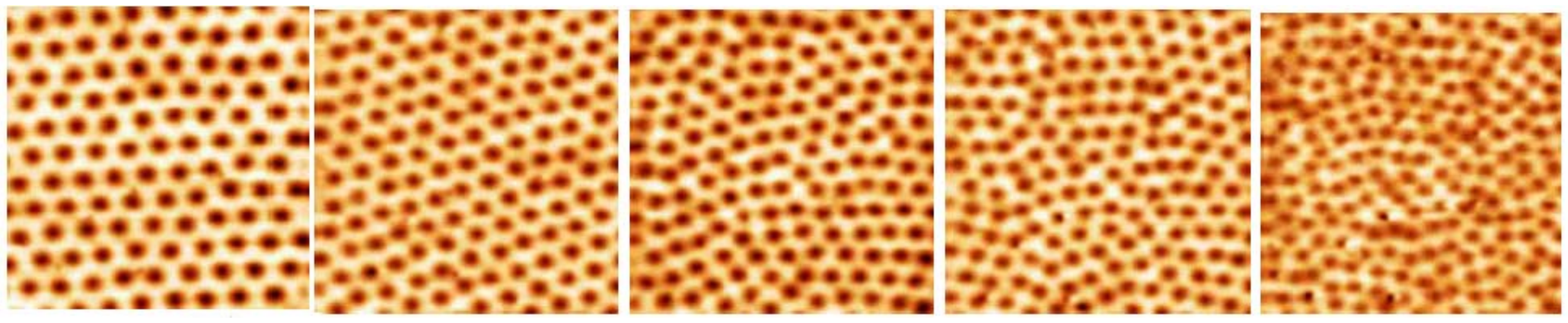}
\vskip -6cm \caption{Four vortex lattice images with increasing magnetic field (from left to right, 1.8 T, 2.3 T, 2.5 T, 2.7 T and 3.3 T, area is in all cases 375 nm$\times$375 nm, a normal tip is used) in a Co doped sample (0.4\% Co). The lattice is observed to disorder at the magnetic field where the peak effect as measured in the magnetization, is highest. Adapted from \protect\cite{Iavarone08}. Copyright (2008) by The American Physical Society.} \label{Fig22}
\end{figure}

On the other hand, the effect of the pinning potential on the lattice was studied in Refs.\cite{Behler94,Troyanovski99,Iavarone08,Karapetrov09} (Fig.\ 22). In Ref.\cite{Behler94}, authors study heavy ion irradiated 2H-NbSe$_2$ and observe the formation of defects in the vortex lattice close to irradiation damaged areas. In Ref.\cite{Iavarone08}, authors observe the vortex lattice by STM and compare with magnetization measurements in 2H-NbSe$_2$ crystals doped with Co and Mn. They correlate increased pinning at the peak effect with vortex lattice disorder. The impurities affect also the interlayer distance and possibly provide some magnetic interaction. The peak effect amplitude has a non-monotonic dependence as a function of the Co concentration. In the sample with largest peak effect, authors observe how the vortex lattice disorder increases.

\begin{figure}[ht]
\includegraphics[width=1\columnwidth,keepaspectratio, clip]{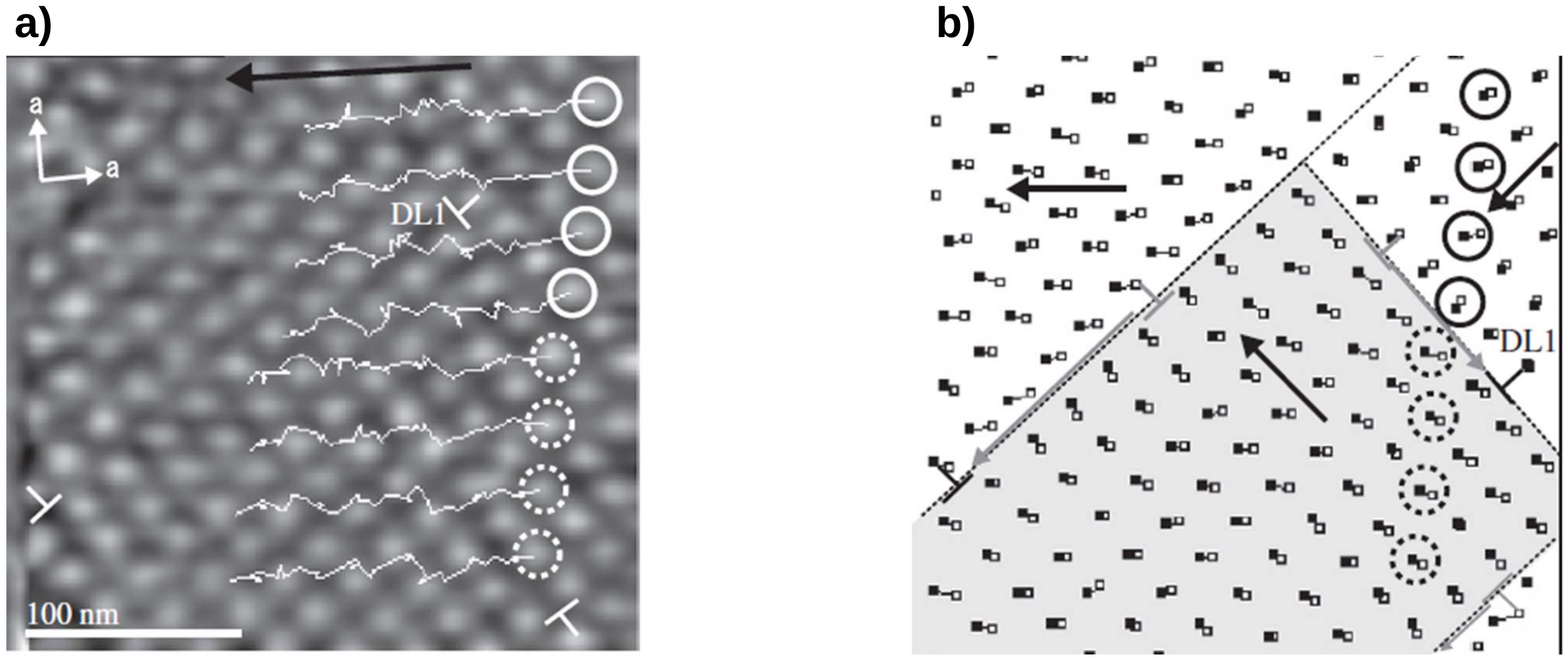}
\vskip -2cm \caption{Vortex creep observed in YNi$_2$B$_2$C. Note the fourfold vortex lattice (T=0.45 K and H=1 T). In a, in addition to the vortex lattice (measured using a normal tip), the underlying crystal lattice, as well as dislocations of the vortex lattice and the trajectories of four vortices are highlighted. In b, bundles are schematically shown, and the direction of motion of each vortex bundle, as well as the gliding planes are schematically represented. Data and images from Ref.\protect\cite{Uchiyama10}.} \label{Fig23}
\end{figure}

\subsubsection{YNi$_2$B$_2$C.} In Ref.\cite{Uchiyama10}, authors make fast imaging at low temperatures in the superconductor YNi$_2$B$_2$C (one image in 15 s at 0.45 K, Fig.\ 23), and produce vortex motion by increasing slowly the applied magnetic field. As discussed above, the vortex lattice is here a square lattice due to non-local effects\cite{deWilde97,Kogan95,Abrahamsen01,Gammel97,Kogan97}. Authors nicely observe vortex creep. Vortices follow fluctuating paths, and move in bundles which are separated just by dislocations of the square lattice. The direction of motion is determined by planes defined by the square symmetry and the gliding planes of the dislocations.

\subsubsection{Chevrel phase SnMo$_6$S$_8$.} An order-disorder transition has been imaged in the compound SnMo$_6$S$_8$ between 2 T and 9 T\cite{Petrovic09}. To examine pinning effects, authors compare vortex imaging to macroscopic specific heat and magnetization experiments. They find that disorder appears in the lattice well below the peak effect. They observe coincidence of a magnetization peak with the specific heat anomaly, without detecting latent heat, and propose the formation of a liquid state just below $H_{c2}(T)$. Vortices are disordered at high fields in a static configuration caused by pinning centers. These materials have a strongly reduced electronic mean free path. The large amount of defects and impurities in the bulk plays a relevant role to create the disordered lattice and in producing pinning close to the peak effect.

\subsubsection{Pnictides.} Further disordered vortex lattices have been observed in the iron pnictide materials. Sr$_{1-x}$K$_x$Fe$_2$As$_2$ shows surfaces with large electronic and gap inhomogeneities. The vortex lattice has been observed at magnetic field of 9 T. It appears strongly disordered, with a large amount of vortex lattice defects and spatial correlations smeared out. Authors compare the disordered lattice observed with ordered lattices appearing in Ba$_{0.6}$K$_{0.25}$Fe$_2$As$_2$\cite{Yin09}. Dopant clustering as well as the associated electronic inhomogeneity, increased due to large size ion mismatch in Sr$_{1-x}$K$_x$Fe$_2$As$_2$ accounts for the disordered lattice. The clean limit was analyzed in detail in Ref.\cite{Hanaguri12} in measurements on LiFeAs from 0.1 T to 11 T. Authors observe a lattice which gradually disorders and, at high magnetic fields, develops square fourfold correlations. Authors associate the differences found with respect to the hexagonal lattice observed in other compounds to the much smaller upper critical field of LiFeAs and thus a softer vortex lattice at the same magnetic fields. The rather strong interaction of vortices with defects is seen in the rich core features found in these materials. How this influences pinning is not yet clear. In these compounds, disorder competes with the underlying crystal symmetry, which also strongly shapes the vortex cores.

\begin{figure}[ht]
\includegraphics[width=1\columnwidth,keepaspectratio, clip]{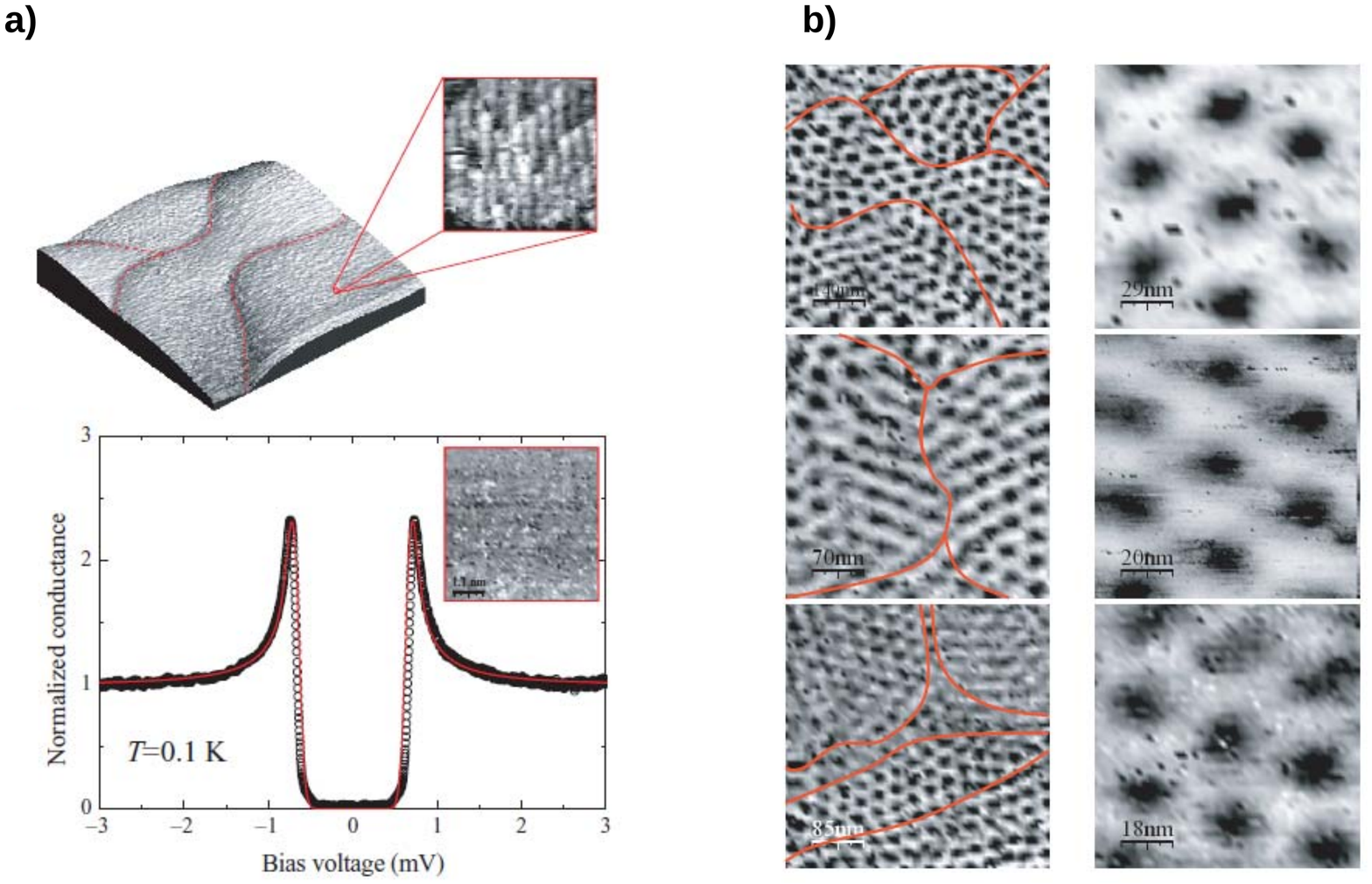}
\vskip -1cm \caption{In a we show the superconducting gap and topography of a W-based focused-ion-beam deposited thin film, measured with a normal tip (area of $1\mu$m $\times$ $1\mu$m). Gap is homogeneous, as exemplified by a conductance map in the inset of bottom panel of a). The conductance map is made at the quasiparticle peak, and the contrast is adjusted to show changes in the quasiparticle peak height of a few per cent. In b) we show several vortex lattice images on different locations of the thin film. Features in the surface topography are highlighted by red lines. Images are taken at 1 T, 2 T and 3 T (from top to bottom). Data and images from Ref.\protect\cite{Guillamon08b}.} \label{Fig24}
\end{figure}

\subsubsection{W based thin films deposited using focused-ion-beam.} In Ref.\cite{Guillamon08b}, it was shown that focused ion beam deposited W-based thin films give homogeneous superconducting gap features at zero field and good tunneling conditions. The vortex images (Fig.\ 24) obtained in these films show that the lattice orientation and vortex positions are determined by the morphology of the surface. It consists of linear surface depressions separating different areas several hundreds of $nm$ in size. The lattice orientation is not fixed by any crystalline orientation, as the film is amorphous, but by the linear features in the topography. The composition is homogeneous and not change along the thickness. Thus, the main pinning mechanism is the vortex length, or film thickness. The energy per unit length of a vortex can be written as 

\begin{equation}
\varepsilon_{L}=\frac{\Phi_{0}^2}{4\pi\mu_0\lambda^2} ln\left(\frac{\lambda}{\xi}\right)
\end{equation}

where $\lambda$ and $\xi$ are respectively the magnetic penetration depth ($\lambda = 850 nm$) and the superconducting coherence length ($\xi=6.25nm$). Thickness variation of 1 nm gives an energy difference of 100 K. This corresponds to weak pinning\cite{Blatter94,Pautrat04}. As the morphology of the surface consists of lines which separate different areas, the pinning centers are linear, defining a patchy landscape of amorphous "droplets" or rounded structures which break and orient the vortex lattice into patches.

Vortex motion experiments were addressed in Ref.\cite{Guillamon11b}. The vortex positions were fixed by the surface topography, and thermal excitations were eliminated by working at 100 mK. The sample was zero field cooled down to 100 mK, and a magnetic field above 1 T was applied to bring the sample to a critical state. The lattice is hexagonal order but is distorted due to variations in the film thickness. Vortex motion is provoked by slightly increasing the magnetic field in steps much smaller than the applied magnetic field $\delta H \ll H$. Vortex displacements are dominated by the surface pinning landscape. This gives bundles including about 10 to 20 vortices.

Note that the situation is different than vortex creep observations in single crystalline compounds, where the underlying crystal defines the symmetry of the lattice within bundles and the crystal defects the size of bundles.

\begin{figure}[ht]
\includegraphics[width=1\columnwidth,keepaspectratio, clip]{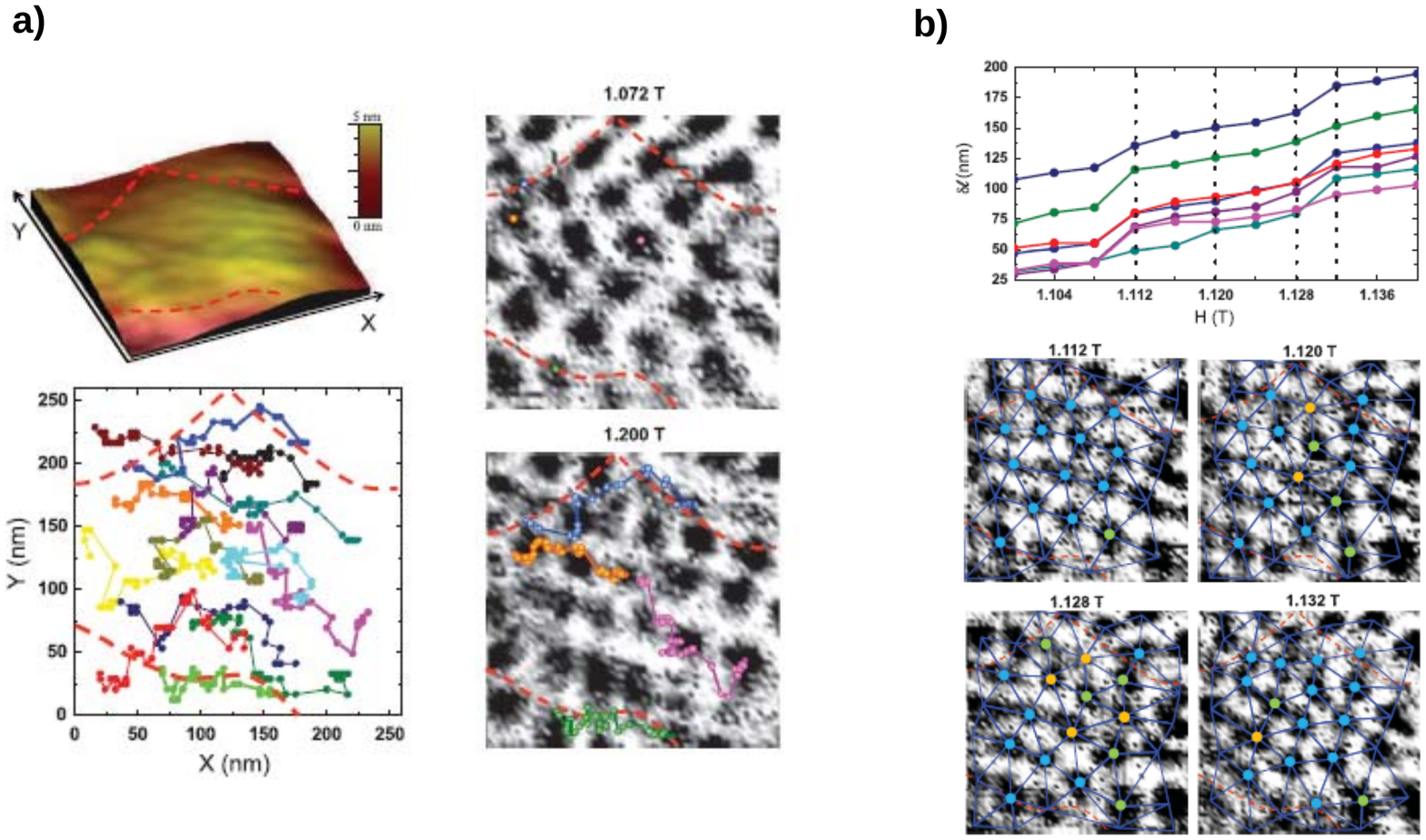}
\vskip -1cm \caption{In a we show the topography (top left panel), initial vortex lattice (top right panel), vortex trajectories (bottom left panel) and final vortex arrangement (bottom right panel). Data are taken using a normal tip of Au, and the images are of size as shown in bottom panel of a, where we highlight the paths followed by a few representative vortices. In b we show the displacement as a function of the magnetic field (top panel) for several vortices. In the bottom panel we show four images, representing the ordered bundle, the stressed bundle with small sized motion of a few vortices in the bundle and stress release with motion of nearly all vortices in the image. Data and images from Ref.\protect\cite{Guillamon11b}. Copyright (2011) by The American Physical Society.} \label{Fig25}
\end{figure}

In the W-based thin film, it is observed that motion of vortices is only related to the pinning landscape. Vortices move along the linear pinning features which separate bundles. A careful vortex by vortex analysis of the bundle in between linear pinning features gives a washboard hexagonal potential which governs the motion inside the bundle. Between configurations relaxed into the washboard potential, two characteristic behaviors are found, continous displacements and jumps. During continous displacement, the lattice disorders, accumulating stress, which is relieved when there is a collective jump (Fig.\ 25). Thus, these data show that the system organizes either in a state with slight disorder inside the bundle, and linear motion, or a state with hexagonal order in the bundle just after an abrupt jump. It is tempting to trace parallels to the random organization model proposed in Refs.\cite{Corte08,Reichhardt09}. Particles self-order themselves into configurations where they avoid irreversible interactions (linear motion in Fig.\ 25), and configurations where they go through plastic jumps (jumps in Fig.\ 25).

In the same W-based thin films, authors further observe thermally induced vortex de-pinning by imaging the changes in zero field cooled vortex arrangements induced with increasing temperature at different magnetic fields\cite{Guillamon09Nat}. For example, at 1T the thermal de-pinning of the lattice was observed at 1.5 K. To obtain this value, a sequence of 145 vortex lattice images was taken while increasing the temperature starting from 0.1K in steps of 0.1K. Fig.\ 26 shows a set of representative vortex maps with the de-pinning transition.

At 0.1K, the lattice is distorted and its morphology is determined by the surface pinning at the linear features observed in the topography. Upon increasing temperature, the disorder in the lattice remains the same up to 1.5 K. At this temperature, vortices start moving and they rearrange to form an ordered hexagonal lattice above 1.6 K. The de-pinning transition is also observed in the temperature dependence of the angular distortion of the lattice (Fig.\ 26). The transition is sharp at this location, and leads to an increase in the radial correlation length $R_c$.

\begin{figure}[ht]
\includegraphics[width=1\columnwidth,keepaspectratio, clip]{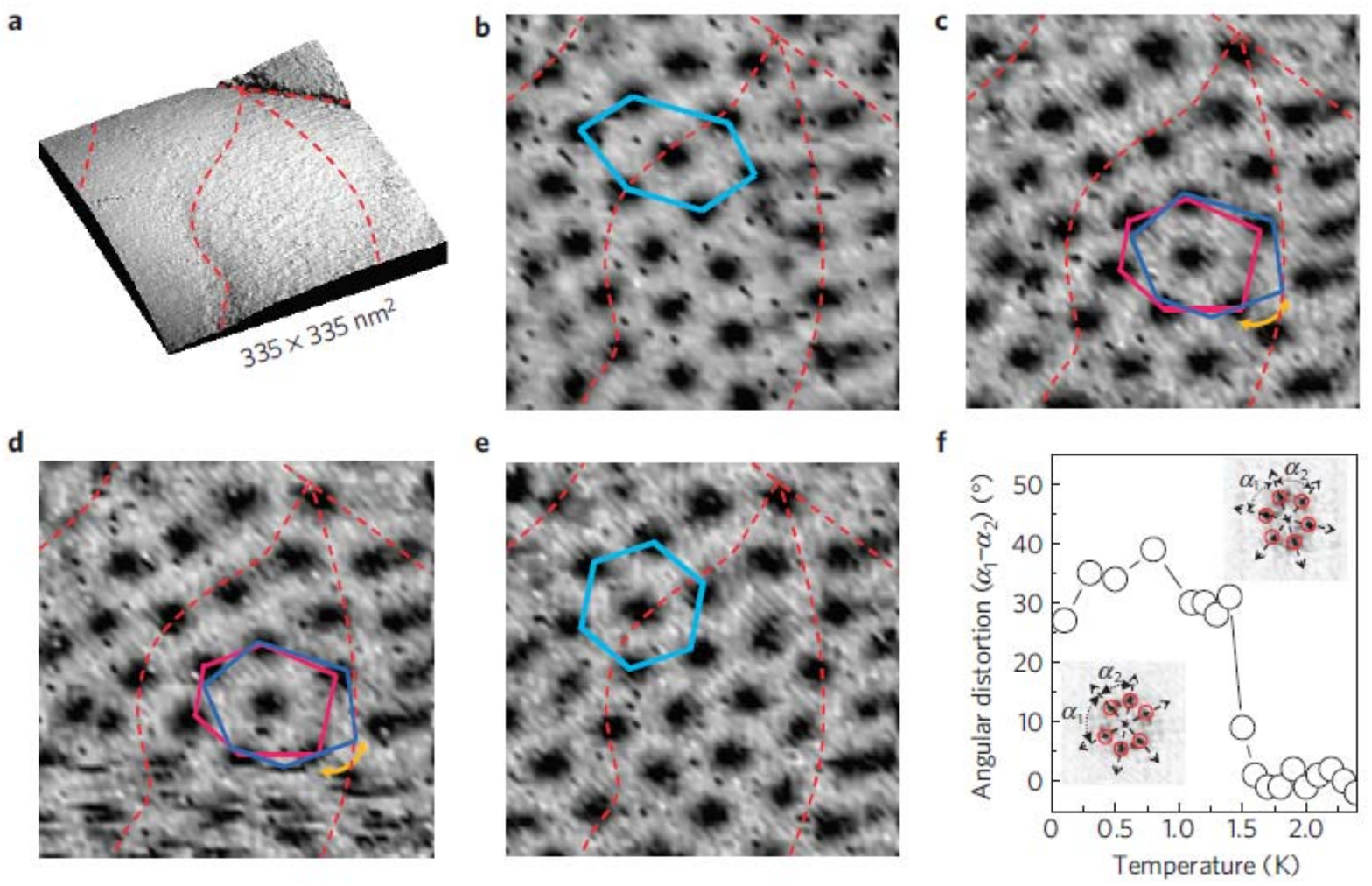}
\vskip -1cm \caption{In a we show the topography on a W based thin film, taken using a normal tip of Au at 1 T. b shows a distorted vortex lattice created after increasing the magnetic field during zero field cooling (at 0.1 K). c and d are images taken at 1.5 K and 1.6 K, respectively. In f we show the lattice angular distortion as measured in the Fourier transforms. Data and images from Ref.\protect\cite{Guillamon09Nat}.} \label{Fig25}
\end{figure}

\subsubsection{Other thin films.} Several authors have observed the vortex lattice in Nb based and in MoGe thin films. To obtain clean surface properties, authors evaporated an Au layer on top of a MoGe thin film\cite{Baarle02,Baarle03}. A disordered vortex lattice was observed. NbN caped with an Au film \cite{Nishizaki03} also shows a disordered structure. More recent measurements in as grown NbN show again disordered vortex lattices\cite{Noat13}. Interestingly, vortices were observed to vanish when studying ultra thin films (of around 2 nm thickness), indicating that the core shows similar tunneling conductance features than the surrounding material\cite{Noat13}. Using deposition techniques in a UHV chamber connected to the STM, several authors have observed vortices in ultra thin films. Authors of Ref.\cite{Karapetrov05} developed a method to fabricate atomically flat superconducting surfaces in thin films and map the vortex distribution. Vortex nucleation has been traced in single crystalline thin film and in nanosized Pb islands\cite{Zhang10,Nishio08,Cren09,Cren11,Ning09,Ning10,Tominaga12,Tominaga13}. In those cases, pinning mechanisms are probably related to impurities, step edges or proximity effect issues.

\subsubsection{Summary.} The collective behavior of vortices has been studied in bulk systems for hexagonal and square lattices, and in thin films with disordered vortex lattices. In bulk systems, motion is observed in large hexagonal or square bundles, influenced by the presence of pinning centers. Bundles move with respect to each other through dislocation gliding at their boundaries. In crystals, a tendency to disorder mixed up with influence of crystal symmetry in intervortex interaction, is observed when increasing temperature and magnetic field and approaching the critical values, leading to a decrease in the size of the ordered bundles. Pinning properties compete with the underlying crystal symmetry.

In amorphous thin films, the surface is the only pinning mechanism and thus the situation is particularly well controlled. Systems with a disordered pinning landscape created through surface corrugation have characteristic length scales allowing for the formation of bundles of 10-20 vortices. Generally, when vortex motion is produced on such a disordered lattice, the system reacts by leaving some vortices fixed. Others are mobile and give net motion\cite{Hinrichsen00,Liu10,Scheidl98,Paltiel00,Tonomura99,Banerjee04,Abulafia96}. Similar phenomena have been observed in many particle systems, and they are termed plastic de-pinning. The appearance of complex flow patterns, and their statistical behavior remains poorly understood. Near the threshold action needed to produce motion, macroscopic experiments show jumps between a state without fluctuations and a state allowing plastic motion\cite{Shaw12,Drocco11}.

Most interesting situation should be created in thin films with a very weak and controlled surface corrugation. In that case, pinning should vanish giving access to uniquely study the compression of a hexagonal lattice at zero temperature. Experiments will reveal fundamental properties of zero temperature phase transitions which cannot be accessed otherwise.

\subsection{Melting of the 2D vortex lattice.}\index{4. Imaging vortex matter!4.2. Collective behavior}

\begin{figure}[ht]
\includegraphics[width=1\columnwidth,keepaspectratio, clip]{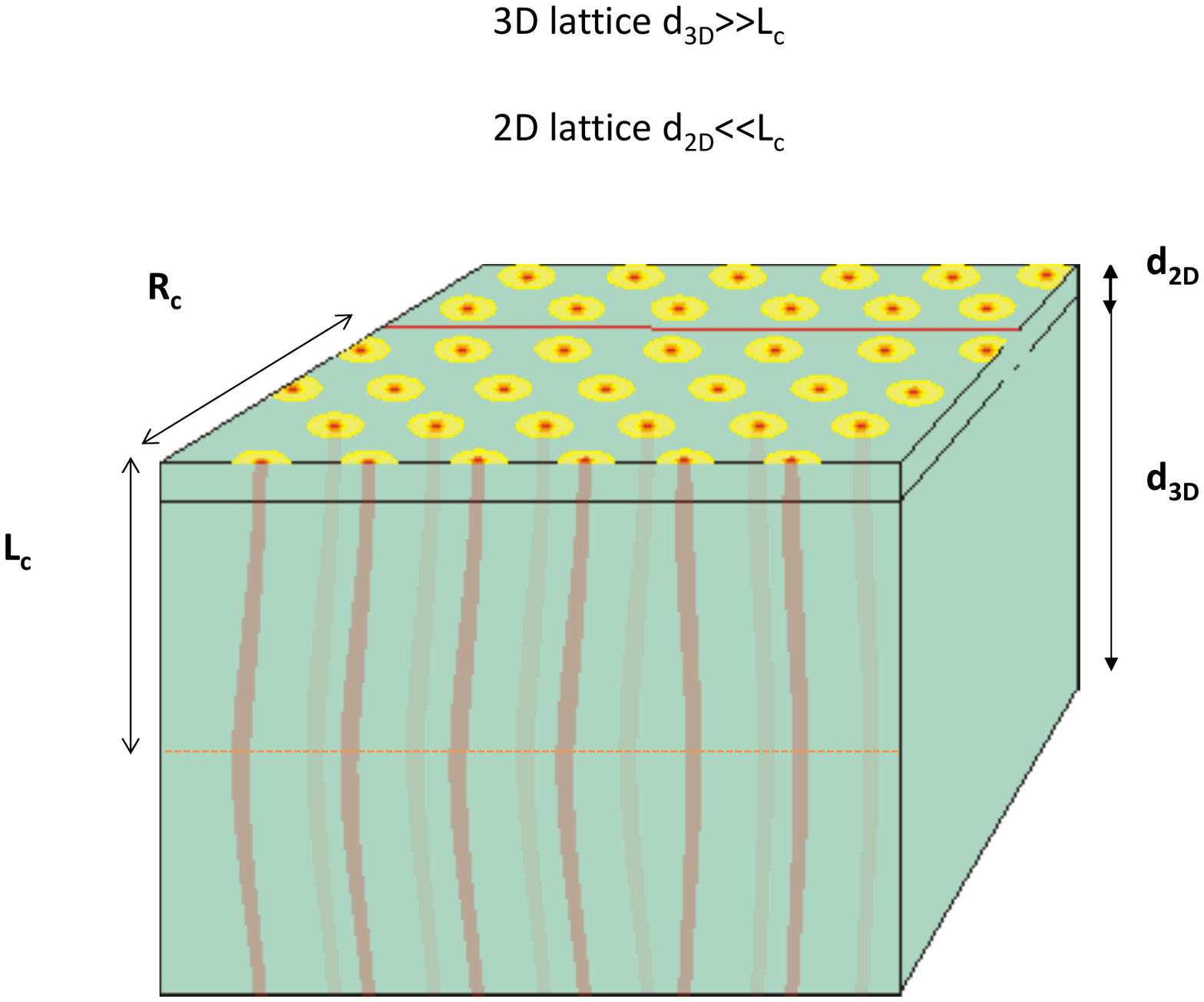}
\vskip -1cm \caption{Thin films of amorphous or disordered material with thickness below 100 nm show in most cases 2D superconducting lattice. The thickness is far below the longitudinal correlation length $L_c$, which is of order of $\lambda$, typically well above several hundred nm. Vortex lattice melting occurs as in a 2D solid.} \label{Fig27}
\end{figure}

\subsubsection{Introduction.} The melting transition has been a central issue of the debate within the experimental and theoretical scientific community since the early days of condensed matter physics\cite{Mott39}. Lindeman criterion successfully describes melting in many systems\cite{Lindemann,Kierfeld04}. Microscopically, the 3D melting transition still lacks of a widely accepted picture. 2D behavior is generally more tractable than 3D, and the phase diagram is fundamentally different\cite{Brinkman82}. Berezinskii and Kosterlitz and Thouless propose that 2D melting goes through the unbinding of pairs of topological defects, dislocations pairs\cite{Berezinskii72,Kosterlitz73}. Halperin, Nelson and Young further describe the melting transition as a continuous two-step process which is known as BKTHNY-theory of the 2D melting\cite{HN78,Young79}.  A relevant point is that long wavelength positional fluctuations make perfect order impossible in 2D, arbitrarily close to absolute zero in temperature. This was firstly proposed by Peierls and Landau during the 30's \cite{Peierls35,Landau37} and formally demonstrated by Mermin in 1967\cite{Mermin68}. Dislocations pairs are present in the 2D solid at all temperatures. When increasing temperature, they eventually start to unbind. The appearance of free dislocations destroys the quasi long range translational order and leads to the formation of an hexatic phase which retains the orientational long range order. The orientational order is lost at higher temperature at the transition between the hexatic and the isotropic liquid phase. This occurs when dislocations separate into their constituent disclinations. Disclinations are known as orientational defects since their presence into the 2D hexagonal lattice breaks the sixfold rotational symmetry. BKTHNY theory of the 2D melting took advantage of the renormalization group technique which allowed describing order-disorder transitions in the 2D $xy$ model\cite{Wilson71,Doussal10}.

2D melting has been reported in helium 4 thin films, in Bose-Einstein condensates, Wigner crystals, colloids, molecular liquid crystals and unconventional plasmas\cite{Herland13,Dalibard06,Schweikhard07,Clade09,Hung11,Waintal06,Zahn99,Chou98,Knapek13,Yazdani93}. The case of 2D vortex lattices is particularly interesting, because it occurs in a macroscopically ordered quantum state and the temperature and vortex density can be varied at will. Macroscopic measurements show temperature dependencies close to the resistive state compatible with BKTHNY\cite{Berghuis90,Yazdani93}.

2D vortex lattices appear in layered compounds with no interlayer coupling\cite{Bending99,Maniv99,Maniv01,Rosenstein10}. In these systems, until now, no STM observations of melting phenomena have been made. Vortices keep confined to the layers, and, although they can move freely within a layer, there is a magnetic interaction among layers\cite{Blatter94,Brandt93}. Thin films of amorphous superconductor provide an excellent testing bed for the 2D vortex lattice. The longitudinal correlation length $L_c$ is of order of the magnetic penetration depth, which is always of several hundreds of nm or above (Fig.\ 27). As the typical thickness is easily far below $L_c$, vortex lattices in many thin films behave as 2D lattices. Furthermore, as described above, the remaining pinning mechanism is the surface corrugation, which can be characterized directly by topographic imaging, and often controlled to some extend during fabrication. The direct visualization by STM of the melting transition in W thin films \cite{Guillamon09Nat} provides therefore new understanding of 2D melting.

\subsubsection{2D vortex lattice melting.}

2D vortex lattices are rather sensitive to thermal and quantum fluctuations as well as to deviation in vortex positions due to pinning\cite{Blatter94,Brandt95}. When increasing temperature, melting can appear. Melting is in competition with de-pinning because it involves a reduction of $R_c$. A useful parameter to understand the relevance of melting in 2D is the Levanyuk-Ginzburg number $LG$. It describes the effect of fluctuations on this phase transition, by comparing the condensation energy in a correlated volume with the critical temperature\cite{Levanyuk59,Ginzburg60}:

\begin{equation}
LG=\frac{1}{2}\left(\frac{k_{B}T_{c}(0)}{4\pi\mu_{0}H_{c}^{2}(0)\xi^{3}(0)}\right)^{2} \approx 10^{-7}\frac{\kappa^{4}T_{c}^{2}(0)}{H_{c2}(0)}
\end{equation}

$LG$ is small in classical superconductors, but it can increase significantly in thin films with a reduced mean free path\cite{Kokubo07}. In the W based FIB deposited thin films, the mean free path is of a few nm and $LG= 1.15$    $10^{-4}$\cite{Guillamon09Nat}. This value is relatively high and comparable to the value found in some high T$_c$ materials, which means that there can be a small region, close to T$_c$, where the 2D vortex lattice melts into a liquid.

\begin{figure}[ht]
\includegraphics[width=1\columnwidth,keepaspectratio, clip]{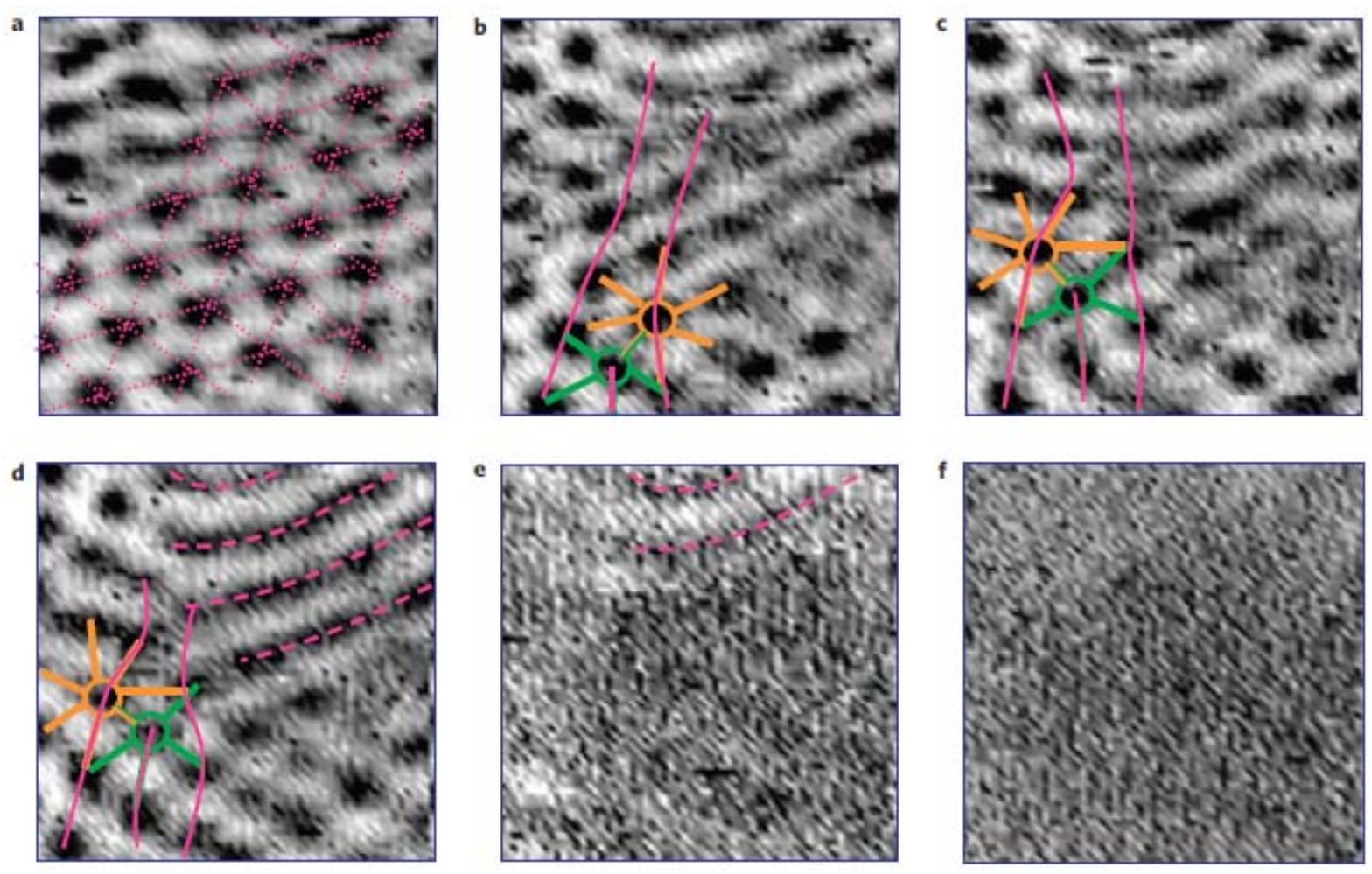}
\vskip -1cm \caption{In a we show an ordered vortex lattice at 1.2 K. When increasing the temperature to 1.9 K, a dislocation, highlighted by the orange and green circles, appears on the bottom of the image. Increasing further the temperature leads to a continous transition into the isotropic liquid (f), going through a smectic-like phase with one-dimensional vortex motion (d and e). Images taken using a normal tip of Au and have a lateral size of 220 nm. Data and images from Ref.\protect\cite{Guillamon09Nat}.} \label{Fig27}
\end{figure}

In Fig.\ 28  we show results at 2T.  Authors identify four different vortex phases (Fig.\ 29). At the lowest temperatures (1.2 K in Fig.\ 28), vortices form a hexagonal ordered 2D vortex solid. At slightly higher temperature a free dislocation appears into the lattice depleting the positional order. This signals the presence of the hexatic phase of BKTHNY theory. Thermally induced motion of vortices is further increased at higher temperatures. Vortices start moving preferentially along one of the high symmetry directions of the lattice and form 1D striped arrangements of moving vortices. This is a new smectic-like phase.

At temperatures much closer to $T_c$, vortices disappear from the images and homogeneous superconducting features, i.e. a slightly decreasing conductance close to zero bias, are found over the whole scanning area. This corresponds to the isotropic vortex liquid. Thermally excited motion is so large that the vortices are washed out at the time scale of imaging. The amplitude of the Bragg peaks in the Fourier transform of the conductance maps decreases when increasing temperature. In the isotropic liquid (Fig.\ 28), it reaches zero, whereas the spatially averaged superconducting signal in the conductance remains finite.

It is particularly remarkable that the smectic-like phase coexists with solid (Fig.\ 28 d) and liquid (Fig.\ 28e) phases. Sliding vortex phases have been predicted close to the melting temperature in different scenaria\cite{Maniv99,Maniv01,Zhuravlev06,Reichhardt11}, and can appear in presence of periodic pinning which competes with vortex repulsion, or as a consequence of the interplay between disorder and the lattice symmetry.

Note that the conductance in the isotropic liquid phase is largely gapless, with a high amount of states at the Fermi level; still, there is a small dip at voltages of order of the gap, which evidences superconducting correlations. The question about the nature of these correlations is interesting, because they are due to Cooper pairing in a highly fluctuating environment.

\begin{figure}[ht]
\includegraphics[width=1\columnwidth,keepaspectratio, clip]{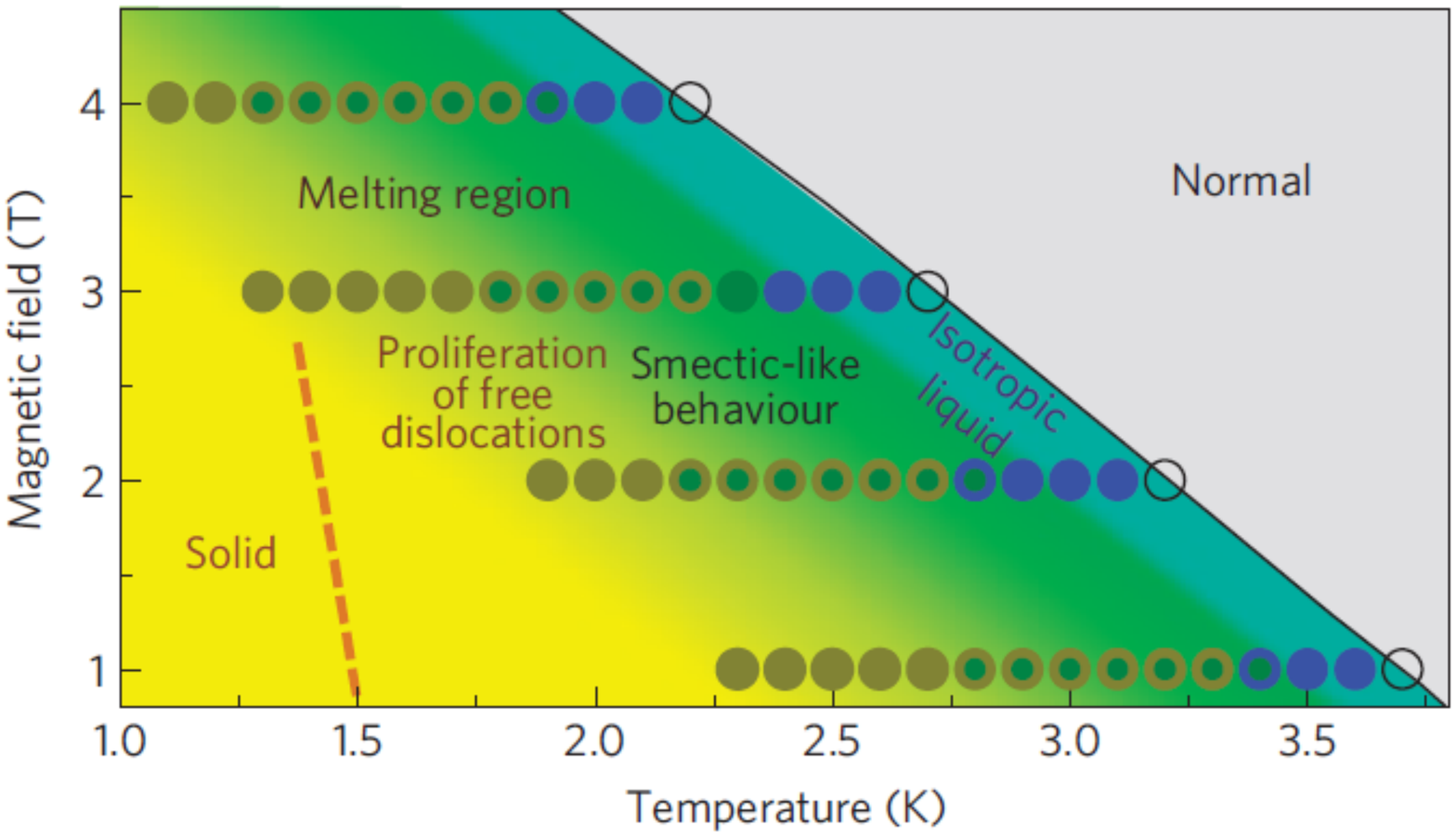}
\vskip -2cm \caption{Phase diagram for the 2D vortex lattice. The low temperature and higher field part of the phase diagram, as well as the current axis, are still unkown. Data and images from Ref.\protect\cite{Guillamon09Nat}.} \label{Fig24}
\end{figure}

In other superconductors, in particular in the cuprate materials, the melting transition has been shown to be first order, with a discontinous jump in the magnetization, transport and other properties\cite{Zeldov95,Paltiel00b}. The continous transition observed in thin films, as opposed to the first order transition of the cuprates, is a consequence of reduced vortex lattice dimensionality and absence of the interlayer coupling found in layered crystalline systems. Monte-Carlo simulations agree with this point\cite{Iaconis10}.

\section{Outlook}\index{6. Outlook}

\subsection{Liquid phases}\index{6. Outlook!6.1. Dynamics and liquid phases}

The direct observation of vortices in liquid phases in Ref.\cite{Guillamon09Nat} suggests imaging the vortex liquid as a future direction of work. In particular, the next step will be to make images involving a large amount of vortices, in order to access exponents regarding positional and orientational correlation. STM will allow addressing the whole phase diagram, in particular crossover regimes at very low temperatures, and possibly gap structures above T$_c$.

Studying vortex physics when increasing disorder in a thin film towards the insulating phase is another promising avenue. Anderson's theorem \cite{A59} shows that s-wave superconductivity is robust to moderate amounts of disorder created by defects or non-magnetic impurities. But if disorder is too strong, the superconductor can be eventually localized into an insulator. Early experiments showed that when the resistance of very thin Pb, Bi and Sn superconducting films increases above the quantum of resistance, it undergoes a zero temperature transition into an insulator\cite{Strongin70}. This behavior has been refined, with a wealth of new effects appearing when changing disorder level or magnetic field. One sound idea is that the transition occurs between pinned vortices with a condensate of Cooper pairs (superconductor) and a vortex condensate with localized Cooper pairs (insulator)\cite{Fisher90,Bruder05}. The latter would be a Bose quantum liquid, which has been addressed through many macroscopic experiments, yet without clear-cut direct microscopic observation. For instance, mirror reversal symmetric curves were observed in Titanium Nitride (TiN) and Indium Oxyde (InO) films, with a critical current in the superconducting and a critical voltage in the insulating state\cite{Baturina07,Vinokur08,Ovadia09}. The presence of a peculiar behavior of the magnetoresistance, which includes a giant peak, has been also associated to charge-vortex duality, as well as experimental evidences for Cooper pairs within the insulating phase\cite{Sherman12}. At zero field, a strong suppression of quasiparticle peaks is observed\cite{Sacepe08,Sacepe10,Sacepe11}.

\subsection{Vortex core and collective behavior under an applied current}\index{6. Outlook!6.1. Dynamics and liquid phases}

Recently, a new scanning mode, termed Current Drive STM has been developed. It allows to make tunneling conductance imaging with an applied current through the sample\cite{Maldonado11,Maldonado13} (Fig.\ 30). This gives the possibility to study the moving lattice. The effect of a supercurrent on a vortex is not just the Lorentz force\cite{Kopnin02,Sonin97}. The Lorentz force pushes vortices perpendicular the current, but vortices feel also forces from the pinning environment, a drag force due to interaction effects of the moving normal quasiparticles in the core, and Magnus force from current flow. Vortices are not expected to move exactly perpendicularly to the applied current, but at an angle depending on the ratio between forces. At a macroscopic level, the frictional longitudinal part of the force leads to dissipation and a longitudinal voltage. The transverse component leads to a Hall effect. Many macroscopic measurements, mainly electric and thermal transport as well as thermodynamic studies, have addressed the problem of vortex motion\cite{Caroli67,Banerjee04,Blatter94,Brandt95,Pautrat12}. Usually, the appearance of a voltage above a $\mu$V, is considered as the point where electrical dissipation sets in. Even if this voltage is small at a macroscopic level, it implies motion of many vortices. An interesting question is to understand how this motion looks like at the level of the individual vortices.

At very low temperatures, the experiments will allow studying vortex jamming when increasing density\cite{Liu10}. Further, one can also switch the current on/off and look at vortex arrangements. In a disordered pinning potential at the right magnetic field, this can unjam and jam the vortex lattice. Both issues are of a wider relevance than vortex physics\cite{Nagel05}.

\begin{figure}[ht]
\includegraphics[width=\columnwidth,keepaspectratio, clip]{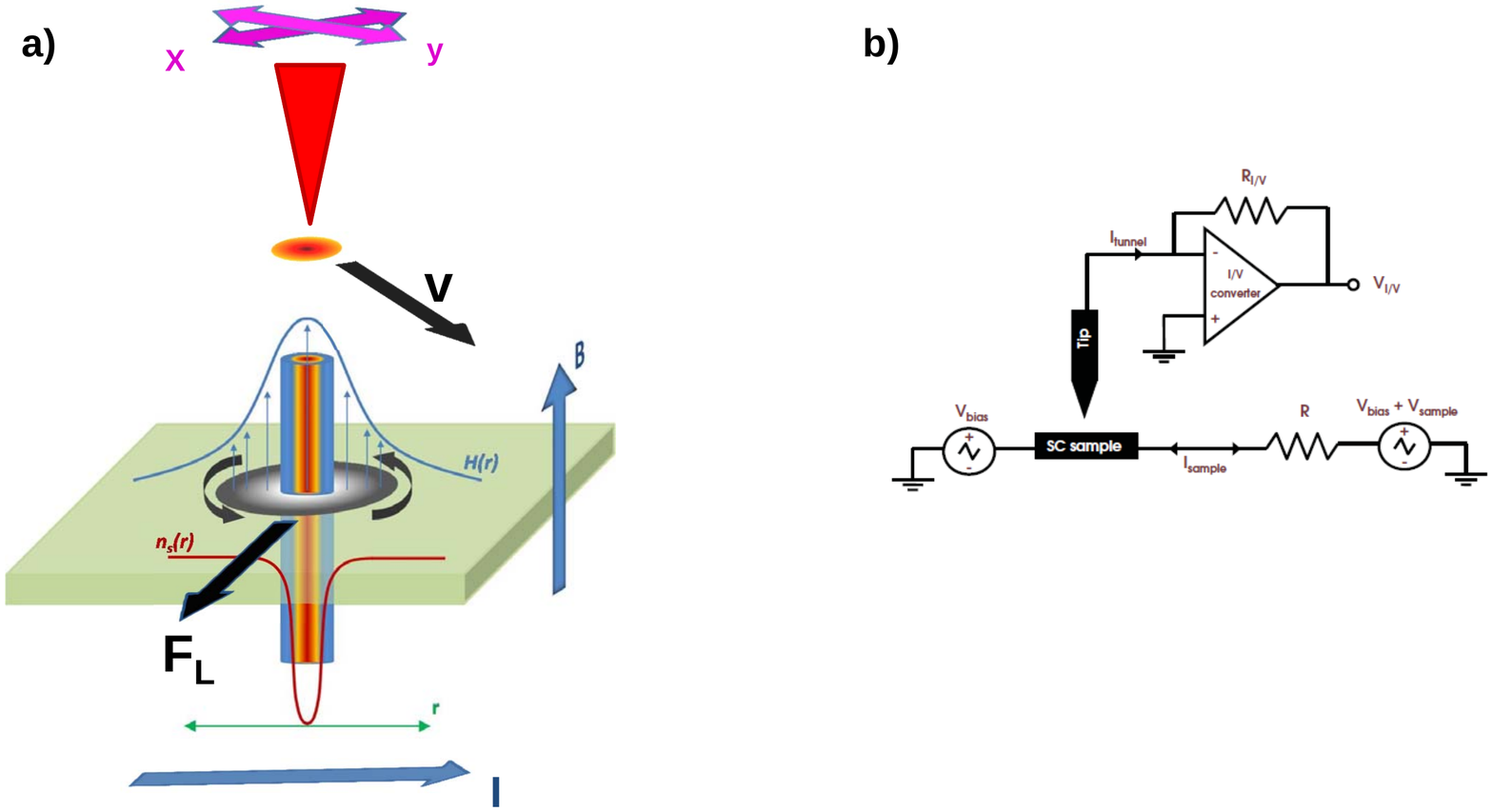}
\vskip -3cm \caption{In a we schematically show forces acting on a moving vortex. A vortex moves at an angle under the action of a current. The Lorentz force acts on the vortex, in addition to pinning. Even in the absence of pinning, longitudinal friction and transverse forces act on the vortices. In b we show schematically the change required to be able to drive a current through the sample. Two bias voltage points, which must be compensated during the ramp are needed to acquire the full current-voltage characteristics, and to maintain always a constant current through the sample. The latter figure is adapted from Ref.\protect\cite{Maldonado11}.} \label{Fig30}
\end{figure}

Another question to address using this technique is the current flow through the vortex. It has been shown that bound states inside the vortex cores can carry a current. To discuss this in simple terms, let us view conduction through a single Josephson junction between two equal superconductors. The Josephson current between both superconductors flows through the establishment of localized Andreev bound states at the junction\cite{Rev.Nic}. These lead, at a finite bias voltage, to the multiple Andreev reflection features characteristic of junctions between two superconductors, which consist of resonances located at $2\Delta/n$, where $n$ is the order of the resonance and $\Delta$ the superconducting gap\cite{RevAgrait,T96,Setal98,Suderow00c,Schmidt}. These resonances are due to virtual bound states consisting of multiple Andreev levels (of order $n$) in the junction, and provoke a finite subgap current and Cooper pair transport across the junction. At zero bias, Josephson Cooper pair transport can be understood as a transport process through these resonances\cite{RevAgrait,Goffman00}. Andreev states inside the cores can be roughly understood in a similar way. These states have been related to the symmetry of the wavefunction\cite{Volovik93,Nakai06,Kaneko12}. Moreover, it has been shown that they can carry a current\cite{Rainer96,Kopnin98}. Recent STM work in Ref.\cite{Maldonado13}, and calculations in Ref.\cite{Berthod13} show that the core localized state is influenced by a small current in pinned vortices, leading to a reduction of the zero bias peak and of the anisotropic core features. The reduction is due to the influence of the current in the superconducting properties between cores, demonstrating that there is an intimate connection between the core and the surrounding superconductor.

On the other hand, the importance of core states in explaining dissipation due to vortex motion was early recognized in Ref.\cite{Bardeen65,Nozieres66,Schmid66,Caroli67} and further worked out in Refs.\cite{Kes90,Stone96,Stone00,Guinea95,Guinea96,Kopnin96,Kopnin00b,Kopnin02,Silaev12b}. It has been shown that there is an intimate connection between core states and the surrounding superconductor\cite{Bardeen72,Heida98,Sonin13,Rainer96}. Core levels influence vortex motion and pinning. In Ref.\cite{Stone96} it is shown that, in presence of a current, the vortex may remain stationary because momentum can flow into the core through Andreev reflection. Scattering between core levels has been described in a simple Boltzmann approximation by a scattering rate $\tau$. The relationship between core level separation $\delta$ and electron scattering $1/\tau$ allows to separate different regimes for dissipation. A more microscopic understanding of the vortex core through imaging will provide new methods to manage vortex dynamics.

\subsection{Multiband superconductivity.}

The discovery of superconductivity in MgB$_2$ in 2001 came just at the point where experimental and theoretical tools were ready to open up the world of Fermi surface complexity into wide spread discussion by many groups\cite{Nagamatsu01}. The clarity of the two-band behavior is unique in MgB$_2$. Similar features have been since then found in many other superconductors, in particular in several iron-pnictides\cite{Zehetmayer13,Hoffman11,Hirschfeld11}. Initially available data gave very clean tunneling spectroscopy measurements in MgB$_2$ with a gap whose size is three times below the value expected from BCS\cite{Rubio01,Martinez03}. This pointed out strongly Fermi surface sheet dependent superconductivity, and sparked the theoretical treatment of this material as a two-band superconductor\cite{Liu01,Bascones01}. Other tunneling spectroscopy data provided smeared superconducting features\cite{Karapetrov01} which were then subsequently resolved into two gap features\cite{Iavarone02,Martinez03}. This and further experiments and theories are reviewed in Refs.\cite{Zehetmayer13,Hoffman11}

The vortex lattice in MgB$_2$ has been observed with STM\cite{Eskildsen02,Eskildsen03}. The lattice has been shown to be hexagonal. At low fields, a state with flux-free areas and vortex clustering in other areas has been observed\cite{Moshchalkov09,Gutierrez12}. The features and geometries have been associated to the two-band superconducting properties in this material, coining the term "1.5-type" superconductivity to highlight the difference with respect to known classical type I (pure Meissner) and II (pure Shubnikov phase) behaviors. In particular, an attractive interaction due to qualitatively different screening behavior of electrons in each of the two main subgroups of bands is been discussed \cite{Silaev12,Babaev09}.

We believe that high field vortex structures in this material have not yet been fully explored. No core states have been observed, and the lattice can show disordering close to the peak effect regime or orientational changes due to Fermi surface symmetry\cite{Zehetmayer13,Chaves11}. For instance, neutron scattering measurements on MgB$_2$ have given evidence for a continuous reorientation transition in the hexagonal vortex lattice with increasing the field perpendicular to the ab-plane\cite{Cubitt03}. This transition has been associated to the field-induced disappearance of superconductivity in the $\pi$-band which is also consistent with a strong decrease of the scattered intensity observed along the transition. In particular, it has been suggested that the hexagonal vortex lattice gradually rotates by 30$^{\circ}$ from an orientation dominated by the screening currents in the $\pi$-band to another favoured by the $\sigma$ band in the field range between 0.5 T and 0.9 T.

\subsection{Topological superconductivity.}

Several theoretical calculations discuss the influence of the topological properties of the superconductor on the Andreev core states\cite{Beenakker13}. A truly 2D interface superconductor is analyzed, with 2D surface electrons behaving as massless Dirac fermions. When such a superconductor is brought on top of a topological insulator, and a vortex is nucleated in the superconductor \cite{Hasan10,Qi11}, the lowest energy state of the vortex core is brought to the Fermi level. Energy quantization becomes $E_n=n\delta$ instead of $E_n=(n+\frac{1}{2})\delta$ (with $\delta\approx\frac{\Delta^2}{E_F}$).  

Related proposals are made for topologically non-trivial superconductors such as chiral $p_x\pm ip_y$ superconductivity. Few materials are known which could show such a superconducting behavior. Ferromagnetic superconductors with triplet pairing show superconductivity at very high magnetic fields\cite{Aoki2001}. The symmetry of the Cooper pair wavefunction is to be determined. The material Sr$_2$RuO$_4$ has been discussed since long as a candidate for a chiral p-wave superconductor\cite{Mackenzie03}. Magnetic phase diagrams, thermal measurements and several angle resolved and phase sensitive experiments are compatible with Fermi surface sheet dependent superconducting gap\cite{Mackenzie03}. Residual values for the electronic density of states close to zero temperature are very small in practically all thermal experiments, and are consistent with the nicely opened gap observed in STM\cite{Suderow09njp}. Vortices compatible with spin triplet correlations have been observed using Scanning SQUID microscopy\cite{Hasselbach05,Hasselbach06}, although no vortices have been observed using STM\cite{Suderow09njp,Firmo13}.

Recent work suggests a different approach to locate topological features at surfaces \cite{Rodrigo12,Rodrigo13}. Authors measure the Josephson effect in a system made of two asymetric conical Pb nanostructures, joined by an atomic size neck with a small amount of conduction channels. In a few cases they find that, above the critical field of Pb, the zero bias current first drops and then increases again. This observation is difficult to reconcile with conventional BCS theory. Above the critical field, superconductivity in the conical nanostructures is gradually suppressed\cite{Petal98,Misko01,Suderow02,Rodrigo04b,Suderow00c}. In Refs.\cite{Rodrigo12,Rodrigo13} it is proposed that a system with a small amount of channels, a small magnetic field induced Zeeman splitting and spin-orbit coupling can lead to gap opening with an odd number of bands crossing the Fermi energy. Further, if the Fermi energy, which depends on the geometry and environment of the junction, is located within the gap, topological superconductivity may arise in one of the conical nanostructures. In that case, at the boundary, a zero energy resonance can form in one of the conical structures. The resonance can eventually hybridize with a conventional N-S Andreev resonance at the other conical structure. In absence of coupling between the two conical structures, the resonance has all the features of Majorana fermion\cite{Leijnse12,Alicea12,Beenakker13,Mourik12}.

The advantage of this proposal is its simplicity, and is thus promising for future work. The control over the environment of the junction can be improved by forming many conical structures, using methods described in Refs.\cite{Petal98,Rodrigo04b}. The work shows that a low amount of separated conducting channels is favorable to the formation of anomalous local resonances.

\subsection{Unconventional superconductivity.}

Other materials with strong correlations such as heavy fermion materials UPt$_3$, CeCu$_2$Si$_2$, UBe$_{13}$ or CeCoIn$_5$ show unconventional reduced symmetry superconductivity with singlet and triplet order parameter proposals with sign changes across different parts of the Fermi surface \cite{Pfleiderer09,FlouquetRoad}. In some heavy fermion materials such as URu$_2$Si$_2$ or PrOs$_4$Sb$_{12}$, the superconducting gap has been characterized by scanning tunneling spectroscopy at zero magnetic field\cite{Maldonado12URu2Si2,Suderow04}. The superconducting gap is nicely opened, although with low energy excitations in PrOs$_4$Sb$_{12}$ and a weak V-shaped form in URu$_2$Si$_2$. The surface of CeCoIn$_5$ shows quasiparticle interference \cite{Zhou13,Allan13} giving band dispersion compatible with unconventional anisotropic superconductivity.  As discussed above, vortices have been observed yet only in CeCoIn$_5$, with a four-fold core which is related to the gap anisotropy.

There are other systems where magnetism coexists with superconductivity such as ErNi$_2$B$_2$C or TmNi$_2$B$_2$C\cite{Canfield03}. The rare earths in these compounds possibly interact through exchange mediated by the superconducting electrons. Atomic resolution, and a structural transition from hexagonal to square lattice was observed in the vortex lattice of TmNi$_2$B$_2$C\cite{Guillamon10}. Gapless superconductivity was observed in ErNi$_2$B$_2$C\cite{Crespo06a}. Although these systems are clearly s-wave superconductors, exotic vortex core behavior can appear due to the long range magnetic order coexisting with superconductivity\cite{Buzdin05}.

\subsection{Nanostructures.}

The observation of vortex lattice in nanostructured superconductors has been made at low magnetic fields. At high fields, the vortex density increases and the intervortex interaction weakens\cite{Blatter94,Brandt95}. The vortex lattice softens and thus jams previously mobile vortex arrangements. This allows to study the interplay between jamming and nanoscale geometry.

\begin{figure}[ht]
\includegraphics[width=\columnwidth,keepaspectratio, clip]{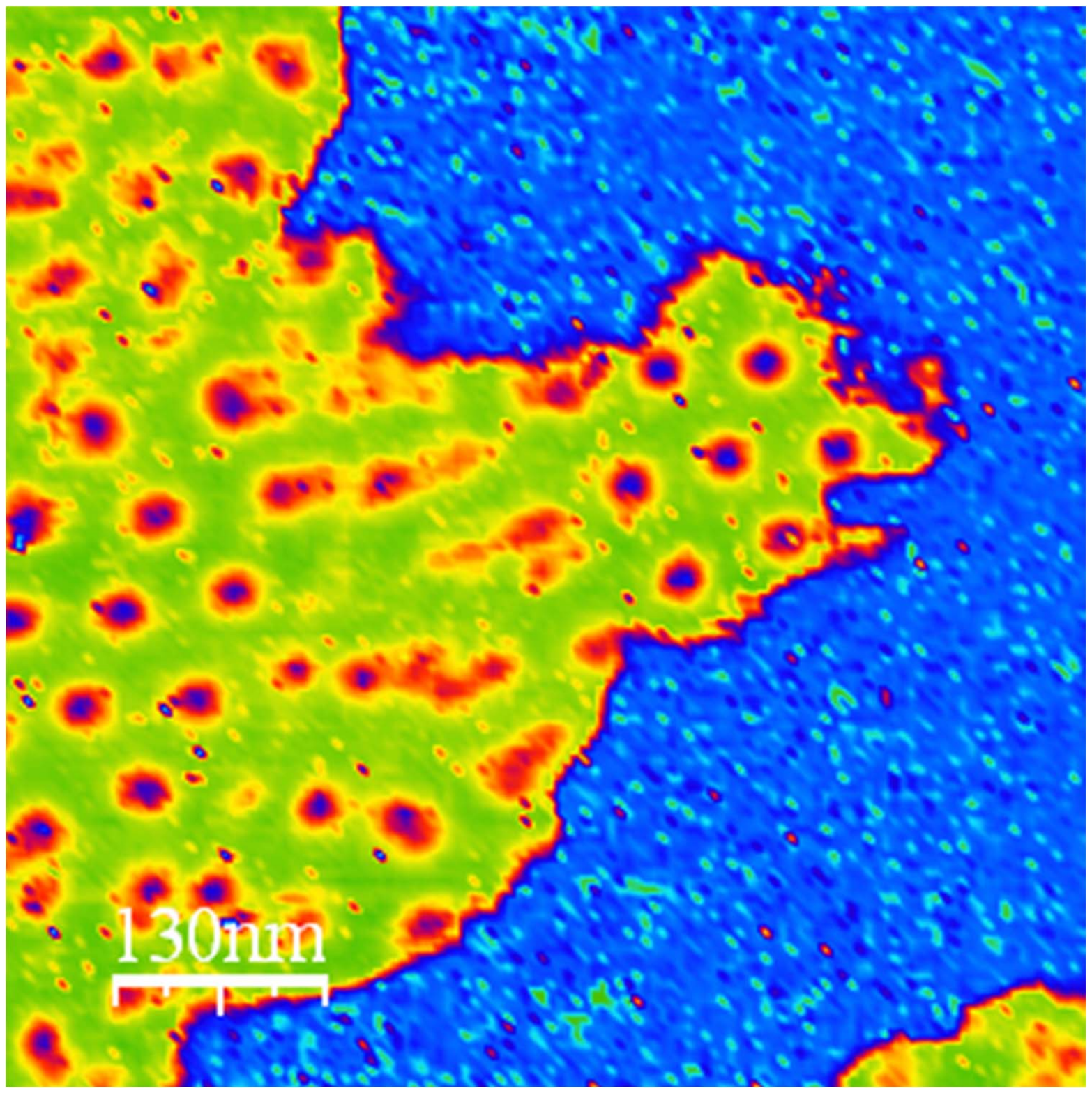}
\vskip -0cm \caption{Vortex lattice image taken at the border of a W-based focused ion beam deposited thin film using a tip of Au. Blue is a normal Au substrate, yellow is the superconductor, and dark blue to red are vortex cores. Note the formation of a perfect hexagon within the superconducting peninsula at the right part of the image. The applied magnetic field is of 0.7 T.} \label{Fig31}
\end{figure}

Technically, there are two challenges to address. First, to obtain nanostructures with enough surface quality allowing to tunnel into them. Second, to be able to locate the nanostructures in-situ and navigate across them. The first challenge has not been solved fully. Tries with nanostructures fabricated using resin based nanofabrication techniques and classical materials such as Al, Pb or Nb have been found to leave traces which make local tunneling difficult. An experiment aiming at studying small superconducting disks was reported in Ref.\cite{Zalalutdinov00}. Some success has been obtained by combining STM and atomic force microscopy, the latter used to image and locate the nanostructures\cite{LeSueur08}. However, the tunneling junction is then not a vacuum junction, which makes it difficult to obtain proper imaging conditions. 

Using deposition of stable materials and resin free nanofabrication allows designing structures with different geometries. Focused ion beam film deposition of superconducting thin films is an interesting route to create nanostructures for STM studies\cite{Guillamon08b}. In Fig.\ 31. we show a vortex lattice image taken on a superconducting nanostructure structure. Using an in-situ positioning mechanism\cite{Suderow11}, we have found an island containing only one single vortex hexagon, which matches perfectly to the size of the island. Proximity junctions were also studied in Ref.\cite{Guillamon09LT}, observing the form of the tunneling conductance curves when crossing a N-S interface. Interestingly, authors observe vortices close to the N-S interface, showing that surface barrier for vortex entry is significantly reduced with a N layer in contact to the surface of the superconductor.

In nanostructures, multiquanta vortices can nucleate under appropriate conditions, as well as vortex-antivortex pairs\cite{BookVictor}. This can significantly alter vortex core shapes. In principle STM is insensitive to the size and direction of the magnetic field. Thus, it is generally assumed that this technique cannot be used to detect these structures. However theoretical calculations show that the core level Andreev structure is sensitive to the phase winding properties around the core\cite{Virtanen99,Berthod05}. Thus, core properties are expected to be different in multiquanta vortices and in vortex-antivortex pairs\cite{Virtanen99,Berthod05,Melnikov,Geurts10}. To observe these features, nanostructuring of clean crystals with pronounced zero bias anomalies at the core are needed. Experiments in 2H-NbSe$_2$ and related materials could lead to the observation of vortex-antivortex differences in the Andreev core level structure.

Nanostructured superconductors are also expected to show interesting effects in vortex arrangements of high density lattices. Recently, a re-entrant superconducting behavior was found using transport measurements when increasing the magnetic field in nanostructures at magnetic fields above a Tesla\cite{Cordoba13}. A suppression of the resistance between 4 and 7 orders of magnitude is observed in a wire and in a perforated thin film (Fig.\ 32). The resistance remains practically zero over a field range as large as a Tesla. These observations lead to resistance variations which go in the same direction as the ubiquitous peak effect features measured in many systems\cite{Blatter94,Brandt95}. But the observed size is well above any previous observations of such an effect. Early work in long superconducting bridges \cite{Benacka77,Takacs78} shows that, in some cases, a Josephson effect re-appears at magnetic fields much higher than those expected from known material's properties, although still an order of magnitude below the field where re-entrance is reported in Ref.\cite{Cordoba13}. To explain the re-entrance of Josephson features, authors of Refs.\cite{Benacka77,Takacs78} propose that the one dimensional chain of vortices is stable to the action of a magnetic field and a current. Calculations are provided to support this proposal. The stability enhances superconducting properties, and can be far stronger than pinning effects. In Ref.\cite{Cordoba13}, the re-entrance of superconductivity is imposed in nanostructures by the presence of a small amount of confined vortices. Vortex mobility, is reduced by confinement and core merging. The origin of the new effect is the establishment of dense vortex packings confined by surface superconductivity. Imaging should address this behavior, through patterning of appropriately sized structures in thin films.  The unexpected reduction of the resistance is a result of the effect of geometry and surface properties of the superconductor on vortex transport. The observations show that vortex confinement can lead to a significant reduction of their motion.

\begin{figure}[ht]
\includegraphics[width=\columnwidth,keepaspectratio, clip]{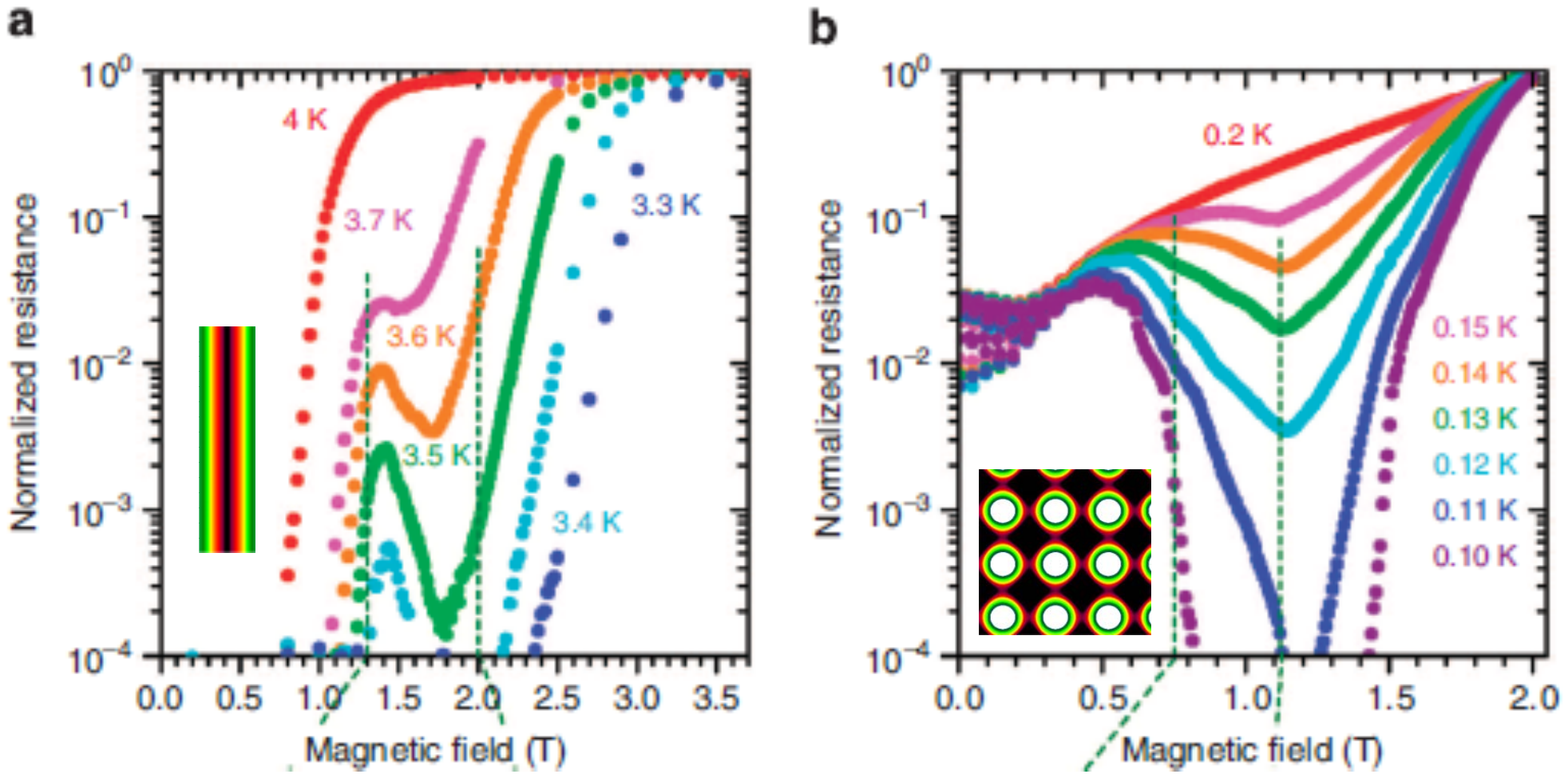}
\vskip -4cm \caption{In a we show the resistance vs magnetic field in a long focused ion beam deposited W nanowire. Note the drop in the resistance by several orders of magnitude. In b we show the drop in the resistance observe in a perforated TiN thin film. In the insets we show the proposed form of the order parameter in higher field part of the re-entrant region. In the wire, the surface remains superconducting (green) and a single vortex row is present at the center of the wire. When the field increases, the contrast between vortices is expected to reduce. In the perforated thin film, the remaining superconducting area also lies on the surface of the superconductor (green). Figure is adapted from Ref.\protect\cite{Cordoba13}.} \label{Fig32}
\end{figure}

Finally, nanostructures arise naturally in single and few layers systems of Pb, grown in UHV conditions. Pb grows as nanosized islands with internal voids and a circular structure whose geometry can be chosen among the many different islands which typically appear when imaging large surfaces. Very nice vortex nucleation and confinement experiments have been made in several experiments in these nanostructures \cite{Nishio08,Cren09,Ning10,Tominaga12,Tominaga13}. Vortex confinement in \cite{Cren09}, for instance, is related to the interplay between vortex nucleation and edge circulating supercurrents. These extremely small structures are starting an interesting line of studies regarding nanosized vortex nucleation and pinning.

\subsection{Van der Waals materials.}

2D superconducting behavior can have anomalous superconducting properties in van der Waals materials. It is an interesting development, which can possibly lead to anomalous vortex features. Therefore, we believe that it is useful to make a few comments about the recent advances in this field.

Layered systems have quasi-2D electronic correlations and often show Fermi surface tubes with practically small out of plane dispersion\cite{Wang12,Geim13}. Among these, the transition metal dichalcogenides, such as 2H-NbS$_2$, 2H-NbSe$_2$, 2H-TaS$_2$ and 2H-TaSe$_2$, have the advantage that they can be produced in high quality single crystalline form and their surface can be easily prepared through exfoliation. Furthermore, band structure calculations have been made in detail and compared to band dispersion photoemission measurements\cite{Rossnagel11,Johannes06,Faraggi13}. While the vortex lattice has been measured in 2H-NbS$_2$ and in 2H-NbSe$_2$, its properties are still unkown in 2H-TaS$_2$ and 2H-TaSe$_2$, where the layer separation is largest. Their low temperature properties have been shown to be rather peculiar. 2H-TaS$_2$ shows in-plane anisotropic charge patterns which could produce interesting superconducting properties\cite{Guillamon11,Wezel12}. 2H-TaSe$_2$ is the material of the series which has mostly marked two-dimensional properties. Further materials showing quasi 2D features are the diantimonides, such as LaSb$_2$ \cite{Budko98,Galvis13b}. The observation of vortices has not yet been achieved in the latter three systems (2H-TaS$_2$, 2H-TaSe$_2$ and LaSb$_2$), as the areas showing superconducting features are relatively small.

\begin{figure}[ht]
\includegraphics[width=\columnwidth,keepaspectratio, clip]{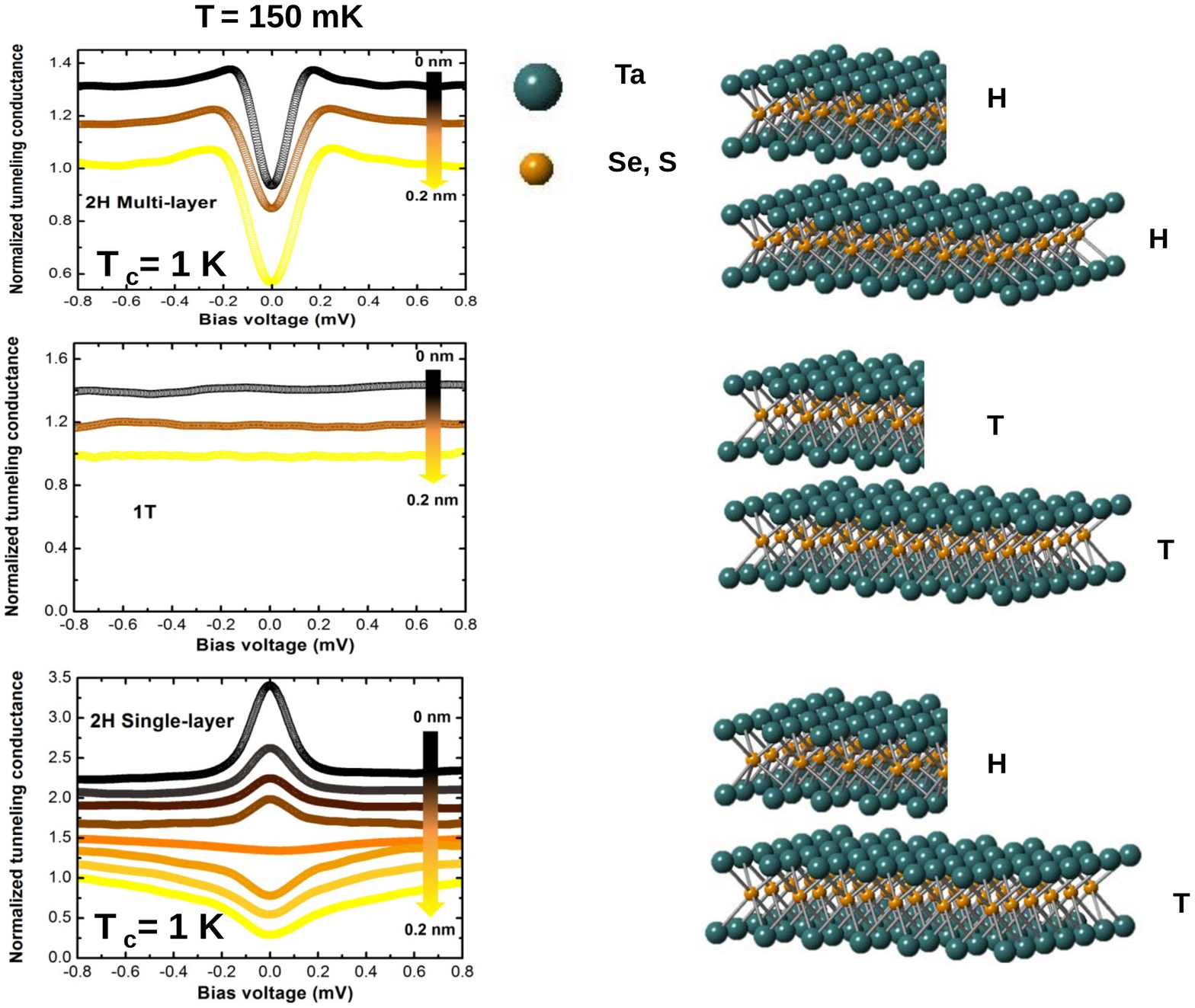}
\caption{Interface superconductivity at the surface of layered TaSe$_2$. At the surface of TaSe$_2$, the Se-TaSe structures (right panels) arrange either in prisms or antiprisms, giving a hexagonal a trigonal unit cells. The coordination of the transition metal atom is trigonal prismatic and octahedral, respectively. Accordingly, electronic properties are strongly modified. Authors find situations with hexagonal layers and large superconducting features (top panels), situations with trigonal layers and no superconductivity at all (middle panel) and situations where the layers have a mixed character (bottom panels). The latter shows the emergence of a remarkable zero bias peak, highlighting anomalous superconducting properties. The zero bias peak evolves along a 0.2 nm path from a Se atom to an intersite, into a shallow dip. Adapted from Ref.\protect\cite{Galvis13}. Copyright (2013) by The American Physical Society.} \label{Fig33}
\end{figure}

Isolated layers consisting of the molecular building blocks of these materials (e.g. TaSe$_2$) will open ways to obtain new superconductors. Their fabrication has been largely discussed in literature and presents some unsolved challenges, as for instance the increase of the reactivity when approaching few layers\cite{Geim13}. Creating heterostructures which isolate the 2D behavior of single layers or build up 2D interface electronic properties allows to solve some of those problems. One way is to exfoliate the material and search for areas with surfaces of markedly different properties. This was accomplished in Ref.\cite{Galvis13}, where authors have been able to find surfaces with single layer superconducting behavior. T$_c$ is considerably increased, from 150 mK to 1 K at the surface (Fig.\ 33). Furthermore, the bulk superconducting gap does not show up. Instead, a strong zero bias anomaly, which is temperature and magnetic field dependent, is observed\cite{Galvis13}. The way such an anomaly would evolve close to or inside a vortex is an intriguing question which may be addressed in future experiments.

\section{Summary}\index{7. Summary}

We have reviewed experiments about the observation of the vortex lattice in real space at high magnetic fields. We have explained the experimental features and advances which have lead to improved resolution in vortex lattice imaging. We discuss imaging of the vortex core, which can be until now uniquely made using STM, and the collective behavior of the lattice.  Challenging lines for future work include imaging zero temperature transitions of the vortex lattice, the vortex liquid at high temperatures and currents, vortex cores in two-dimensional metals and nanostructures, and Fermi surface features in vortex cores of unconventional and multiband superconductors. Addressing these questions will provide new concepts for vortex pinning, and thus improve functionality of superconducting devices.

\section{Acknowledgments}
The microscopy work made in Madrid would not have been possible without the excellent staff of our UAM's support workshops, SEGAINVEX. We acknowledge discussions with F. Guinea, P.C. Canfield, J.M. De Teresa, R. Ibarra, V. Vinokur, S. Bannerjee, A.I. Buzdin, J. Flouquet, J.P. Brison, T. Baturina, T. Puig, X. Obradors, J.L. Vicent, P. Wahl, M.A. Ramos and M. Garc\'ia-Hern\'andez. We acknowledge J.A. Galvis for help in the preparation of some of the graphs shown here. We also acknowledge collaboration and discussions with A. Maldonado, V. Crespo, R.F. Luccas, M.R. Osorio, A. Fente and E. Herrera-Vasco. This work was supported by the Spanish MINECO (Consolider Ingenio Molecular Nanoscience CSD2007-00010 program and FIS2011-23488) and by the Comunidad de Madrid through program Nanobiomagnet. Victor Moshchalkov convinced us to start working on vortex matter. We are grateful for this, and acknowledge particularly the presently running COST MP1201 action.

\section{References}

%\bibliography{Lastbib_noTitle}

\end{document}